\documentclass[%
 reprint,
 amsmath,amssymb,
 aps,
 superscriptaddress,
 prb,
]{revtex4-2}

\usepackage{graphicx}% Include figure files
\usepackage{dcolumn}% Align table columns on decimal point
\usepackage{bm}% bold math
\usepackage{hyperref}% add hypertext capabilities
\begin{document}

\title{Scalable Simulation of Strongly Correlated Electron-Phonon Systems via Non-Gaussian Matrix Product States}% Force line breaks with \\

\author{Siyuan Jiang}
\affiliation{CAS Key Laboratory of Theoretical Physics, Institute of Theoretical Physics, Chinese Academy of Sciences, Beijing 100190, China}
\affiliation{School of Physical Sciences, University of Chinese Academy of Sciences, Beijing 100049, China}
\author{Tao Shi}
\email{tshi@itp.ac.cn}
\affiliation{CAS Key Laboratory of Theoretical Physics, Institute of Theoretical Physics, Chinese Academy of Sciences, Beijing 100190, China}
\affiliation{School of Physical Sciences, University of Chinese Academy of Sciences, Beijing 100049, China}

\date{\today}% It is always \today, today,

\begin{abstract}
    We investigate strongly correlated electron-phonon (e-ph) systems via a non-Gaussian matrix product state method. By combining non-Gaussian states with matrix product states, our method efficiently characterizes the intractable entanglement between strongly correlated electrons and phononic modes of unbounded Hilbert space, enabling scalable simulations across broad parameter regimes. In one-dimensional generalized Hubbard--Holstein (HH) models, we identify a pronounced tendency toward phase separation (PS), an instability relevant to recent angle-resolved photoemission spectroscopy observations on doped cuprate chain. In two-dimensional HH models, we construct the phase diagram at half-filling featuring a metallic phase emerging from the competition between non-local phonon-mediated attraction and local Hubbard repulsion. Upon doping, we elucidate the role of soft phonons in stabilizing stripe phases. In the antiferromagnet, the stabilization of the fully filled stripe is attributed to a local retardation effect, wherein the charge order is pinned by phonons, leading to a diminished response to spin fluctuations. In the doped charge-density-wave regime, a novel bipolaronic stripe phase with an enlarged unit cell is stabilized via a non-local retardation effect, where long-range phonon-mediated interactions suppress PS. Our work establishes a systematic route to decoding the e-ph interplay that is crucial for superconductivity.
\end{abstract}

\maketitle

%\tableofcontents
\section{Introduction}

% 1 — **Big picture: why phonons matter now.**
Phonons play an indispensable role in the physics of superconductivity. In cuprates, phonon signatures are ubiquitous across various experimental probes. Examples include isotope effects in scanning tunneling microscopy~\cite{leeInterplayElectronLattice2006} and superfluid density measurements~\cite{tallonIsotopeEffectSuperfluid2005}, dispersion kinks and replica features in angle-resolved photoemission spectroscopy (ARPES)~\cite{lanzaraEvidenceUbiquitousStrong2001,cukReviewElectronPhonon2005,shenMissingQuasiparticlesChemical2004,heRapidChangeSuperconductivity2018}, and phonon softening in Raman and neutron scattering~\cite{thomsenUntwinnedSingleCrystals1988,reznikElectronPhononCoupling2006}. These observations underscore the coexistence of substantial electron-phonon (e-ph) entanglement with strong electronic correlations. Notably, a recent ARPES study has revealed an anomalously strong nearest-neighbor attraction~\cite{chenAnomalouslyStrongNearneighbor2021}, directly implicating low-energy physics shaped by phonons~\cite{wangPhononMediatedLongRangeAttractive2021,quSpintripletPairingInduced2022,wangZerotemperaturePhasesTwodimensional2020,tangTracesElectronPhononCoupling2022,zhaoChebyshevPseudositeMatrix2023,thomasTheoryElectronPhononInteractions2025,chenRoleElectronphononCoupling2023,caiHightemperatureSuperconductivityInduced2025,wangRobustDwaveSuperconductivity2025}. Extending beyond equilibrium, ultrafast optical pump-probe experiments have reported transient responses indicative of superconductivity above the critical temperature~\cite{faustiLightInducedSuperconductivityStripeOrdered2011,huOpticallyEnhancedCoherent2014,nicolettiMagneticFieldTuningLightInduced2018,creminPhotoenhancedMetastableCaxis2019,buzziPhotomolecularHighTemperatureSuperconductivity2020}. While the nature of this light-induced state remains elusive~\cite{zhangPhotoinducedMetastableState2018,zhangLightinducedNewCollective2018,sunTransientTrappingMetastable2020,buzziPhotomolecularHighTemperatureSuperconductivity2020,boschiniCollapseSuperconductivityCuprates2018,lemonikTransportSpectralSignatures2019,patelLightinducedEnhancementSuperconductivity2016,michaelParametricResonanceJosephson2020}, it offers a complementary perspective on the role of strong e-ph coupling in superconductivity. Similar effects in iron-based~\cite{leeInterfacialModeCoupling2014,xiangHightemperatureSuperconductivityFeSe2012,liWhatMakesTc2016,suzukiPhotoinducedPossibleSuperconducting2019} and nickelate-based~\cite{talantsevDebyeTemperatureElectronphonon,zhanCooperationElectronPhononCoupling2025,liDistinctUltrafastDynamics2025} superconductors suggest the broad importance of phonons across diverse correlated systems.

% 2 — **Status quo and computational bottlenecks.**
These experimental observations demand a framework that treats electronic correlations and phononic fluctuations on equal footing. Such an approach must incorporate the intricate e-ph entanglement while simultaneously reconciling the complexities intrinsic to both fermions and bosons. Yet, even in the absence of phonons, solving the many-body problem for strongly correlated electrons remains a central challenge~\cite{dagottoCorrelatedElectronsHightemperature1994,keimerQuantumMatterHightemperature2015,qinHubbardModelComputational2022,arovasHubbardModel2022,leblancSolutionsTwoDimensionalHubbard2015}. The breakdown of traditional mean-field or perturbative approaches beyond the weak-coupling limit~\cite{kozikNonexistenceLuttingerWardFunctional2015,gunnarssonBreakdownTraditionalManyBody2017,reitnerAttractiveEffectStrong2020} necessitates various advanced numerical approaches. These methods have successfully elucidated experimental discoveries across weak-to-strong coupling regimes~\cite{darmawanStripeSuperconductingOrder2018,zhengStripeOrderUnderdoped2017,changSpinChargeOrder2010,corbozStripesTwodimensionalModel2011,huangStripeOrderPerspective2018,xuCoexistenceSuperconductivityPartially2024,kokaljBadmetallicBehaviorDoped2017,huangStrangeMetallicityDoped2019,jiangSuperconductivityDopedHubbard2019,liTangentSpaceApproach2023}. However, incorporating phonons not only compounds the intractability of correlated electrons but also introduces distinct complexities. Quantum Monte Carlo (QMC) simulations~\cite{blankenbeclerMonteCarloCalculations1981} are generically hampered by the sign problem~\cite{lohNUMERICALSTABILITYSIGN2005,troyerComputationalComplexityFundamental2005}. The inclusion of soft phonons further imposes substantial autocorrelation times~\cite{weberDirectedLoopQuantumMonte2017}, thereby significantly increasing the computational cost. In a distinct paradigm, wavefunction-based methods such as exact diagonalization (ED) operate in the full many-body Hilbert space, whose dimension scales exponentially with system size. This complexity is exacerbated by the unbounded local Hilbert space of the phonons. To address this, tensor network states (TNS)~\cite{orusTensorNetworksComplex2019,ciracMatrixProductStates2021,schollwoeckDensitymatrixRenormalizationGroup2011,verstraeteRenormalizationAlgorithmsQuantumMany2004,whiteDensityMatrixFormulation1992,zhangDensityMatrixApproach1998} compress the Hilbert space using structured low-rank approximations~\cite{whiteDensityMatrixFormulation1992,kochDynamicalLowRankApproximation2007,evenblyGaugeFixingCanonical2018,zhangDensityMatrixApproach1998,guoCriticalStrongCouplingPhases2012}. Nevertheless, the associated optimization remains computationally demanding and susceptible to local minima~\cite{stolppComparativeStudyStateoftheart2021,knorzerSpinHolsteinModelsTrappedIon2022,orusTensorNetworksComplex2019}, particularly in regimes characterized by dense near-degeneracies or pronounced phonon excitations.

Recently, the non-Gaussian state (NGS) approach~\cite{shiVariationalStudyFermionic2018,hacklGeometryVariationalMethods2020} has emerged as a complementary paradigm to overcome these limitations. This method has proven highly efficient across diverse domains, including quantum impurity problems~\cite{PhysRevB.97.155156,ashidaVariationalPrincipleQuantum2018,ashidaSolvingQuantumImpurity2018,ashidaEfficientVariationalApproach2019,ashidaQuantumRydbergCentral2019,dolgirevEmergenceSharpQuantum2021,quEfficientVariationalApproach2022,weiKondoImpurityAttractive2025,shiUltrafastMolecularDynamics2020}, lattice gauge theory (LGT)~\cite{salaVariationalStudyU12018}, spin glasses~\cite{schindlerVariationalAnsatzGround2022}, and e-ph systems~\cite{shiUltrafastMolecularDynamics2020, shiVariationalStudyFermionic2018,shiVariationalApproachManyBody2020,knorzerSpinHolsteinModelsTrappedIon2022}. 
Specifically for e-ph interactions, it faithfully captures the strong e-ph entanglement with appropriate canonical transformations acting on bosonic and fermionic Gaussian states. The NGS method yields accurate descriptions of equilibrium and non-equilibrium physics~\cite{knorzerSpinHolsteinModelsTrappedIon2022,shiVariationalApproachManyBody2020,shiUltrafastMolecularDynamics2020,shiVariationalStudyFermionic2018} with a two-fold computational advantage~\cite{knorzerSpinHolsteinModelsTrappedIon2022}: (i) it reproduces results from brute-force density matrix renormalization group (DMRG)~\cite{whiteDensityMatrixFormulation1992} at a significantly lower computational cost; (ii) it generates physically informed initial states that guide DMRG in regimes susceptible to local minima. Still, an unbiased description of strongly correlated electrons remains necessary. Most recently, a combination with ED has enabled reliable, albeit small-scale, investigations of Hubbard--Holstein (HH) models at zero temperature~\cite{wangZerotemperaturePhasesTwodimensional2020,wangPhononMediatedLongRangeAttractive2021,wangFluctuatingNatureLightEnhanced2021} and under non-equilibrium driving~\cite{wangFluctuatingNatureLightEnhanced2021}. 

% 3 — **This work: bridging the gap with a hybrid variational framework.**
Despite substantial progress, studying strongly correlated e-ph systems remains challenging at the large scales required to resolve competing instabilities across various ordering vectors. To overcome these challenges, we introduce a hybrid method that combines NGS with matrix product state (MPS)~\cite{schollwoeckDensitymatrixRenormalizationGroup2011,ciracMatrixProductStates2021}, i.e., NGS-MPS. This approach integrates compact parametrization with robust co-optimization workflow. The parametrization retains the non-Gaussian transformation to capture the e-ph entanglement; in the transformed frame, weakly correlated phonons are captured by a bosonic Gaussian state, while the strongly correlated electrons are encoded in an MPS. The workflow leverages a robust initialization based on pure NGS optimization to bootstrap alternating updates between NGS flows and MPS sweeps. Crucially, the method scales efficiently, enabling simulations at a computational cost comparable to that of standard fermionic MPS. Furthermore, the effective electronic Hamiltonian can be derived analytically. This feature grants direct access to the microscopic mechanisms driving phase transitions, which we exploit to interpret the rich physics of generalized HH models.

For generalized HH models in one dimension (1D), our results for phonon-mediated attraction are benchmarked against NGSED studies within the realistic soft-phonon regime relevant to a recent ARPES experiment~\cite{wangPhononMediatedLongRangeAttractive2021}. Beyond quantitative agreement, our calculations uncover a pronounced tendency toward phase separation (PS) in large-scale systems. Extending the analysis to 2D, we reproduce NGSED results~\cite{wangZerotemperaturePhasesTwodimensional2020,wangFluctuatingNatureLightEnhanced2021} on small clusters and push our simulations to previously inaccessible scales by NGSED. At half-filling, we map out the phase diagram, identifying an intermediate metallic phase sandwiched between the antiferromagnetic (AFM) and charge-density-wave (CDW) phases. The emergence of this phase is driven by the competition between non-local phonon-mediated attraction and local Hubbard repulsion, consistent with quantum Monte Carlo (QMC) studies~\cite{costaPhaseDiagramTwodimensional2020,weberTwodimensionalHolsteinHubbardModel2018,nowadnickCompetitionAntiferromagneticChargeDensityWave2012}. Furthermore, we dissect the interplay between phonons and stripe order in the computationally intractable regime of $1/8$ doping. Upon doping the AFM parent compound, we demonstrate that softer phonons further stabilize the fully filled stripe phase through a retardation effect: these phonons effectively pin the static charge order, while progressively reducing screening of the rapid spin fluctuations. Conversely, doping the CDW parent compound reveals a distinct form of retardation: long-range phonon-mediated interactions suppress PS that would otherwise be driven by local pinning~\cite{ohgoeCompetitionSuperconductingAntiferromagnetic2017}, thereby stabilizing a novel bipolaronic stripe phase characterized by an enlarged axial period of 16 lattice sites.

The organization of this paper is as follows. The section~\ref{sec:formalism} introduces the hybrid NGS-MPS method, detailing the variational ansatz and the self-consistent optimization workflow. The section~\ref{sec:applications} presents ground-state results for generalized HH models, progressing from 1D~(Sec.~\ref{sec:1D}) to specific 2D regimes at half-filling~(Sec.~\ref{sec:2D_half_filled}) and $1/8$ doping~(Sec.~\ref{sec:2D_1/8_doping}). Finally, Sec.~\ref{sec:conclusion_outlook} concludes with a summary and outlook.

\section{\label{sec:formalism}Formalism}
We focus on the ground states of a generic e-ph system defined by the Hamiltonian
\begin{equation}\label{eq:diagonal_e-ph_H}
        H = H_{\mathrm{e}} + H_{\mathrm{ph}} + H_{\mathrm{e}\text{-}\mathrm{ph}} + H_{\mathrm{e}\text{-}\mathrm{e}}. \\
\end{equation}
The kinetic term $H_{\mathrm{e}} = \sum_{ij\sigma} t_{ij} c_{i\sigma}^\dagger c_{j\sigma}$ describes electronic hopping with real amplitudes $t_{ij}=t_{ji}$. Here, $c_{j\sigma}^\dagger$ ($c_{j\sigma}$) is the creation (annihilation) operator of an electron with spin $\sigma$ in $\{\uparrow,\downarrow\}$ at site $j$. The free phonon propagation is governed by $H_{\mathrm{ph}}= \sum_{ll'}\omega_{ll'}b_l^\dagger b_{l'}$, where $b_l^\dagger$ ($b_l$) is the phonon creation (annihilation) operator at site $l$. We take the real symmetric matrix $\omega$ (with elements $\omega_{ll'}$) to be positive definite, ensuring a lower-bounded spectrum of phonons. The diagonal e-ph interaction~\cite{HOLSTEIN1959325} takes the form:
\begin{equation}
    H_{\mathrm{e}\text{-}\mathrm{ph}} = \sum_{lj\sigma}g_{lj}(b_l + b_l^\dagger)n_{j\sigma}.
\end{equation}
Here, $n_{j\sigma}=c^{\dagger}_{j\sigma}c_{j\sigma}$ is the density operator of the spin-$\sigma$ electron. The real symmetric matrix $g$ (with elements $g_{lj}$) encodes the e-ph coupling strength. Finally, the electron-electron (e-e) interaction $H_{\mathrm{e}\text{-}\mathrm{e}}$ can involve combinations of density and spin operators that commute with the local density operator $\sum_{\sigma}n_{j\sigma}$. Standard terms include the Hubbard repulsion and the superexchange interaction in the $t$-$J$ model.

A faithful description of these systems across broad parameter regimes requires capturing both the strong entanglement arising from the e-ph coupling and e-e interactions, and the retardation effects induced by phonon dynamics. To this end, we construct the hybrid variational method that integrates NGS with MPS. This approach scales efficiently to thousands of sites in 1D and $100 \times 4$ cylinders, granting access to the long-range correlations necessary to resolve competing instabilities. We leverage this capability to elucidate how retardation shifts the balance of these competitions, resolving PS in 1D and identifying the stabilized metallic and stripe phases in 2D.

\subsection{Variational Ansatz}
We formulate our variational ansatz in real space to secure two advantages: (i) physically, it captures inhomogeneous spatial fluctuations, enabling the identification of instabilities involving large unit cells; and (ii) technically, it ensures full compatibility with the MPS framework, which is intrinsically optimized for open boundary conditions (OBC)~\cite{schollwoeckDensitymatrixRenormalizationGroup2011,ciracMatrixProductStates2021,evenblyGaugeFixingCanonical2018}.

The variational ansatz
\begin{equation}\label{eq:ansatz}
    |\Psi(\tilde{\lambda},\Delta_R,\Gamma_R,\xi)\rangle = e^{S(\tilde{\lambda})} |\Psi_\mathrm{b}(\Delta_R,\Gamma_R)\rangle |\Psi_{\mathrm{e}}(\xi)\rangle 
\end{equation}
is constructed by entangling a bosonic Gaussian state $|\Psi_\mathrm{b}\rangle$ with an electronic state $|\Psi_{\mathrm{e}}\rangle$ through a non-Gaussian transformation $U_{S}=e^{S}$. The generator is defined as
\begin{equation}
    S(\tilde{\lambda}) = i \sum_{lj \sigma} p_{l} \tilde{\lambda}_{lj \sigma} n_{j\sigma},
\end{equation}
where the phonon canonical momentum $p_l = i(b_l^\dagger - b_l)$ couples to the electron density $n_{j\sigma}$ via the variational parameters $\tilde{\lambda}_{lj\sigma}$. The bosonic Gaussian state
$|\Psi_\mathrm{b}(\Delta_R,\Gamma_R)\rangle$ is fully characterized by the displacement vector $\Delta_R=\langle R\rangle_\mathrm{b}$ and the covariance matrix $\Gamma_R=\langle\{\delta R,\delta R^{T}\}\rangle_\mathrm{b}/2$ of the quadrature fluctuations $\delta R = R - \Delta_R$~\cite{shiVariationalStudyFermionic2018}. Here, $R=(x_1,...,x_N, p_1, ..., p_N)^T$ and $x_l = b_l + b_l^\dagger$. Moreover, the electronic state $|\Psi_{\mathrm{e}}(\xi)\rangle$ is an MPS parameterized by its constituent tensors $\xi$. An MPS with bond dimension $D$ is constructed within an adaptively optimized many-body basis of dimension $D$~\cite{schollwoeckDensitymatrixRenormalizationGroup2011,whiteDensityMatrixFormulation1992}. We quantify its reliability via the truncation error, defined as the 2-norm distance between the state and its projection onto the optimized basis~\cite{schollwoeckDensitymatrixRenormalizationGroup2011}. Throughout this work, expectation values $\langle\cdots\rangle$ are evaluated with respect to the full variational state $|\Psi\rangle$, as exemplified in Appendix~\ref{app:transformed_expectations}, while evaluations under the electron (phonon) state are denoted by $\langle\cdots\rangle_\mathrm{e(b)}$.

By construction, our ansatz becomes asymptotically exact in both the adiabatic and anti-adiabatic limits, where the phonon frequency approaches zero and infinity, respectively. The variational parameters $\tilde{\lambda}$ smoothly interpolate between these limits. $U_S$ incorporates non-local e-ph entanglement arising from finite phonon frequencies~\cite{alexandrovPolaronsAdvancedMaterials2007,weberTwodimensionalHolsteinHubbardModel2018}, phonon dispersion~\cite{zoliNonlocalElectronphononCorrelations2005}, and long-range e-ph coupling~\cite{spencerEffectElectronphononInteraction2005}. In the transformed frame, the Gaussian state $|\Psi_\mathrm{b}(\Delta_R,\Gamma_R)\rangle$ efficiently accounts for residual phononic fluctuations. Notably, the Hamiltonian preserves time-reversal symmetry (TRS). Assuming the ground state respects this symmetry, $\Delta_p=0$ and the off-diagonal blocks of $\Gamma_R$ vanish, i.e., $\langle \delta x\delta p^{T}\rangle_\mathrm{b}=\langle \delta p\delta x^{T}\rangle^\dagger_\mathrm{b}=0$. Furthermore, for the pure state, the symplectic condition, 
\begin{equation*}
    \Gamma_R \begin{pmatrix}
        0 & \mathbb{I} \\ 
        -\mathbb{I} & 0
    \end{pmatrix} \Gamma_R = \begin{pmatrix}
        0 & \mathbb{I} \\ 
        -\mathbb{I} & 0
    \end{pmatrix},
\end{equation*}
implies the relation $\Gamma_{xx}=(\Gamma_{pp})^{-1}$ between the diagonal blocks $\Gamma_{xx}=\langle \delta x\delta x^{T}\rangle_\mathrm{b}$ and $\Gamma_{pp}=\langle \delta p\delta p^{T}\rangle_\mathrm{b}$. In practice, we determine $\Gamma_{xx}$ implicitly via this relation to enhance computational efficiency and numerical stability. Unlike phonons, electrons reside in a finite local Hilbert space spanned by $\{|0\rangle_\mathrm{e},c^{\dagger}_{j\uparrow}|0\rangle_\mathrm{e},c^{\dagger}_{j\downarrow}|0\rangle_\mathrm{e},c^{\dagger}_{j\uparrow}c^{\dagger}_{j\downarrow}|0\rangle_\mathrm{e} \}$. However, repulsive e-e interactions generally induce strong electronic correlations. To accurately capture these correlations, we employ an MPS $|\Psi_{\mathrm{e}}\rangle$, exploiting the area-law scaling of entanglement characteristic of ground states~\cite{pasqualecalabreseEntanglementEntropyQuantum2004,hastingsAreaLawOnedimensional2007,eisertColloquiumAreaLaws2010,verstraeteMatrixProductStates2006,ciracMatrixProductStates2021}.

\subsection{\label{sec:optimization}Optimization}

The ground state is obtained by minimizing the variational energy,
\begin{equation}
    E(\tilde{\lambda}, \Delta_R,\Gamma_R,\xi) = \langle\Psi|H| \Psi\rangle,
\end{equation}
through a self-consistent optimization workflow. The workflow comprises four main stages: initialization, a self-consistent loop of alternating optimization, convergence assessment, and final validation of the ground state.

\paragraph*{Initialization.}To mitigate the risk of becoming trapped in local minima, we employ an initialization scheme that generates multiple distinct seeds. In this stage, the electronic state is parametrized as a fermionic Gaussian state, fully characterized by its covariance matrix $\Gamma_f=\langle C C^\dagger \rangle$. Here, $C=(c_1,...,c_N,c^\dagger_1,...,c^\dagger_N)^T$ collects the creation and annihilation operators for the $N_f$ modes. We perform a pure NGS flow by simultaneously solving the equations of motion (EoMs) for all variational parameters (see Appendix~\ref{app:EoM}). Using the optimized $\Gamma_f$, we construct the Gaussian MPS $|\Psi_\mathrm{e}(\xi)\rangle$ (see Appendix~\ref{app:gaussianMPS}). The resulting configuration $(\tilde{\lambda},~\Delta_x,~\Gamma_{pp},~\xi)$ seeds the self-consistent loop. This initialization is particularly effective in systems exhibiting mean-field-like behavior, such as those in ordered phases or PS, where conventional methods relying on artificial pinning fields may lack control. While this scheme is generally robust, specific cases may necessitate alternative strategies, such as preconditioned states (e.g., Appendix C in Ref.~\cite{wangZerotemperaturePhasesTwodimensional2020}) or fully randomized configurations.

\paragraph*{The self-consistent loop.}We refine the variational parameters through an iterative procedure. Each iteration comprises two sequential steps. 

First, we optimize the MPS. The effective electronic Hamiltonian $H_\mathrm{e}^\mathrm{eff} = \langle\Psi_\mathrm{b}| U_{S}^{\dagger} H U_{S}|\Psi_\mathrm{b}\rangle$ is derived by averaging out the phononic degrees of freedom in the transformed frame. As detailed in Appendix~\ref{app:EoM}, $H_\mathrm{e}^\mathrm{eff}$ takes the analytical form:
\begin{align}\label{eq:H_eff}
    H_{\mathrm{e}}^{\mathrm{eff}} =&E_{\mathrm{ph}}(\Gamma _{pp},\Delta_{x})+\sum_{ij\sigma }\tilde{t}_{ij\sigma }c_{i\sigma }^{\dagger }c_{j\sigma}+\sum_{lj\sigma }(\Delta _{x})_{l}\tilde{g}_{lj\sigma }n_{j\sigma }^{f} \nonumber \\
    &+\frac{1}{2}\sum_{jj'\sigma \sigma'} V_{jj'\sigma\sigma'} n_{j\sigma }n_{j'\sigma'}+H_{\mathrm{e}\text{-}\mathrm{e}}. 
\end{align}
Here, the phonon energy is
\begin{equation}\label{eq:E_ph}
    E_\mathrm{ph} = \frac{1}{4}\operatorname{tr}[\omega(\Gamma_{pp}+\Gamma^{-1}_{pp}-2\mathbb{I})] + \frac{1}{4}\Delta_x^T\omega\Delta_x.
\end{equation}
The polaron hopping amplitudes are renormalized due to the phonon dressing effect:
\begin{equation}\label{eq:t_eff}
    \tilde{t}_{ij\sigma} = t_{ij}\exp(-\frac{1}{2}\sum_{ll'} \Lambda_{l,ij\sigma}(\Gamma_{pp})_{ll'}\Lambda_{l',ij\sigma}),
\end{equation}
with $\Lambda_{l,ij\sigma} = \tilde{\lambda}_{li\sigma} - \tilde{\lambda}_{lj\sigma}$. The reduced e-ph coupling is
\begin{equation}\label{eq:reduced_g}
    \tilde{g}_{lj\sigma}=g_{lj} - \sum_{l'}\omega_{ll'}\tilde{\lambda}_{l'j\sigma},
\end{equation} 
where $\tilde{\lambda}$ describes extended phonon clouds, minimizing the e-ph entanglement by balancing interaction and kinetic effects~\cite{wangZerotemperaturePhasesTwodimensional2020}. This induces a non-local phonon-mediated density-density interaction:
\begin{equation}\label{eq:V_eff}
    V = 2\tilde{\lambda}^T\omega\tilde{\lambda} - 2g^T\tilde{\lambda} - 2\tilde{\lambda}^Tg.
\end{equation}
With $H_\mathrm{e}^\mathrm{eff}$ established, we construct its MPO representation to optimize the MPS $|\Psi_\mathrm{e}(\xi)\rangle$. The optimization strategy of MPS evolves over the self-consistent loop: we initially employ imaginary-time evolution (ITE)~\cite{haegemanUnifyingTimeEvolution2016} with a time step $\delta\tau$ as a robust warm-up, switching to DMRG for final convergence. Although DMRG converges faster than ITE, a premature switch risks locking the optimization into a local minimum, particularly for large systems with symmetries.

Second, we optimize the NGS. With the optimized MPS, we evaluate the electron correlations $\mathcal{G}_{ij\sigma}=\langle c_{i\sigma}^\dagger c_{j\sigma}\rangle_\mathrm{e}$ and $\mathcal{C}_{i\sigma,j\sigma'}=\langle \delta n_{i\sigma} \delta n_{j\sigma'}\rangle_\mathrm{e}$, which drive the NGS flow equations:
\begin{subequations}
    \begin{align}
        \label{eq:Δx}
        \partial_\tau\Delta_x = &2\partial_\tau \tilde{\lambda}\langle n\rangle_\mathrm{e} -\Gamma_{pp}^{-1}(\omega\Delta_x + 2\tilde{g}\langle n\rangle_\mathrm{e}), \\
        \label{eq:Γp}
        \partial_{\tau} \Gamma_{pp} = &\omega - \Gamma_{pp}(\omega - \kappa)\Gamma_{pp}, \\
        \label{eq:λ}
        \partial_\tau \tilde{\lambda}_{lj\sigma} = &\sum_{ij'\sigma'} \Lambda_{l,ij'\sigma'} \tilde{t}_{ij'\sigma'}\mathcal{G}_{ij'\sigma} \mathcal{C}^{-1}_{i\sigma,j\sigma'} + (\Gamma_{pp}^{-1}\tilde{g})_{lj\sigma}.
    \end{align}
\end{subequations}
These equations are obtained by projecting ITE onto the tangent space of the NGS parameters $\Delta_x$, $\Gamma_{pp}$, and $\tilde{\lambda}$ (see Appendix~\ref{app:EoM}). In the steady-state limit ($\tau\to\infty$), the right-hand side of the EoMs vanishes. From Eq.~\eqref{eq:Δx}, we obtain the optimal displacement, $\Delta_x=-2\omega^{-1}\tilde{g}\langle n\rangle_\mathrm{e}$. The displacement in the untransformed frame is thus $\langle x\rangle = \Delta_x + 2\tilde{\lambda}\langle n\rangle_\mathrm{e} = -2\omega^{-1}g\langle n\rangle_\mathrm{e}$. This result captures the enhancement of phonon displacement driven by softer modes or stronger coupling, in quantitative agreement with local basis optimization (LBO) results~\cite{tangInfluenceExtendedInteractions2024}. Crucially, substituting the optimal displacement back into Eq.~\eqref{eq:H_eff} allows us to recast the effective electronic Hamiltonian. By absorbing the linear coupling term ($\Delta^T_x \tilde{g} n$) and the phonon energy ($\Delta^T_x \omega \Delta_x$) into the interaction in Eq.~\eqref{eq:V_eff}, we derive a unified phonon-mediated interaction $H_{\mathrm{int}}^{\mathrm{eff}}$:
\begin{equation} \label{eq:V_eff_final}
H_{\mathrm{int}}^{\mathrm{eff}} = \frac{1}{2} \sum_{jj'}(\mathcal{V}^\mathrm{inst}_{jj'} n{_j}n_{j'} + \sum_{\sigma\sigma'}\mathcal{V}^\mathrm{ret}_{j\sigma j'\sigma'} \delta n_{j\sigma}\delta n_{j'\sigma'}),
\end{equation}
where $n_j=\sum_\sigma n_{j\sigma}$ denotes the total local occupancy and $\delta n_{j\sigma}=n_{j\sigma}-\langle n_{j\sigma}\rangle$ represents the density fluctuations. This decomposition explicitly separates the instantaneous attraction ($\mathcal{V}^\mathrm{inst}=-2g^T\omega^{-1}g$) from the retardation-induced correction ($\mathcal{V}^\mathrm{ret}=2\tilde{g}^T\omega^{-1}\tilde{g}$). The weight of the correction varies with phonon frequencies, interpolating between $0$ (at anti-adiabatic limit) and $-\mathcal{V}^\mathrm{inst}$ (at adiabatic limit). The physical consequences of this competition are detailed in Sec.~\ref{sec:2D_1/8_doping}. The electronic feedback enters Eq.~\eqref{eq:Γp} via
\begin{equation}\label{eq:kappa}
    \kappa_{ll'} = 2\sum_{ij\sigma}
    \Lambda_{lij\sigma}\Lambda_{l'ij\sigma}
    \tilde{t}_{ij\sigma}\langle c_{i\sigma}^\dagger c_{j\sigma}\rangle_\mathrm{e},
\end{equation}
rendering the phonon effectively dispersive and squeezed. Integrating these EoMs over a time step $\delta\tau$ completes one iteration.

\paragraph*{Convergence and Workflow Control.}We monitor convergence via key observables: the total energy $E$, the NGS parameters $(\Delta_x$, $\Gamma_{pp}$, $\tilde{\lambda})$, and the electron correlations $(\mathcal{G}_{ij\sigma}$, $\mathcal{C}_{i\sigma,j\sigma'})$. The loop terminates when all quantities converge within a predefined threshold. Divergence in these quantities typically signals a physical instability---such as phonon softening at a CDW transition---where the bosonic Hamiltonian loses positive definiteness. Such issues usually stem from a biased initial state or insufficient MPS bond dimension, requiring reinitialization or increased bond dimension.

\paragraph*{Ground State Validation.}We validate the ground state by comparing energies of resulting states across distinct initializations. In practice, maintaining a truncation error of $10^{-5}$ during MPS optimization ensures reliability. A reasonable initial state typically yields convergence to a total-energy precision of $10^{-6}$ in fewer than 10 iterations. 

\section{\label{sec:applications}Applications}
\begin{figure*}[htbp]
    \includegraphics[width=\textwidth]{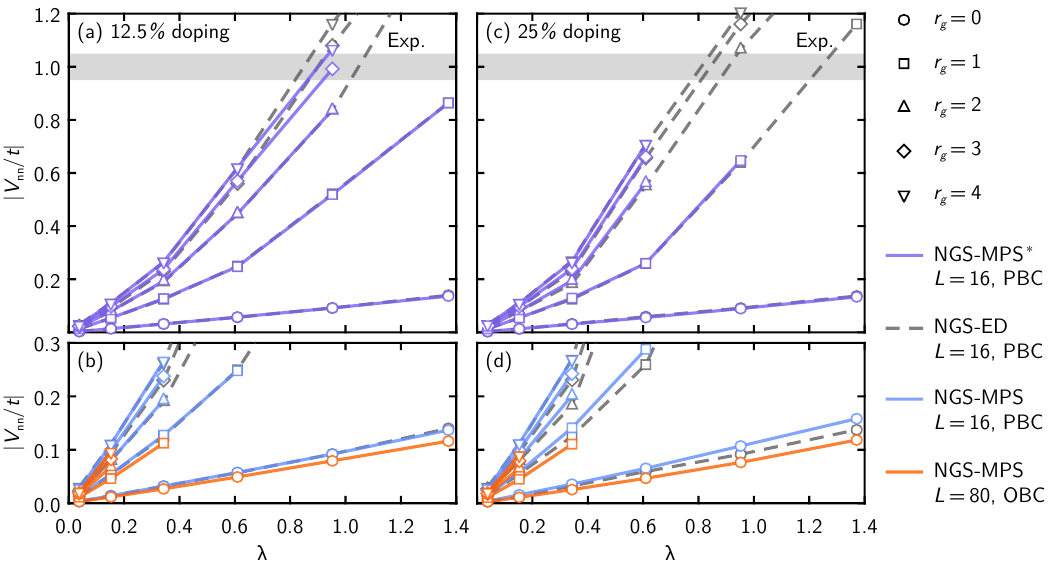}
    \caption{Nearest-neighbor phonon-mediated attraction $V_\mathrm{nn}$ versus effective coupling $\lambda=g_{0}^{2}/(\omega t)$ in the 1D generalized Hubbard--Holstein model at $\omega=0.12t$, $u=8$ and $g_{lj}=g_0/\sqrt{1+(l-j)^2}$. Data are shown for the longest interaction ranges $r_{g}=0\text{--}4$ (distinguished by markers) at hole dopings of $12.5\%$ ((a) and (b)) and $25\%$ ((c) and (d)). (a), (c) $\text{NGS-MPS}^*$ results on an $L=16$ ring (purple) benchmarked against NGSED (gray). The results match perfectly below the experimentally inferred threshold of $|V_\mathrm{nn}/t|\sim 1$ (gray shaded band). (b), (d) NGS-MPS results on $L=16$ rings (blue) and $L=80$ open chains (orange). The absence of data points at large $\lambda$ indicates the onset of phase separation. This instability manifests at weaker couplings in the $L=80$ system,, demonstrating that boundary effects artificially obscure thermodynamic instabilities in small clusters.}
    \label{fig:induced_V_g+}
\end{figure*}
We apply the NGS-MPS method to generalized HH models, governed by the Hamiltonian:
\begin{equation}
    \begin{aligned}
        H = &-t\sum_{\langle ij\rangle\sigma}c_{i\sigma}^\dagger c_{j\sigma} + U\sum_{i}n_{i\uparrow}n_{i\downarrow} \\
        &+ \omega\sum_{l}b_l^\dagger b_l + \sum_{lj\sigma}x_lg_{lj}n_{j\sigma},
    \end{aligned}
\end{equation}
where the nearest-neighbor hopping strength $t$ sets the unit of energy, $U$ denotes the Hubbard repulsion, and $\omega$ is the bare frequency of the Einstein phonons. The term $g_{lj}$ represents a finite-range Holstein coupling that depends on the relative distance $|l-j|$. We define the dimensionless parameters $u=U/t$ and $\lambda=g^2/(\omega t)$. In the anti-adiabatic limit, the non-Gaussian transformation reduces to the Lang--Firsov transformation~\cite{langKineticTheorySemiconductors1963}, with local parameters $\tilde{\lambda}_{lj} = (g/\omega)\delta_{lj}$.

Our real-space MPS simulations explicitly preserve $\mathrm{SU}(2)_\text{spin} \otimes \mathrm{U}(1)_\text{charge}$ symmetry, preventing spurious symmetry breaking and significantly reducing computational costs~\cite{Li_FiniteMPS_jl,Devos_TensorKit_jl_A_Julia_2025}.

\subsection{\label{sec:1D}Results in 1D}
We investigate the e-ph system motivated by the experiment~\cite{chenAnomalouslyStrongNearneighbor2021}, where the hopping amplitude $t=600~\mathrm{meV}$ and the dimensionless parameter is $u=8$~\cite{wangPhononMediatedLongRangeAttractive2021}. We first benchmark against the NGSED approach~\cite{wangPhononMediatedLongRangeAttractive2021}, adopting a phonon frequency $\omega/t=0.12$ and a long-range coupling profile $g_{lj}=g_0/\sqrt{1+(l-j)^2}$ truncated at a distance $r_g$ ($0 \le r_g \le 4$). NGSED enforces translational symmetry by optimizing in momentum space. In contrast, we employ two real-space implementations to assess the role of symmetry breaking: (i) a restricted ansatz ($\text{NGS-MPS}^*$), which imposes translational invariance on the NGS but allows the MPS to break it; and (ii) an unrestricted ansatz (NGS-MPS), where all variational parameters are free to capture inhomogeneous spatial fluctuations.

Figure~\ref{fig:induced_V_g+} displays the benchmark results for the induced nearest-neighbor attraction $V_\mathrm{nn}$, with elements $(V_\mathrm{nn})_{j,j+\delta} =\mathcal{V}^\mathrm{inst}_\mathrm{j,j+\delta} + \mathcal{V}^\mathrm{ret}_\mathrm{j,j+\delta}$ [Eq.~\eqref{eq:V_eff_final}], at hole doping levels of $12.5\%$ (Figs.~\ref{fig:induced_V_g+}(a) and \ref{fig:induced_V_g+}(b)) and $25\%$ (Figs.~\ref{fig:induced_V_g+}(c) and (d)). For a small system ($L=16$ with PBC), $\text{NGS-MPS}^*$ quantitatively reproduces the NGSED results for $|V_\mathrm{nn}/t|$ below the experimentally inferred threshold of $\sim 1$. Beyond this threshold, the absence of stable solutions signals the onset of PS. Figures~\ref{fig:induced_V_g+}(b) and \ref{fig:induced_V_g+}(d) broadens the analysis using the unrestricted NGS-MPS ansatz on both $L=16$ rings and larger $L=80$ open chains. Here, PS emerges at smaller $|V_\mathrm{nn}/t|$, indicating that inhomogeneous spatial fluctuations enhance this instability. Furthermore, the pronounced PS tendency in the larger system ($L=80$) confirms that boundary effects suppress PS in finite-size simulations~\cite{knorzerSpinHolsteinModelsTrappedIon2022}, underscoring the necessity of large-scale capabilities.

\begin{figure*}[htbp]
    \centering
    \includegraphics[width=\textwidth]{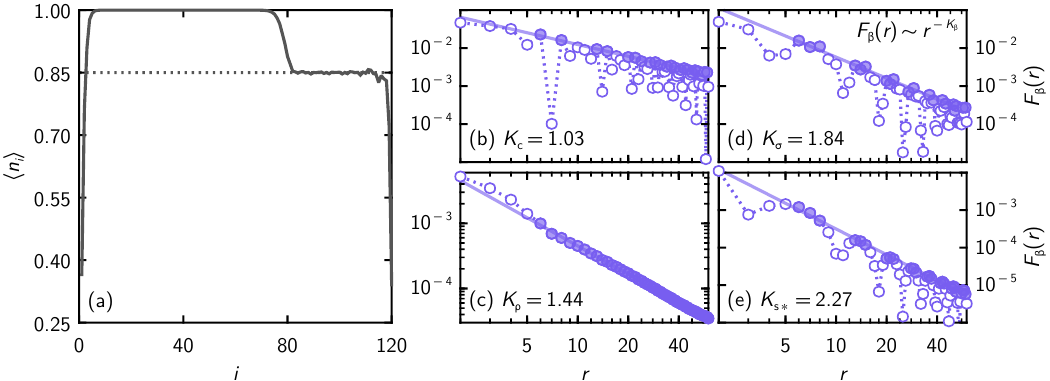}
    \caption{Ground-state properties of the generalized Hubbard--Holstein model on an $L=120$ chain at $\omega/t=0.2$, $u=8$, and $g_{lj}/t=0.3\delta_{lj} + 0.15\delta_{l,j\pm1}$. (a) Spatial profile of the charge density $\langle n_i\rangle$ for the $6.7\%$-doped system. The ground state exhibits macroscopic phase separation, segregating into a half-filled antiferromagnetic domain and a hole-rich domain with a local doping of approximately $15\%$ (the dotted reference line indicates $\langle n_i\rangle=0.85$). (b)--(e) Real-space correlations evaluated for the homogeneous $15\%$-doped phase on an $L=120$ chain, capturing the intrinsic bulk properties of the hole-rich domain shown in (a). The panels display correlations in the (b) single-particle ($\langle c^\dagger_ic_j\rangle$), (c) charge-density-wave ($\langle\delta n_i\delta n_j\rangle$), (d) spin-density-wave ($\langle\mathbf{S}_i\cdot\mathbf{S}_j\rangle$), and (e) extended-singlet Cooper pairing ($\langle P^\dagger_{\mathrm{s\ast},i}P_{\mathrm{s\ast},j} \rangle$) channels. All correlation functions exhibit an algebraic decay $F_\beta(r) \sim r^{-K_\beta}$ characteristic of a Luttinger liquid; the extracted Luttinger exponents $K_\beta$ are indicated in each respective panel.}
    \label{fig:1D_LBO_PS_and_corr}
\end{figure*}
Enabled by this scalability, we proceed to analyze ground-state correlations in large-scale systems. We benchmark our approach against a recent LBO study~\cite{tangTracesElectronPhononCoupling2022}, adopting the parameters $\omega/t=0.2$ and a truncated coupling $g_{lj}/t=0.3\delta_{lj} + 0.15\delta_{l,j\pm1}$. For the $6.7\%$-doped generalized HH model on an $L=120$ chain, we find a ground-state energy per site of $E/N=-2.003t$. As illustrated by the charge density distribution in Fig.~\ref{fig:1D_LBO_PS_and_corr}(a), this ground state exhibits macroscopic PS. Specifically, it segregates into two distinct domains corresponding to a half-filled AFM phase and a $15\%$-doped Luttinger liquid (LL) phase. To characterize correlations in the hole-rich region, we simulate a homogeneous $15\%$-doped system at the same total size ($L=120$), as shown in Fig.~\ref{fig:1D_LBO_PS_and_corr}(b)--\ref{fig:1D_LBO_PS_and_corr}(e). The correlations display the power-law decay characteristic of the LL, $F_\beta(r)\sim r^{-K_\beta}$. We extract the Luttinger exponents for the single-particle ($K_c$), CDW ($K_\rho$), spin-density-wave (SDW) ($K_\sigma$), and Cooper (extended-singlet pairing, $K_\mathrm{s*}$) channels. Boundary effects are mitigated by averaging correlation functions over five central sites, $i\in\{L/4-2,\dots,L/4+2\}$, and restricting separations to $r=|i-j|\leq L/2$.

In the Cooper channel, we examine the extended-singlet pairing operator. The singlet pair annihilation operator on a generic bond connecting sites $i$ and $j$ reads:
\begin{equation}
    b_{i,j} = \frac{1}{\sqrt{2}} (c_{i\uparrow} c_{j\downarrow} - c_{i\downarrow} c_{j\uparrow}).
\label{eq:bond_singlet}
\end{equation} 
Using this definition, the extended-singlet pairing operator on nearest-neighbor bonds in 1D is given by:
\begin{equation}\label{eq:1D_pair_def}
    P_{\mathrm{s\ast},i} = b_{i,i+1}.
\end{equation}
All channels exhibit algebraic decay, where the slowest decay of the four-point correlations in the CDW channel indicates the dominant density fluctuation.

Our analysis indicates that soft phonons ($\omega/t=0.2$) facilitate local lattice distortions that significantly lower the total energy, thereby driving the system toward PS. Consistent with this mechanism, the charge profile shows that PS persists across the doping range from $0\%$ to $15\%$, encompassing the $10\%$ doping level previously identified as a stable LL phase in the LBO study~\cite{tangTracesElectronPhononCoupling2022}. We independently corroborate this instability via a Maxwell construction~\cite{shiVariationalApproachManyBody2020} under PBC using the restricted $\text{NGS-MPS}^*$ ansatz (see Appendix~\ref{app:1D_PS_maxwell}). The proliferation of near-degenerate states in such soft-phonon regimes necessitates a rigorous treatment of the low-frequency phonon sector in both LBO and NGS-MPS calculations.

\subsection{Results in 2D}
Extending the method to 2D captures the essential geometry of layered cuprates, albeit at the cost of substantially increased computational complexity. We analyze the model on four-leg cylinders with axial lengths sufficiently large to resolve a rich variety of instabilities. Simulations on these cylinders remain computationally tractable while capturing competing orders in 2D, most notably unidirectional charge-density-wave (stripe) order~\cite{zhengStripeOrderUnderdoped2017} and $d$-wave superconducting correlations~\cite{chungPlaquetteOrdinaryWave2020}. We benchmark NGS-MPS against NGSED on a $4\times 4$ torus (see Appendix.~\ref{app:benchmark_2D}), finding quantitative agreement in pairing correlations and structure factors (both spin and charge) at half-filling~\cite{wangZerotemperaturePhasesTwodimensional2020}, as well as local moments at $1/8$ doping~\cite{wangFluctuatingNatureLightEnhanced2021}.

\subsubsection{\label{sec:2D_half_filled}Half-filled case}
\begin{figure}[t]
    \centering
    \includegraphics[width=\columnwidth]{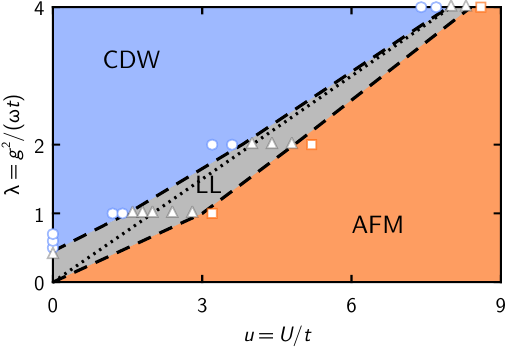}
    \caption{Phase diagram of the half-filled Hubbard--Holstein model at a phonon frequency of $\omega=5t$, computed on a $48\times4$ cylinder in the $(u, \lambda)$ plane. Three regimes are identified: a $(\pi,\pi)$ charge-density-wave (CDW) phase (blue), an intermediate metallic phase corresponding to a 2D analog of the Luttinger liquid (LL) phase (gray), and a $(\pi,\pi)$ antiferromagnetic (AFM) phase (orange). Dashed lines indicate the approximate phase boundaries, whereas the dotted line marks the phase boundary ($u=2\lambda$) in the anti-adiabatic limit.}
    \label{fig:f1_phase_diagram}
\end{figure}

For the half-filled HH model at $\omega=5t$, we present the phase diagram in Fig.~\ref{fig:f1_phase_diagram}. The anti-adiabatic limit provides a theoretical baseline (dotted line, $u=2\lambda$). In this limit, the system maps onto an effective Hubbard model ($U \to U-2\lambda t$), predicting a quantum phase transition at $u=2\lambda$. This line separates the $(\pi,~\pi)$-AFM phase for $u>2\lambda$~\cite{CINI2001451} from degenerate ground states comprising a $(\pi,~\pi)$-CDW and an $s$-wave superconducting phase for $u<2\lambda$. This CDW-SC degeneracy stems from an emergent $\mathrm{SU}(2)$ pseudospin symmetry in the charge sector~\cite{yangPairingOffdiagonalLongrange1989,zhangPseudospinSymmetryNew1990}. Our NGS-MPS calculations reveal that fluctuations of finite-frequency phonons fundamentally reshape the phase diagram: as delineated by the dashed lines, they broaden the transition to give rise to an intermediate metallic phase~\cite{nowadnickCompetitionAntiferromagneticChargeDensityWave2012,costaPhaseDiagramTwodimensional2020,weberTwodimensionalHolsteinHubbardModel2018} (gray region).

We interpret this intermediate metallic regime as a 2D analog of the LL phase, distinguished by intertwined quasi-long-range orders. To characterize this phase, we examine equal-time correlation functions $F_\beta(r_x)$ along the axial direction. These correlations exhibit the power-law decay $F_\beta(r_x)\sim r_x^{-K_\beta}$ characteristic of LL physics. In addition to $F_c,~F_\rho$ and $F_\sigma$ analyzed in Sec.~\ref{sec:1D}, we compute pairing correlations $F_\beta(\mathbf{i},\mathbf{j}) = \langle P^\dagger_{\beta,\mathbf{i}} P_{\beta,\mathbf{j}} \rangle$ in the Cooper channel. The singlet pairing operators for the various symmetry channels $\beta$ are given by:
\begin{subequations}
    \begin{align}
        P_{\mathrm{s},\mathbf{i}} &= b_{\mathbf{i},\mathbf{i}}, \label{eq:2D_s} \\
        P_{\mathrm{d},\square,\mathbf{i}} &= b_{\mathbf{i},\mathbf{i}+\hat{y}} - b_{\mathbf{i}+\hat{y},\mathbf{i}+2\hat{y}}, \label{eq:plaq_d} \\
        P_{\mathrm{s\ast},\mathbf{i}} &= b_{\mathbf{i},\mathbf{i}+\hat{y}} + b_{\mathbf{i},\mathbf{i}+\hat{x}}, \label{eq:2D_s*} \\
        P_{\mathrm{d},x^2-y^2,\mathbf{i}} &= b_{\mathbf{i},\mathbf{i}+\hat{y}} - b_{\mathbf{i},\mathbf{i}+\hat{x}}. \label{eq:2D_d_x2y2}
    \end{align}
\end{subequations}
Equation~\eqref{eq:2D_s} represents the on-site $s$-wave pairing. Eq.~\eqref{eq:plaq_d} denotes the plaquette $d$-wave pairing, which involves a sign change between adjacent rungs along the circumferential direction. Finally, Eqs.~\eqref{eq:2D_s*} and \eqref{eq:2D_d_x2y2} correspond to the standard extended $s$-wave and $d_{x^2-y^2}$-wave symmetries, respectively. As detailed in Appendix~\ref{app:half_filled_2D}, the extended $s$-wave and $d_{x^2-y^2}$-wave pairings do not develop quasi-long-range order in this intermediate metallic regime and are thus excluded from further analysis. To mitigate finite-size effects, we average $F_\beta(r_x)$ over five reference sites along the axial direction, $x_0 \in {L_x/4-2,\dots,L_x/4+2}$, and over the narrow circumferential direction $y$ of the cylinder. For each $(x_0, y)$, correlations are evaluated up to separations $r_x \le L_x/2$. Therefore, $F_\beta(r_x)$ probes the central bulk region of the system, thereby reducing boundary effects.

\begin{figure*}[t]
    \centering
    \includegraphics[width=\textwidth]{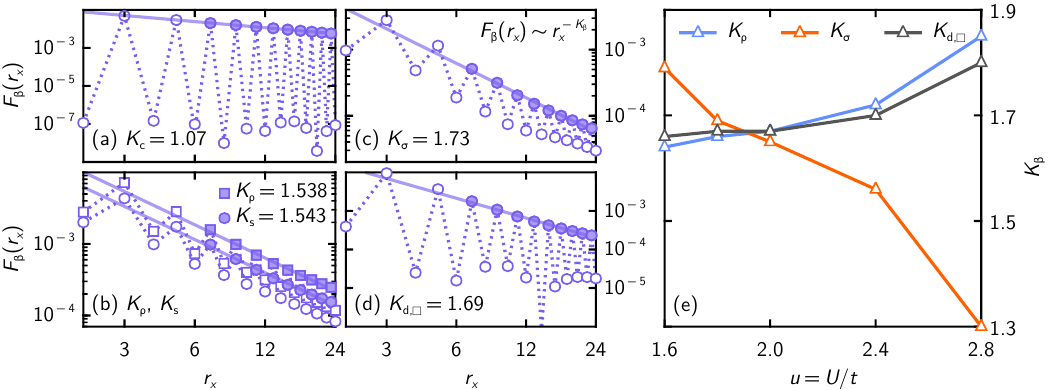}
    \caption{Correlation functions and Luttinger exponents for the half-filled Hubbard--Holstein model on a $48 \times 4$ cylinder. (a)--(d) Spatial decay of correlation functions along the axial direction, $F_\beta(r_x)$, in the Holstein limit ($u=0$) at an electron-phonon coupling of $\lambda=0.4$, extrapolated to the $D\to\infty$ limit. The panels display correlations in the (a) single-particle ($\langle c^\dagger_{\mathbf{i}} c_{\mathbf{j}} \rangle$), (b) charge-density-wave ($\langle \delta n_{\mathbf{i}} \delta n_{\mathbf{j}} \rangle$) alongside $s$-wave pairing ($\langle P^\dagger_{\mathrm{s},\mathbf{i}} P_{\mathrm{s},\mathbf{j}} \rangle$), (c) spin-density-wave ($\langle \mathbf{S}_{\mathbf{i}} \cdot \mathbf{S}_{\mathbf{j}} \rangle$), and (d) plaquette $d$-wave pairing ($\langle P^\dagger_{\mathrm{d},\square,\mathbf{i}} P_{\mathrm{d},\square,\mathbf{j}} \rangle$) channels. Notably in (b), $K_\mathrm{s}$ is marginally lower than $K_\rho$, reflecting the proximity to the $\mathrm{SU}(2)_\text{charge}$ symmetry restoration in the high-frequency limit. (e) Luttinger exponents $K_\rho$ (blue), $K_\sigma$ (orange), and $K_{\mathrm{d},\square}$ (gray) as a function of the Hubbard repulsion $u$, evaluated along the $\lambda=1$ cut within the intermediate regime of Fig.~\ref{fig:f1_phase_diagram}.}
    \label{fig:corrs_and_Ks_combined}
\end{figure*}

We next discuss the CDW, AFM, and LL regimes in Fig.~\ref{fig:f1_phase_diagram}. In the Holstein limit ($u=0$), the existence of a nonzero critical $\lambda_\rho$ for the onset of a CDW insulator remains debated~\cite{costaPhaseDiagramTwodimensional2020,ohgoeCompetitionSuperconductingAntiferromagnetic2017,karakuzuSuperconductivityChargedensityWaves2017,weberTwodimensionalHolsteinHubbardModel2018}. On the $48\times 4$ cylinder, we find a finite CDW onset at $\lambda_\rho\sim 0.4$, consistent with variational Monte Carlo (VMC) studies~\cite{ohgoeCompetitionSuperconductingAntiferromagnetic2017, karakuzuSuperconductivityChargedensityWaves2017}. Figures~\ref{fig:corrs_and_Ks_combined}(a)--\ref{fig:corrs_and_Ks_combined}(d) reveals that among the four-point correlations at $\lambda=0.4$, CDW fluctuation dominates, followed closely by $s$-wave pairing. This near-degeneracy reflects the emergent pseudospin symmetry of the anti-adiabatic limit discussed above. The small value of $\lambda_\rho$ indicates that increasing e-ph coupling rapidly stabilizes CDW order, restricting the metallic phase to a narrow window. In the strongly repulsive regime ($u\gg 2\lambda$), the system forms an AFM-Mott insulator, where phonons induce only quantitative modifications, similar to the doped case discussed in Sec.~\ref{sec:2D_1/8_doping}.

We identify an intermediate metallic regime proximate to the line $u=2\lambda$. On the narrow cylinder, this regime appears as a 2D analogue of the LL phase, characterized by intertwined fluctuations in the charge, spin, and plaquette $d$-wave pairing channels. The quasi-1D geometry provides a tractable setting to investigate these competing tendencies, which serves as precursors to the behavior in the full 2D limit~\cite{stoudenmireStudyingTwoDimensionalSystems2012}. We attribute the emergence of this regime to the delicate balance between charge and spin ordering tendencies (see Appendix~\ref{app:half_filled_2D} for a discussion of geometric frustration as an additional mechanism). Figure~\ref{fig:corrs_and_Ks_combined}(e) shows the variation of the Luttinger exponents along the cut $\lambda=1$; here, charge and spin correlations dominate in different parameter regions, with a crossover around $u\sim 2$. The plaquette $d$-wave pairing remains subdominant throughout (Table~\ref{tab:K_λ2_λ4}).
\begin{table}[b]
    \caption{Luttinger exponents $K_\rho$, $K_\sigma$, and $K_\mathrm{d,\square}$ at the representative points along the $\lambda=2$ and $\lambda=4$ cuts of Fig.~\ref{fig:f1_phase_diagram}.}
    \begin{ruledtabular}
        \begin{tabular}{c | c c c c c}
            & \multicolumn{5}{c}{$(u,~\lambda)$} \\
            $K_\beta$ & $(4,~2)$ & $(4.4,~2)$ & $(4.8,~2)$ & $(8,~4)$ & $(8.3,~4)$ \\
            \hline
            $K_\rho$ & $1.68$ & $1.75$ & $2.08$ & $1.71$ & $1.97$ \\
            $K_\sigma$ & $1.64$ & $1.49$ & $1.02$ & $1.63$ & $1.28$ \\
            $K_\mathrm{d,\square}$ & $1.68$ & $1.70$ & $1.9$ & $1.72$ & $1.82$ \\
        \end{tabular}
    \end{ruledtabular}
    \label{tab:K_λ2_λ4}
\end{table}

Furthermore, calculations on $48 \times 4$ systems with bond dimensions $D\sim 2\times10^4$ support continuous transitions from the AFM to the LL regime and from LL to the CDW regime as the e-ph coupling increases at finite $U$. This observation aligns with QMC results~\cite{costaPhaseDiagramTwodimensional2020,nowadnickCompetitionAntiferromagneticChargeDensityWave2012,weberTwodimensionalHolsteinHubbardModel2018}. We find no evidence of phase coexistence. Instead, the rapid growth of the required bond dimension and the smooth evolution of the charge and spin order parameters are characteristic signatures of proximity to a critical point~\cite{eisertColloquiumAreaLaws2010,pollmannTheoryFiniteEntanglementScaling2009,schollwoeckDensitymatrixRenormalizationGroup2011}. Tracing the dotted line ($u=2\lambda$) in Fig.~\ref{fig:f1_phase_diagram}, the intermediate regime shrinks at stronger couplings. This narrowing follows from the reduction in the effective polaronic hopping strength (Table~\ref{tab:u=2λ_t_eff}), corresponding to an increasing polaron mass. This suppression of the kinetic energy hinders the formation of the LL state, which relies on charge mobility. Eventually, the LL phase disappears in the large $\lambda$-limit, and the transition between the AFM and CDW phases becomes first order, consistent with VMC results~\cite{karakuzuSuperconductivityChargedensityWaves2017}.  
\begin{table}[b]
    \caption{Renormalized polaronic hopping amplitudes for a central bulk site along the $u=2\lambda$ line.}
    \begin{ruledtabular}
        \begin{tabular}{c | c c c c}
            & \multicolumn{4}{c}{$(u,~\lambda)$} \\
            $\tilde{t}_{c,c+\hat{\alpha}}$ \footnote{Here, $\tilde{t}_{c,c+\hat{\alpha}}$ denotes the dressed hopping amplitude of the central site $c$ in the axial ($\hat{\alpha}=\hat{x}$) and circumferential ($\hat{\alpha}=\hat{y}$) directions.} & $(0,~0)$ & $(2,~1)$ & $(4,~2)$ & $(8,~4)$ \\
            \hline
            $\hat{\alpha} = \hat{x}$ & $1.000$ & $0.930$ & $0.862$ & $0.730$ \\
            $\hat{\alpha} = \hat{y}$ & $1.000$ & $0.929$ & $0.861$ & $0.730$ \\
        \end{tabular}
    \end{ruledtabular}
    \label{tab:u=2λ_t_eff}
\end{table}

Finally, we note that the 4-leg geometry lies in a crossover regime between 1D and 2D. Finite-width effects manifest primarily in two ways. First, in the spin sector, the AFM phase is gapped for even widths but gapless for odd widths, in accordance with the Haldane conjecture~\cite{HALDANE1983464,haldaneNonlinearFieldTheory1983}. In the 2D limit, however, the AFM phase supports gapless Goldstone magnons~\cite{chakravartyDimensionalCrossoverQuantum1996}. Consequently, our finite-width calculations likely underestimate the robustness of AFM ordering. Second, the 4-leg cylinder supports a subdominant plaquette $d$-wave pairing channel~\cite{chungPlaquetteOrdinaryWave2020} (Fig.~\ref{fig:corrs_and_Ks_combined}(d)), which competes with both CDW and AFM orders but is absent in both 4-leg ladders and the 2D limit. Consistent with this picture, the CDW order parameter is sizeable at $\lambda=0.4$ on a 4-leg ladder (see Appendix~\ref{app:half_filled_2D}), leading us to expect $\lambda_\rho\to 0$ in the 2D limit. This expectation is consistent with finite-temperature QMC results~\cite{costaPhaseDiagramTwodimensional2020,hohenadlerDominantChargeDensity2019,weberTwodimensionalHolsteinHubbardModel2018}. 
\begin{figure*}[t]
    \centering
    \includegraphics[width=\textwidth]{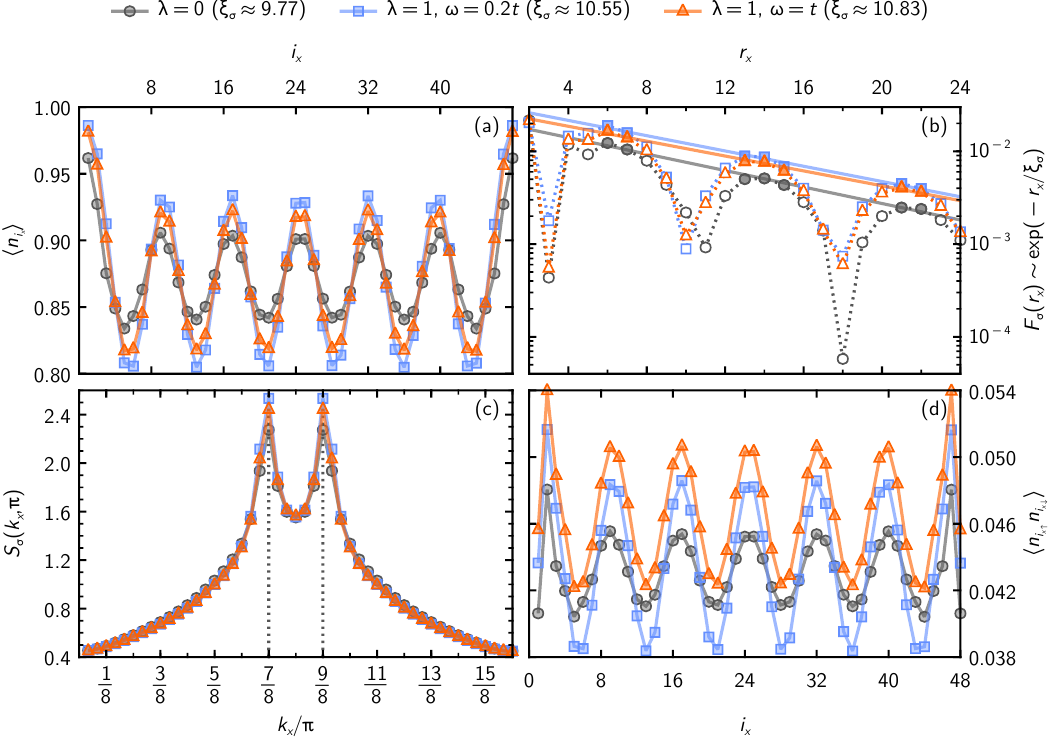}
    \caption{Ground-state properties of the fully filled stripe phase in the $1/8$-doped Hubbard--Holstein model on a $48 \times 4$ cylinder at $u=8$. (a) Charge density profile $\langle n_{i_x} \rangle$ along the axial direction. (b) Real-space spin correlations $\langle \mathbf{S}_{i_x,c_y}\cdot\mathbf{S}_{j_x,c_y}\rangle$ along the axial direction, extrapolated to the $D\to\infty$ limit and exhibiting exponential decay. (c) Spin structure factor $S_\sigma(k_x, \pi)$, with two peaks at $k_x=\pi\pm\frac{1}{8}\pi$. (d) Double occupancy profile $\langle n_{i_x\uparrow}n_{i_x\downarrow} \rangle$ along the cylinder axis $i_x$. Results are compared across the Hubbard limit ($\lambda=0$, gray circles) and the Hubbard--Holstein model ($\lambda=1$) with phonon frequencies $\omega=0.2t$ (blue squares) and $\omega=t$ (orange triangles). Compared to the Hubbard limit, the inclusion of phonons enhances both the charge modulation amplitude in (a) and the spin correlation length $\xi_\sigma$ in (b). However, lowering the phonon frequency further amplifies the charge modulation but suppresses the gain in the spin correlation length. Concurrently, as depicted in (d), the global amplitude of the double occupancy increases with phonon frequency, whereas its spatial modulation amplitude decreases.}
    \label{fig:combined_dopedAFM}
\end{figure*}
Collectively, these considerations suggest that the intermediate metallic regime will shrink and shift toward the CDW sector as $L_y \to \infty$.

\subsubsection{\texorpdfstring{$\frac{1}{8}$ hole doped case}{1/8 hole doped case}}
\label{sec:2D_1/8_doping}
Upon doping the half-filled parent compound to $p=1/8$, the system enters the regime of intertwined orders central to high-temperature superconductivity~\cite{fradkinColloquiumTheoryIntertwined2015,tranquadaCuprateSuperconductorsViewed2020}. Numerical investigation of this regime poses severe challenges, stemming from the sign problem and an extremely dense spectrum of competing low-energy states~\cite{qinHubbardModelComputational2022,arovasHubbardModel2022}. Here, we employ NGS-MPS to explicitly resolve the interplay among bipolarons, stripe order, and phase separation in the soft-phonon regime ($\omega \leq t$).

Even in the pure Hubbard limit (without phonons), simulating the $1/8$-doped 4-leg cylinder imposes severe computational demands. While the ground state at strong coupling ($u \sim 6$--$10$) is widely identified as a fully filled stripe phase~\cite{arovasHubbardModel2022, qinHubbardModelComputational2022, jiangGroundStatePhase2020,zhengStripeOrderUnderdoped2017}, the intermediate-coupling regime ($u\leq4$) remains unsettled due to the presence of numerous nearly degenerate states, including stripe phases with varying periodicities~\cite{arovasHubbardModel2022, qinHubbardModelComputational2022, jiangGroundStatePhase2020,zhengStripeOrderUnderdoped2017}. To navigate this complexity, we employ an initialization strategy based on Gaussian states with hole-line stripes of varying periodicities, motivated by the understanding that stripe formation is primarily charge driven~\cite{zacharStripesFormationAntiphase2002,zaanenCURRENTIDEASORIGIN1998}. While AFM correlations are not encoded in the initial seed, they can emerge during the subsequent NGS-MPS optimization. This procedure facilitates a systematic comparison among competing stripe states, enabling identification of the true ground state and preventing entrapment in local minima associated with mixed periodicities. For the strongly repulsive Hubbard model, we find that the anti-phase fully filled stripe~\cite{zhengStripeOrderUnderdoped2017} is stabilized within a few sweeps. This state features a period-$8$ charge modulation (Fig.~\ref{fig:combined_dopedAFM}(a)) characterized by the wave vector $\mathbf{Q}_\text{charge}=(\pm\pi/4,~0)$, where the holes form domain walls. Short-range AFM singlets preferentially form on the rungs, leading to exponentially decaying spin correlations along the axial direction (Fig.~\ref{fig:combined_dopedAFM}(b)). The associated spin modulation wave vector, $\mathbf{Q}_\text{spin}=(\pi\pm\pi/8,~\pi)$, satisfies the mutual commensurability condition $\mathbf{Q}_\text{charge}=2\mathbf{Q}_\text{spin}$ (Fig.~\ref{fig:combined_dopedAFM}(c))~\cite{zaanenCURRENTIDEASORIGIN1998,arovasHubbardModel2022}. 

In the regime of weak e-ph coupling, the fully filled stripe characteristic of the pure Hubbard model undergoes primarily quantitative modifications. To illustrate an intuitive physical picture, we consider two asymptotic limits. In the anti-adiabatic limit, phonons screen the Hubbard repulsion, thereby enhancing charge fluctuations and spin correlations~\cite{jiangGroundStatePhase2020,zhengStripeOrderUnderdoped2017}. Conversely, in the adiabatic limit, phonons behave as a classical field that adapts to and pins the stripe order~\cite{provilleMobileBipolaronsAdiabatic1998,provilleSmallBipolarons2dimensional1999}. These qualitative trends persist at finite phonon frequencies. Given that the low-energy physics of the doped antiferromagnet is dictated by the strong onsite Hubbard repulsion, the impact of phonons is most significant through their modification of the local effective interaction. Accordingly, we analyze the onsite component of the phonon-mediated interaction $H_{\mathrm{int}}^{\mathrm{eff}}$ [Eq.~\eqref{eq:V_eff_final}]. We decompose the local terms at site $j$ into three distinct contributions, 
\begin{equation*}
    \frac{1}{2} (\mathcal{V}^\mathrm{inst}_{jj} n_j^2 + \sum_{\sigma\sigma'}\mathcal{V}^\mathrm{ret}_{j\sigma j\sigma'} \delta n_{j\sigma}\delta n_{j\sigma'} ) \equiv h^{(1)}_j + h^{(2)}_j + h^{(3)}_j,
\end{equation*}
where the three terms are given by:
\begin{subequations}
   \begin{align}
     h^{(1)}_j &= -2 v_j n_{j\uparrow}n_{j\downarrow}, \label{eq:Hubbardlike_attraction} \\
     h^{(2)}_j &= - 2(\lambda-v_j) (\langle n_j\rangle n_j - \frac{1}{2}\langle n_j\rangle^2), \label{eq:self_trapping} \\ 
     h^{(3)}_j &= - v_j n_j. \label{eq:chemical_potential_shift}
   \end{align}
\end{subequations}
Here, we have utilized the property $\tilde{\lambda}_{lj\uparrow}=\tilde{\lambda}_{lj\downarrow}\equiv\tilde{\lambda}_{lj}$ imposed by $\mathrm{SU}(2)_\mathrm{spin}$ symmetry. The frequency-dependent coupling $v_j=\sum_l(2g\tilde{\lambda}_{lj} - \omega\tilde{\lambda}^2_{lj})$ interpolates between $0$ (as $\omega\to0$) and $\lambda$ (as $\omega\to\infty$). This decomposition explicitly illustrates the local competition governed by retardation: the on-site attraction $h^{(1)}_j$ favors local pairing fluctuations, while the self-trapping potential $h^{(2)}_j$, with its negative coefficient $-2(\lambda-v_j)<0$, tends to pin the charge order. The last term $h^{(3)}_j$ shifts the chemical potential.

For the representative parameters $(\lambda = 1, u = 8)$, we observe that both charge order (Fig.~\ref{fig:combined_dopedAFM}(a)) and spin correlations (Fig.~\ref{fig:combined_dopedAFM}(b)) are enhanced relative to the pure Hubbard model. However, these enhancements exhibit distinct dependencies on the phonon frequency. Specifically, lowering the phonon frequency enhances the amplitude of the charge modulation, while simultaneously suppressing the gain in the spin correlation length. This behavior stems from the trade-off inherent in Eqs.~\eqref{eq:Hubbardlike_attraction} and \eqref{eq:self_trapping}: as the frequency lowers, the pinning strength $2(\lambda-v_j)$ grows while the attraction strength $2v_j$ diminishes. Since the attraction term effectively screens the bare Hubbard repulsion, i.e., $U_{\text{eff},j} \approx U - 2v_j$, this reduction raises the energy cost for the virtual doublon-holon excitations that mediate AFM interaction~\cite{chaoCanonicalPerturbationExpansion1978}. As a result, the local superexchange strength $J \propto t^2/U_{\text{eff}}$ is suppressed compared to that in the high phonon frequency regime (see Appendix~\ref{app:estimation_doubleOccupancy} for details). This suppression dampens the screening-induced enhancement of spin correlations, leaving the system increasingly controlled by the pinning potential—--a local manifestation of the retardation effect induced by soft phonons.

We further substantiate this picture by examining the double occupancy, $\langle n_{i\uparrow} n_{i\downarrow} \rangle$, which serves as a direct probe of the intermediate states in the relevant high-energy sector. Perturbative analysis reveals that $\langle n_{i\uparrow} n_{i\downarrow} \rangle$ scales with the singlet projector $\langle \frac{1}{4}n_i n_{i+\delta} - \mathbf{S}_i \cdot \mathbf{S}_{i+\delta} \rangle$, which quantifies the probability of finding neighboring electrons in a singlet configuration. This dependence arises because, to leading order, the Pauli exclusion principle restricts virtual hopping to the singlet channel. Since the formation of a singlet pair is contingent upon the simultaneous occupancy of both sites, the spatial inhomogeneity of the double occupancy is primarily inherited from the static charge order. Meanwhile, the global amplitude of the double occupancy scales as $1/U_{\mathrm{eff}}^2$, reflecting the system's capacity to mediate spin correlations. Consistent with our analysis, Fig.~\ref{fig:combined_dopedAFM}(d) demonstrates that increasing the phonon frequency suppresses the spatial modulation of the double occupancy while enhancing its overall intensity. These trends align with the low-energy behavior in the charge and spin sectors, corroborating our microscopic interpretation of the interplay between retardation effects and stripe physics in the doped AFM insulator.

In the regime of strong e-ph coupling, the attraction strength [Eq.~\eqref{eq:Hubbardlike_attraction}], $2v_j$, surpasses that of the Hubbard repulsion $U$. To establish the baseline, we first consider the anti-adiabatic limit, where the system maps onto an attractive Hubbard model. Upon doping, the mapped model exhibits a smooth crossover—--governed by the attraction strength—--from a BCS superconductor to a Bose-Einstein condensate (BEC) of tightly bound local pairs~\cite{micnasSuperconductivityNarrowbandSystems1990}. Away from the limit of $\omega\to\infty$, phonons introduce retardation effects that modify this picture through both local and non-local mechanisms. Locally, as the system approaches the adiabatic limit, the reduction of $v_j$ suppresses the pairing fluctuations essential for $s$-wave superconductivity. Concomitantly, the self-trapping potential [Eq.~\eqref{eq:self_trapping}] becomes dominant. While this potential favors charge ordering, its pronounced concave dependence on density may drive the system toward a negative compressibility $\partial^2_\nu \mathcal{E} < 0$~\cite{shiVariationalApproachManyBody2020,ohgoeCompetitionSuperconductingAntiferromagnetic2017}, where the ground state energy $\mathcal{E}$ varies with the filling $\nu$. Consistent with this mechanism, a previous VMC study has reported the ground state of the adiabatic Holstein model exhibits PS between a half-filled $(\pi,~\pi)$ CDW insulator and a paramagnetic metal~\cite{ohgoeCompetitionSuperconductingAntiferromagnetic2017}.

\begin{figure}[t]
    \includegraphics[width=\columnwidth]{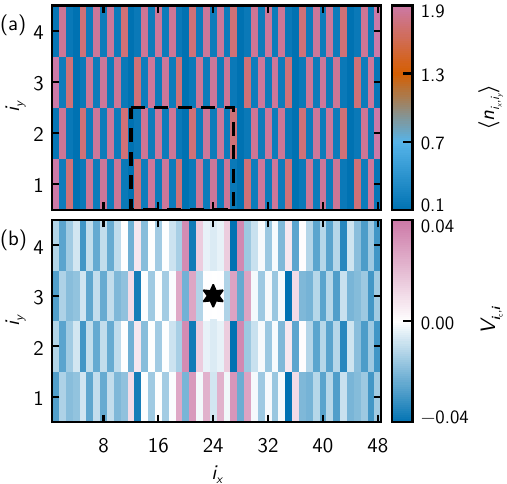}
    \caption{Stabilization of the bipolaronic stripe phase via long-range interactions in the doped charge-density-wave regime. (a) Real-space charge density map $\langle n_{i_x,i_y} \rangle$ on a $48 \times 4$ cylinder at $1/8$ doping ($\lambda=2, u=2, \omega=t$). The system spontaneously forms a commensurate stripe with an enlarged $16 \times 2$ unit cell (outlined by a dashed black rectangle). (b) Spatial profile of the effective phonon-mediated density-density interaction $V_{\mathbf{i}_c, \mathbf{i}}$ [Eq.~\eqref{eq:V_eff}] emanating from a central reference site $\mathbf{i}_c=(24, 2)$ (marked by a black hexagram). The interaction exhibits long-range attraction and repulsion that persist well beyond the local self-trapping potential. These non-local interactions frustrate the global segregation of holes, arresting the macroscopic phase separation and thereby stabilizing the periodic structure observed in (a).}
    \label{fig:bipolaronic_stripe_mechanism}
\end{figure}

Nevertheless, the local analysis is incomplete outside the asymptotic limits of $\omega \to 0$ and $\omega \to \infty$. Unlike the doped AFM regime dominated by the onsite Hubbard repulsion, the doped CDW regime is controlled by the phonon-mediated interaction [Eq.~\eqref{eq:V_eff_final}], which extends well beyond the onsite term. This long-range interactions may frustrate the macroscopic segregation driven by the local potential~\cite{EMERY1993597}. Faithfully capturing this non-local retardation effect fundamentally requires an unbiased treatment of spatial inhomogeneity. Previous variational studies, limited to unit cells of size up to $12\times 2$, claimed that potential charge-ordered phases are preempted by PS at finite frequencies~\cite{karakuzuStripeCorrelationsTwodimensional2022,ohgoeCompetitionSuperconductingAntiferromagnetic2017,karakuzuSuperconductivityChargedensityWaves2017}. In contrast, we discover that a commensurate bipolaronic stripe phase with a large unit cell of $16\times 2$ emerges spontaneously, devoid of any \emph{ad hoc} constraints on the unit-cell size. Figure~\ref{fig:bipolaronic_stripe_mechanism}(a) depicts the charge density map for representative parameters $(u=2, \lambda=2)$. Similar stripe structures persist across a broad parameter range, including $\lambda=2$ ($u\le2$) and $\lambda=4$ ($u\le7$). This bipolaronic stripe comprises domains with a checkerboard arrangement of onsite bipolarons, separated by hole-rich domain walls. We verify the stability of this state via a positive local compressibility~\footnote{Here we calculate it using a central second-order finite-difference scheme, $\partial^2_\nu f\sim E(N_e+2) + E(N_e-2) - 2E(N_e)$. The choice of $\Delta N_\mathrm{e}=2$ preserves the system in the spin-singlet sector and thereby mitigates finite-size effects.}. To exclude metastability, we also confirm that the stripe phase is energetically favorable compared to the candidate macroscopic PS. We interpret such phase as an ``arrested" PS~\cite{EMERY1993597}, where the long-range phonon-mediated interaction (Fig.~\ref{fig:bipolaronic_stripe_mechanism}(b)) prevents global segregation and lock the charge density into stripe order.

Paralleling our observations in 1D (see Sec.~\ref{sec:1D}), these findings reconfirm that unbiased resolution of spatial fluctuations is critical for capturing the physics of soft phonons. Finally, the observation that the stripe's axial period ($16$ sites) corresponds to twice the inverse doping ($2/\delta = 16$) suggests a lock-in mechanism driven by commensurability energy~\cite{bakCommensuratePhasesIncommensurate1982,schulzCriticalBehaviorCommensurateincommensurate1980}. This implies that in the thermodynamic limit, incommensurate charge orders may emerge at irrational doping levels~\cite{ohgoeCompetitionSuperconductingAntiferromagnetic2017}.

\section{\label{sec:conclusion_outlook}Conclusion and Outlook}
We have demonstrated the scalability and effectiveness of the hybrid NGS-MPS method for treating systems with both e-e and e-ph interactions that are computationally intractable at the length scales required to resolve competing orders. By encoding non-local e-ph entanglement into non-Gaussian transformations, the method bridges the scale gap: in the transformed frame, it renders the unbounded local Hilbert space of weakly correlated phonons tractable via Gaussian states, while ensuring a faithful description of strongly correlated electrons through MPS. Paired with this formulation is a self-consistent optimization workflow robust against local minima, enabling scalable simulations with only moderate computational overhead relative to pure fermionic MPS optimization. In this work, we target the ground states of models with diagonal e-ph coupling~\cite{HOLSTEIN1959325}, demonstrating the versatility of this method through extensive applications in both 1D and 2D.

Our validation against NGSED~\cite{wangPhononMediatedLongRangeAttractive2021} on generalized HH rings yields quantitative agreement regarding the anomalous phonon-mediated attraction, a finding pivotal to interpreting recent ARPES observations~\cite{chenAnomalouslyStrongNearneighbor2021}. Moreover, the pronounced tendency toward PS observed in larger systems highlights the necessity of studying extended system sizes to accurately capture long-wavelength physics. NGS-MPS calculations that break translational symmetry yield lower energy states than their translationally invariant counterparts ($\text{NGS-MPS}^*$). The energetic preference for PS confirms that faithfully resolving spatial fluctuations is critical for capturing density-wave instabilities. A comparative study against LBO further identifies a broad range of doping levels where PS dominates over the previously predicted LL phase~\cite{tangTracesElectronPhononCoupling2022}, as substantiated by both the macroscopically segregated density profile and Maxwell construction~\cite{shiVariationalApproachManyBody2020}. These results highlight the method's reliability in soft-phonon regimes. This regime features extensive near-degeneracies compounded by extended phonon dressing clouds exhibiting large displacement and squeezing, challenging wavefunction-based methods.

Extending the method to 2D geometries, we reproduce NGSED results~\cite{wangZerotemperaturePhasesTwodimensional2020,wangFluctuatingNatureLightEnhanced2021} on small clusters. We further scale our simulations to previously inaccessible system sizes. At half-filling, we establish the phase diagram for HH models on a $48\times 4$ cylinder, identifying the metallic phase intervening between the CDW and AFM phases. Access to various correlation functions and to the effective Hamiltonian allows us to trace the microscopic origin of this intermediate regime. It arises from the interplay between non-local phonon-mediated attraction at finite frequencies and local Hubbard repulsion, while narrowing at strong coupling due to polaronic dressing. Furthermore, the emergence of a plaquette $d$-wave instability on 4-leg cylinders, alongside sizeable charge orders on 4-leg ladders for accessible couplings, implies a vanishing critical coupling for CDW onset in the Holstein limit. Combined with the even-odd spin gap effects associated with the Haldane conjecture~\cite{HALDANE1983464,haldaneNonlinearFieldTheory1983}, these finite-width effects suggest that the intermediate regime will shrink and shift toward the CDW sector as $L_y \to \infty$, consistent with QMC results~\cite{costaPhaseDiagramTwodimensional2020,weberTwodimensionalHolsteinHubbardModel2018,nowadnickCompetitionAntiferromagneticChargeDensityWave2012}.

In the computationally demanding regime of $1/8$-doping, we elucidate the role of low-frequency phonons ($\omega \leq t$) in stabilizing stripe orders. Doping the AFM parent compound stabilizes the fully filled stripe phase, characterized by a preferential enhancement of charge correlations over spin correlations. We attribute this asymmetry to a local retardation effect: while softer phonons effectively pin the static charge order, their ability to screen and thereby enhance the rapid spin fluctuations progressively weakens. Using perturbation theory, we demonstrate that the measured double occupancy corroborates this picture: its global intensity tracks the progressively weakening screening response, while its spatial modulation reflects the robustly pinned charge order. Conversely, the stabilization mechanism becomes fundamentally non-local upon doping the CDW insulator. In this regime, while the local self-trapping potential can drive PS, the long-range phonon-mediated interactions suppress this macroscopic segregation~\cite{EMERY1993597}. Such competition stabilizes a novel $16\times 2$ bipolaronic stripe that energetically preempts the PS reported in VMC studies restricted to smaller unit cells~\cite{karakuzuStripeCorrelationsTwodimensional2022,ohgoeCompetitionSuperconductingAntiferromagnetic2017,karakuzuSuperconductivityChargedensityWaves2017}. These findings reinforce the necessity of an unbiased treatment of spatial fluctuations to capture the retardation effects induced by soft phonons. Finally, the matching of the stripe's axial period ($16$ sites) to twice the inverse doping ($2/\delta$) indicates a lock-in mechanism driven by commensurability energy~\cite{bakCommensuratePhasesIncommensurate1982,schulzCriticalBehaviorCommensurateincommensurate1980}, suggesting that incommensurate charge orders may emerge at irrational dopings in the thermodynamic limit~\cite{ohgoeCompetitionSuperconductingAntiferromagnetic2017}.

The NGS-MPS approach opens several promising avenues for future research, encompassing both novel physical regimes and methodological advancements. An immediate extension involves incorporating additional physical ingredients, such as phonon dispersion~\cite{zoliNonlocalElectronphononCorrelations2005,costaPhononDispersionCompetition2018,knorzerSpinHolsteinModelsTrappedIon2022}, next-nearest-neighbor hopping~\cite{jiangSuperconductivityDopedHubbard2019,pavariniBandStructureTrendHoleDoped2001,linTwodimensionalHubbardModel1987}, and long-range Coulomb interactions~\cite{EMERY1993597}. Beyond the current class of systems, designing novel non-Gaussian transformations or recently developed superpositions of Gaussian states~\cite{zhangAttractiveRepulsiveAngulons2025,quVariationalApproachDynamics2025} offers a pathway to capture more intricate forms of e-ph entanglement. This evolution provides access to broader system classes, such as LGT~\cite{benderVariationalMonteCarlo2023} and models with off-diagonal e-ph coupling~\cite{wangRobustDwaveSuperconductivity2025,caiHightemperatureSuperconductivityInduced2025}. In parallel, upgrading the fermionic solver provides a natural route toward scalable simulations in higher dimensions. Promising candidates include projected entangled-pair states (PEPS)~\cite{verstraeteRenormalizationAlgorithmsQuantumMany2004} and neural quantum states (NQS)~\cite{carleoSolvingQuantumManybody2017}. Moreover, the method extends naturally to finite-temperature~\cite{shiVariationalApproachManyBody2020,liTangentSpaceApproach2023} and non-equilibrium dynamics~\cite{wangFluctuatingNatureLightEnhanced2021,shiVariationalStudyFermionic2018,hacklGeometryVariationalMethods2020}, providing a versatile tool for exploring quantum matter both in and out of equilibrium.

Apart from its role as a standalone solver, NGS-MPS exhibits strong interoperability with other state-of-the-art numerical methods. It can generate physically informed initial states to bootstrap more computationally intensive calculations, such as LBO~\cite{zhangDensityMatrixApproach1998,stolppComparativeStudyStateoftheart2021,brocktMatrixproductstateMethodDynamical2015,guoCriticalStrongCouplingPhases2012}. More fundamentally, non-Gaussian transformations map the Hamiltonian to a vastly compressed and physically relevant many-body basis, thereby extending the reach of wavefunction-based methods. This potent reduction of the effective Hilbert space is exemplified in spin-boson models~\cite{guoCriticalStrongCouplingPhases2012}, where even a naive Lang--Firsov transformation~\cite{langKineticTheorySemiconductors1963} increases the accessible local phonon cutoff by eight orders of magnitude. This complementary approach not only enables technically unbiased methods to probe exotic emergent correlations but also enriches our physical understanding of complex e-ph interactions.

% $'$ denotes the math prime. Othereise the single quote as in text.
% the paragraph after the equation in widetext have no indent if no blank line between.

\begin{acknowledgments}
S.J. thanks Qiaoyi Li for insightful discussions regarding the numerical simulations. The MPS computations were performed using the open-source package \texttt{FiniteMPS.jl}~\cite{Li_FiniteMPS_jl}, utilizing \texttt{TensorKit.jl}~\cite{Devos_TensorKit_jl_A_Julia_2025} as the underlying tensor backend. This work was supported by National Key Research and Development Program of China (Grant No. 2021YFA0718304), by the NSFC (Grants No.12525413, No.12135018, and No.12047503), and by CAS Project for Young Scientists in Basic Research (Grant No. YSBR-057).
\end{acknowledgments}

\appendix
\section{\label{app:tech_details}Methodological Implementation and Theoretical Derivations}
This appendix supplements the main text with detailed derivations and technical implementations. We begin with the MPS construction of fermionic Gaussian states and the derivation of the EoMs for NGS parameters. We subsequently provide analytical expressions for expectation values with respect to the full variational wavefunction. Finally, we employ second-order perturbation theory on the effective electronic Hamiltonian to estimate the double occupancy in the strong-coupling limit.

\subsection{Brief Introduction to Fermionic Gaussian MPS}\label{app:gaussianMPS}
This section outlines the construction of the MPS representation for fermionic Gaussian states~\cite{fishmanCompressionCorrelationMatrices2015,shiVariationalStudyFermionic2018}, which we employ to generate robust initial seeds for the optimization workflow. A fermionic Gaussian state~\cite{shiVariationalStudyFermionic2018} is defined as $|\Psi_\mathrm{f}\rangle=U_\mathrm{f}|0\rangle_\mathrm{f}$, where $|0\rangle_\mathrm{f}$ is the vacuum and $U_\mathrm{f}$ is a unitary transformation quadratic in the fermionic creation and annihilation operators:
\begin{equation}
    U_\mathrm{f} = e^{i\theta} e^{\frac{i}{2}C^\dagger \xi C}.
\end{equation}
Here, $\xi$ is Hermitian. Utilizing the Baker-Campbell-Hausdorff (BCH) formula, the transformation $U^\dagger_\mathrm{f} C U_\mathrm{f} = \mathcal{U}_\mathrm{f} C$ yields a linear map defined by the unitary matrix $\mathcal{U}_\mathrm{f} = e^{i\xi}$. This transformation mixes the creation and annihilation operators among the $N_\mathrm{f}$ modes. 

For states with a definite particle number $N_\mathrm{f}$, a more convenient form is $|\Psi_\mathrm{f}\rangle = U_\mathrm{f}|N_\mathrm{f}\rangle$, where $|N_\mathrm{f}\rangle$ is a product state with $N_\mathrm{f}$ occupied modes. In this form, the particle and hole sectors decouple. The basis transformation thus reduces to an $\mathrm{SU}(N_\mathrm{f})$ rotation, permitting factorization into a sequence of consecutive local $\mathrm{SU}(2)$ rotations. This decomposition enables the efficient construction of the MPS representation~\cite{fishmanCompressionCorrelationMatrices2015,itensor}: each local rotation acts on two adjacent modes, indexed by $j$ and $j+1$. In second-quantized form, this rotation takes the form
\begin{equation}
R(\theta_j) = \exp(i\theta_j\sum_{\sigma}(c^\dagger_{j,\sigma} c_{j+1,\sigma} + h.c.)).
\end{equation}
We remark that the scalar operator (spin-$0$) $\sum_{\sigma}(c^\dagger_{j,\sigma} c_{j+1,\sigma} + h.c.)$ acts on the reduced subspace respecting $\mathrm{SU}(2)_\text{spin}\otimes\mathrm{U}(1)_\text{charge}$ symmetry~\cite{Devos_TensorKit_jl_A_Julia_2025} and the exponential-times-vector on the adjacent tensor ($j$-th and $j+1$-th) can be efficiently performed~\cite{doi:10.1137/100788860}.

In 1D, strong quantum fluctuations prevent the spontaneous breaking of global continuous symmetries at zero temperature. Consequently, superconducting order parameters vanish ($\langle cc\rangle=0$) and $\mathrm{SU}(2)_\text{spin}$ symmetry is preserved. In 2D, we still optimize the MPS within the $\mathrm{SU}(2)_\text{spin}\otimes \mathrm{U}(1)_\text{charge}$ symmetry sector for computational efficiency and faithful symmetry preservation. In our simulations, Hartree-Fock wavefunctions provide reliable and efficient initial seeds. Fermionic Gaussian states may be employed for specialized applications, where the corresponding MPS construction requires explicit symmetry restoration~\cite{jinMatrixProductStates2022, bertschSymmetryRestorationHartreeFockBogoliubov2012}.

\subsection{Derivation of EoMs}\label{app:EoM}
This appendix details the derivation of the EoMs for the NGS parameters. The derivation is based on the McLachlan variational principle~\cite{McLachlan01011964}. We project ITE,
\begin{equation}\label{eq:imag_evolution}
    d_\tau|\Psi(\tau)\rangle=-(H-\langle H\rangle)|\Psi(\tau)\rangle,
\end{equation}
onto the tangent space of the variational manifold~\cite{shiVariationalStudyFermionic2018,hacklGeometryVariationalMethods2020}:
\begin{equation}
    \langle V_\nu|d_\tau \xi^\mu|V_\mu\rangle = -\langle V_\nu|\delta H|\Psi\rangle,
\end{equation}
where $|V_\mu\rangle = Q_\Psi\partial_{\mu}|\Psi\rangle=(1-|\Psi\rangle\langle\Psi|)\frac{\partial |\Psi\rangle}{\partial \xi^\mu}$ is the tangent vector. Note that all physical states reside in the projective Hilbert space, ensuring $\langle\Psi|V_\mu\rangle=0$~\cite{hacklGeometryVariationalMethods2020}. The above equation constitutes the projected gradient descent, minimizing the local error in state norm $||d_\tau|\Psi\rangle + (H-\langle H\rangle)|\Psi\rangle||_F$~\cite{hacklGeometryVariationalMethods2020}:
\begin{equation}\label{eq:projected_gradDescent}
    d_\tau\xi_\mu=-\sum_{\nu}G_{\mu\nu}\frac{\delta \langle H\rangle}{\delta\xi_{\nu}},
\end{equation}
where $G_{\mu\nu}=(2\Re\langle V_\mu|V_{\nu}\rangle)^{-1}$ and $\langle V_\mu|V_{\nu}\rangle$ is the Gram matrix defined on the tangent space.

We choose the variational state as in Eq.~\eqref{eq:ansatz}:
\begin{equation}
        |\Psi\rangle = U_S(\tilde{\lambda})|\Psi_\mathrm{b}(\Delta_R,\Gamma_R)\rangle|\Psi_\mathrm{e}\rangle,
\end{equation}
where $U_S(\tilde{\lambda})=e^{i \sum_{lj \sigma} p_{l} \tilde{\lambda}_{lj \sigma} n_{j\sigma}}$. We note that the electron state $|\Psi_{e}\rangle$ is not parameterized here; instead, we first derive flow equations exclusively for $\tilde{\lambda},~\Delta_R$ and $\Gamma_R$. The Hamiltonian [Eq.~\eqref{eq:diagonal_e-ph_H}] in the transformed frame $\bar{H} = U_S^\dagger H U_S$ reads: 
\begin{align}
    \bar{H} =& H_\mathrm{ph} + \sum_{ij\sigma}\Xi_{ij\sigma}t_{ij}c_{i\sigma}^\dagger c_{j\sigma} + \sum_{ln\sigma}x_l\tilde{g}_{ln}n_{n\sigma} \nonumber\\
    & +\frac{1}{2}\sum_{jj'\sigma \sigma'}n_{j\sigma }n_{j'\sigma'}V_{jj'\sigma\sigma'} + H_{\mathrm{e}\text{-}\mathrm{e}},
\end{align}
with $\Xi_{ij\sigma}=e^{-i\sum_l p_l \Lambda_{l,ij\sigma}}$ and $\Lambda_{l,mn\sigma} = \tilde{\lambda}_{lm\sigma}-\tilde{\lambda}_{ln\sigma}$. The reduced e-ph coupling $\tilde{g}$ and the induced phonon-mediated interaction $V$ is defined in Eq.~\eqref{eq:reduced_g} and Eq.~\eqref{eq:V_eff}, respectively.

The bosonic Gaussian state takes the form:
\begin{equation}
    \begin{aligned}
        |\Psi_\mathrm{b}(\Delta_R,\Gamma_R)\rangle &= U_\mathrm{b}(\Delta_R,\Gamma_R)|0\rangle_\mathrm{b}, \\
        &=e^{\frac{i}{2}R^T\Omega\Delta_R}e^{- \frac{i}{4}R^T\xi_\mathrm{b}R}|0\rangle_\mathrm{b},
    \end{aligned}
\end{equation}
where $\Delta_R^T$ is the displacement vector, $\Gamma_R$ is the covariance matrix, $|0\rangle_\mathrm{b}$ is the bosonic vacuum, and $\xi_\mathrm{b} = \begin{pmatrix}\xi^{xx} & \xi^{xp} \\ \xi^{px} & \xi^{pp}\end{pmatrix}$. The transformation $U^\dagger_\mathrm{b}R U_\mathrm{b}=S_\mathrm{b} R+\Delta_R$ yields a linear map defined by the displacement vector and the symplectic matrix $S_\mathrm{b} = e^{\Omega \xi_\mathrm{b}}$~\cite{shiVariationalStudyFermionic2018}. The covariance matrix for a pure state is expressed in terms of $S_\mathrm{b}$:
\begin{equation}
    \Gamma_R = S_\mathrm{b}S_\mathrm{b}^T.
\end{equation}

We project the time derivative on the left hand side (LHS) of Eq.~\eqref{eq:imag_evolution} onto the tangent space:
\begin{equation}\label{eq:timeDerivative_tangent}
    Q_\Psi\partial_\tau|\Psi(\tau)\rangle = Q_\Psi U_{S}(OU_\mathrm{b}+\partial_\tau U_\mathrm{b})|0\rangle_\mathrm{b}|\Psi_\mathrm{e}\rangle,
\end{equation}
where $O=U_S^{-1}\partial_\tau U_S$. We expand $U^\dagger_\mathrm{b}OU_\mathrm{b}$ in the normal ordering form:
\begin{equation} \label{eq:normalOrderingExpansion_O_NGSMPS}
    U_{\mathrm{b}}^{\dagger} O U_{\mathrm{b}} = \langle O\rangle+\frac{1}{2}R^T S_\mathrm{b}^T O_{\Delta}+\delta O.
\end{equation}
Here, the expectation value of $O$ on $|\Psi_\mathrm{b}\rangle|\Psi_\mathrm{e}\rangle$ reads:
\begin{equation*}
    \langle O\rangle = i \sum_{lj\sigma}(\Delta_p)_l (\partial_\tau \tilde{\lambda}_{lj\sigma})\langle n_{j\sigma}\rangle_\mathrm{e}.
\end{equation*}
The coefficient of the linear term reads:
\begin{equation*}
    (O_{\Delta})_{l} = 2\frac{\delta\langle O\rangle_\mathrm{GS}}{\delta(\Delta)_{l}} = \begin{pmatrix}
        \mathbf{0}_l \\
        2i \sum_{j\sigma}(\partial_\tau \tilde{\lambda}_{lj\sigma})\langle n_{j\sigma}\rangle_\mathrm{e}
    \end{pmatrix}.
\end{equation*}
The term beyond quadratic normal order reads:
\begin{equation*}
    \delta O = i \sum_{l, m \sigma}(R^T S_\mathrm{b}^T)_{p;l} \partial_\tau \tilde{\lambda}_{l, m \sigma}\delta n_{m\sigma}.
\end{equation*}
We note that the substitution $\Delta_p=0$ (imposed by TRS) is performed only after the functional differentiation. We write Eq.~\eqref{eq:timeDerivative_tangent} out explicitly~\cite{shiVariationalStudyFermionic2018}:
\begin{equation*}
    \begin{split}
        Q_\Psi\partial_\tau|\Psi(\tau)\rangle = \, & U_{S} U_{\mathrm{b}}\biggl[\frac{1}{2}R^TS_\mathrm{b}^T (O_{\Delta} -\sigma^{y}\partial_\tau\Delta_R) \\
        & + \frac{1}{4}i:R^T S_{b}^{T} \sigma (\partial_\tau S_{b}) R: + \delta O\biggr]|0\rangle_\mathrm{b}|\Psi_\mathrm{e}\rangle,
    \end{split}
\end{equation*}
from which we identify the tangent basis comprised by three vectors: $U_{S} U_{\mathrm{b}}b_j^\dagger |0\rangle_\mathrm{b}\otimes|\Psi_\mathrm{e}\rangle$, $U_{S} U_{\mathrm{b}}b_j^\dagger b_{j'}^\dagger |0\rangle_\mathrm{b}\otimes|\Psi_\mathrm{e}\rangle$, and $U_{S} U_{\mathrm{b}}b_j^\dagger |0\rangle_\mathrm{b}\otimes\delta n_{j'\sigma}|\Psi_\mathrm{e}\rangle$.

The transformed Hamiltonian can also be thrown into the normal ordering form: 
\begin{equation*}
    U_{\mathrm{b}}^{\dagger} \bar{H} U_{\mathrm{b}}=\langle H\rangle+\frac{1}{2} R^T S_\mathrm{b}^T h_{\Delta}+\frac{1}{4}: R^T S_\mathrm{b}^T h_\mathrm{b} S_\mathrm{b} R:+\delta h.
\end{equation*}
Here, the constant term reads:
\begin{align} \label{eq:expectation_H}
    \langle \bar{H}\rangle =& E_\mathrm{ph} + \sum_{mn\sigma}[\tilde{t}_{mn} \langle c_{m\sigma}^\dagger c_{n\sigma} \rangle_\mathrm{e} + (\Delta_x)_m\tilde{g}_{mn\sigma}\langle n_{n\sigma}\rangle_\mathrm{e}] \nonumber\\
    & + \frac{1}{2}\sum_{mn\sigma\sigma'}V_{mn\sigma\sigma'} \langle n_{m\sigma} n_{n\sigma'}\rangle_\mathrm{e} + \langle H_{\mathrm{e}\text{-}\mathrm{e}} \rangle_\mathrm{e}.
\end{align}
where the phononic energy $E_\mathrm{ph}$ and the renormalized hopping amplitude $\tilde{t}_{mn\sigma}=\langle \Xi_{mn\sigma}\rangle_\mathrm{b}t_{mn}$ is defined in Eq.~\eqref{eq:E_ph} and Eq.~\eqref{eq:t_eff}, respectively. Moreover, the effective electronic Hamiltonian, defined in Eq.~\eqref{eq:H_eff}, can be identified from Eq.~\eqref{eq:expectation_H}. The coefficient of the linear term reads:
\begin{equation}
    h_{\Delta} = \begin{pmatrix}
        \omega_{ln}(\Delta_x)_n +2\tilde{g}_{ln\sigma}\langle n_{n\sigma}\rangle_\mathrm{e} \\
        \omega_{ln}(\Delta_p)_n-2i\Lambda_{lmn\sigma}\tilde{t}_{mn\sigma}\langle c_{m\sigma}^\dagger c_{n\sigma}\rangle_\mathrm{e}
    \end{pmatrix}.
\end{equation}
The coefficient of the quadratic term reads:
\begin{equation}
    h_R = 4 \frac{\delta\langle\bar{H}\rangle}{\delta (\Gamma_R)} = \begin{pmatrix}
        \omega_{ll'} & \\
        & \omega_{ll'}-\kappa_{ll'}
    \end{pmatrix}, \\
\end{equation}
with $\kappa_{ll'}$ defined in Eq.~\eqref{eq:kappa}. Moreover, the higher-order fluctuation reads:
\begin{equation}
    \begin{aligned}
    \delta h =& \delta_1: c_{m\sigma}^\dagger c_{n\sigma}: + \delta_2\delta n_{n\sigma} + \delta h', \\
    \delta_1 =& -i(S_\mathrm{b}R)_{p;l}^T \Lambda_{lmn\sigma}\tilde{t}_{mn\sigma}, \\
    \delta_2 =& (S_\mathrm{b}R)_{x;l}^T\tilde{g}_{ln\sigma},
    \end{aligned}
\end{equation}
where $\delta h'$ denotes the fluctuation beyond the tangent space. Based on the normal-ordered Hamiltonian and the tangent vectors, the EoMs for
$\Delta_x$, $\Gamma_{pp}$, and $\tilde{\lambda}$ [Eqs.~\eqref{eq:Δx}--\eqref{eq:λ}]
follow directly~\cite{shiVariationalStudyFermionic2018}. We note that the variational manifold is non-Kahler~\cite{hacklGeometryVariationalMethods2020} and we take the real parts during projection. TRS dictates that quantities such as $\Im\langle c_{j\sigma}^\dagger c_{m\sigma}\rangle_\mathrm{e}$ and $\Im\langle \delta n_{j\sigma'}:c_{n\sigma}^\dagger c_{m\sigma}:\rangle_\mathrm{e}$ to vanish, ensuring the consistency of the resulting EoMs. Otherwise, imaginary parts generally give constraint and may not admit consistent stationary solutions~\cite{hacklGeometryVariationalMethods2020,shiVariationalStudyFermionic2018}.

In the initialization stage, the electronic state is described by a fermionic Gaussian state $|\Psi_\mathrm{e} (\Gamma_\mathrm{f})\rangle = e^{\frac{i}{2}C^\dagger \xi C}|0\rangle_\mathrm{e}$. The state is fully characterized by the covariance matrix $\Gamma_\mathrm{f}$:
\begin{equation} \label{eq:covariance_fermion}
    \begin{split}
        \Gamma_\mathrm{f} &= \langle C C^\dagger \rangle = U_\mathrm{f}
        \begin{pmatrix}
            \mathbf{1} & 0 \\
            0 & 0
        \end{pmatrix}
        U_\mathrm{f}^\dagger \\
        &= \begin{pmatrix}
            \langle c c^\dagger \rangle_\mathrm{e} & \langle c c \rangle_\mathrm{e} \\
            \langle c^\dagger c^\dagger \rangle_\mathrm{e} & \langle c^\dagger c \rangle_\mathrm{e}
        \end{pmatrix}
        = \begin{pmatrix}
            \mathbf{1} - \Gamma_1^T & \Gamma_2 \\
            \Gamma_2^* & \Gamma_1
        \end{pmatrix}.
    \end{split}
\end{equation}
Here, the normal and anomalous correlations are defined as $\Gamma_1=\langle c^\dagger c\rangle_\mathrm{e}$ and $\Gamma_2=\langle c c\rangle_\mathrm{e}$, respectively. The flow equation for $\Gamma_\mathrm{f}$ derived below is integrated together with Eqs.~\eqref{eq:Δx}--\eqref{eq:λ}. We note that the normal-ordering expansion is now respect to the vacuum $|0\rangle_\mathrm{e}$ instead of $|\Psi_\mathrm{e}\rangle$. Equation~\eqref{eq:normalOrderingExpansion_O_NGSMPS} now reads:
\begin{align}
    \tilde{O} = & U_{\mathrm{GS}}^{\dagger} O U_{\mathrm{GS}}, \\
    = & \langle O\rangle_{\mathrm{GS}}+\frac{1}{2}R^T S_\mathrm{b}^T O_{\Delta}+\frac{i}{2}: C^\dagger U^\dagger_\mathrm{f} O_\mathrm{f} U_\mathrm{f} C:+\delta O.\nonumber
\end{align}
where $U_{\mathrm{GS}}=U_\mathrm{b}U_\mathrm{f}$. The additional term quadratic in fermionic creation/annihilation operators reads:
\begin{equation*}
    (O_\mathrm{f})_{n\sigma} = 2\frac{\delta\langle O\rangle_\mathrm{GS}}{\delta\langle n_{n\sigma}\rangle} = 2i \sum_{l}\Delta^p_l (\partial_\tau \tilde{\lambda}_{ln\sigma}).
\end{equation*}
The corresponding new tangent vector $\frac{1}{2}:C^{\dagger}U^{\dagger}_\mathrm{f}(O_\mathrm{f}U_\mathrm{f}+\partial_\tau U_\mathrm{f}) C:$ is orthogonal to the  existing tangent vectors, leaving the derivation of Eqs.~\eqref{eq:Δx}--\eqref{eq:λ} unaffected. Additionally, Wick's theorem is now available for factorizing the density-density correlation, $\langle n_{m\sigma} n_{n\sigma'}\rangle_\mathrm{e}$, which shows up in Eq.~\eqref{eq:λ}. As for the normal-ordering expansion of $U_{\mathrm{GS}}^{\dagger} \bar{H} U_{\mathrm{GS}}$, we have:
\begin{equation*}
    \begin{aligned}
        U_{\mathrm{GS}}^{\dagger} \bar{H} U_{\mathrm{GS}} = &\langle \bar{H}\rangle_{\mathrm{GS}}+\frac{1}{2} R^T S_\mathrm{b}^T h_{\Delta}+\frac{1}{4}: R^T S_\mathrm{b}^T h_\mathrm{b} S_\mathrm{b} R:\\
        &+\frac{1}{2}: C^\dagger U_\mathrm{f}^\dagger h_\mathrm{f} U_\mathrm{f} C:+\delta h.
    \end{aligned}
\end{equation*}
Here, the constant term, $\langle \bar{H}\rangle_{\mathrm{GS}}$, is the mean-field energy. The additional mean-field electronic Hamiltonian contains the Bogoliubov contribution $\Delta$,
\begin{equation}
    \Delta_{mn\sigma\sigma'} = -2\bar{V}_{mn\sigma\sigma'}\langle c_{m \sigma} c_{n\sigma'}\rangle_\mathrm{e}, \\
\end{equation}
where $\bar{V}_{mn\sigma\sigma'} = V_{mn\sigma\sigma'}+\frac{1}{2}U\delta_{mn}\delta_{\sigma\neq\sigma'}$, and the Hartree-Fock contribution $\mathcal{E}$,
\begin{equation}
    \begin{aligned}
        \mathcal{E}_{mn\sigma\sigma'} = - & 2\bar{V}_{mn \sigma\sigma'}\langle c_{n \sigma'}^{\dagger} c_{m \sigma}\rangle_\mathrm{e} + \tilde{t}_{mn\sigma}\delta_{\sigma \sigma'} \\
        + & \Bigl[\sum_{l}(\Delta_x)_l \tilde{g}_{lm\sigma}-\mu+V_{mm\sigma\sigma}\\
        & \hspace{10pt}+2\sum_{j \sigma''}\bar{V}_{jn \sigma''\sigma'}\langle n_{j\sigma''}\rangle_\mathrm{e}\Bigr] \delta_{mn}\delta_{\sigma \sigma'}.
    \end{aligned}
\end{equation} 
Here the chemical potential $\mu$ is nonzero only when the Bogoliubov contribution is included ($\langle cc\rangle_\mathrm{e}\neq0$), in which case a Legendre transformation is employed to fix the particle number. We determine $\mu$ self-consistently by requiring $\operatorname{tr}(\partial_{\tau}\Gamma_1)=0$. This construction is flow-embedded, i.e., the gradient flow is additionally confined to the surface conserving the particle number.

We write out the EoM for the electronic covariance matrix [Eq.~\eqref{eq:covariance_fermion}], $\partial_\tau \Gamma_{f} = \{h_f, \Gamma^2_{f}\} - 2 \Gamma_{f} h_f \Gamma_{f}$~\cite{shiVariationalStudyFermionic2018}, in terms of the Hartree-Fock and Bogoliubov components explicitly:
\begin{subequations}
    \begin{align}
        \label{eq:HartreeFock}
        \partial_{\tau} \Gamma_1 = &\mathcal{D}_1 + \mathcal{D}^\dagger_1, \\
        \label{eq:Bogoliubov}
        \partial_{\tau} \Gamma_2 = &\mathcal{D}_2 - \mathcal{D}^T_2,
    \end{align}
\end{subequations}
where we have utilized $\Gamma_\mathrm{f}^2=\Gamma_\mathrm{f}$ for pure states. The chemical potential is $\mu = -\operatorname{tr}(\mathcal{D}_1 + \mathcal{D}^\dagger_1)/(4||\Gamma_2||^2)$. $\mathcal{D}_1$ and $\mathcal{D}_2$ take the forms:
\begin{align*}
    \mathcal{D}_1 &= \Delta^{\dagger} \Gamma_2 - \mathcal{E}^{T} \Gamma_1 - \Gamma_2^{\dagger} \mathcal{E} \Gamma_2 + \Gamma_1 \mathcal{E}^T \Gamma_1 - 2 \Gamma_1 \Delta^{\dagger} \Gamma_2, \\
    \mathcal{D}_2 &= -\Delta \Gamma_1 - \mathcal{E} \Gamma_2 - \Gamma_2 \Delta^{\dagger} \Gamma_2 + \Gamma_1^{T} \Delta \Gamma_1+ 2 \Gamma_1^{T} \mathcal{E} \Gamma_2 + \frac{1}{2}\Delta.
\end{align*}

\subsection{\label{app:transformed_expectations}Expectation values on the full variational wavefunction}
As mentioned in the main text, expectation values with respect to the full variational state [Eq.~\eqref{eq:ansatz}] are denoted by $\langle\cdots\rangle$, while those with respect to the electronic wave function $|\Psi_\mathrm{e}\rangle$ are denoted by $\langle\cdots\rangle_\mathrm{e}$. We express key observables $\langle\cdots\rangle$ in terms of $\langle\cdots\rangle_\mathrm{e}$ and the associated dressing factors. Note that the density operators are invariant under the unitary transformation $U_S$. Accordingly, related observables, such as the particle number $\langle n_{i\sigma}\rangle$ and density-density correlation $\langle n_{i\sigma} n_{j\sigma'}\rangle$, do not acquire additional dressing factors.

In contrast, the single-particle correlation acquires a polaronic dressing prefactor:
\begin{equation}
    \begin{aligned}
        \langle c_{i\sigma}^\dagger c_{j\sigma}\rangle &= \langle e^{-i\sum_l p_l \Lambda_{l,ij\sigma}}\rangle_\mathrm{ph} \langle c_{i\sigma}^\dagger c_{j\sigma}\rangle_\mathrm{e} \\
        &= \exp(-\frac{1}{2}\sum_{ll'} \Lambda_{l,ij\sigma}(\Gamma_{pp})_{ll'}\Lambda_{l',ij\sigma}) \langle c_{i\sigma}^\dagger c_{j\sigma}\rangle_\mathrm{e}.
    \end{aligned}
\end{equation}
The superconducting pair correlation is evaluated similarly. The corresponding dressing factor for the four-point correlator $\langle c_{i+\delta_\alpha,\sigma_1}^\dagger c_{i,\sigma_2}^\dagger c_{j,\sigma_3}c_{j+\delta_\beta,\sigma_4}\rangle$ is 
\begin{equation}
    F_{ij}^{\alpha\beta} = \exp(-\frac{1}{2}\sum_{ll'} \bar{\Lambda}_{l} (\Gamma_{pp})_{ll'} \bar{\Lambda}_{l'}),
\end{equation}
where the effective coupling vector is defined as $\bar{\Lambda}_{l} \equiv \tilde{\lambda}_{l,j+\delta_\beta,\sigma_4} + \tilde{\lambda}_{l,j,\sigma_3} - \tilde{\lambda}_{l,i+\delta_\alpha,\sigma_1} - \tilde{\lambda}_{l,i,\sigma_2}$, with $\delta_{\alpha,\beta} \in \{\hat{x}, \hat{y}\}$. Different pairing symmetries can be constructed as defined in the main text [Eqs.~\eqref{eq:bond_singlet}--\eqref{eq:2D_d_x2y2}].

\subsection{Perturbative Estimation on the Double Occupancy}\label{app:estimation_doubleOccupancy}
\begin{figure*}[htbp]
    \includegraphics[width=\textwidth]{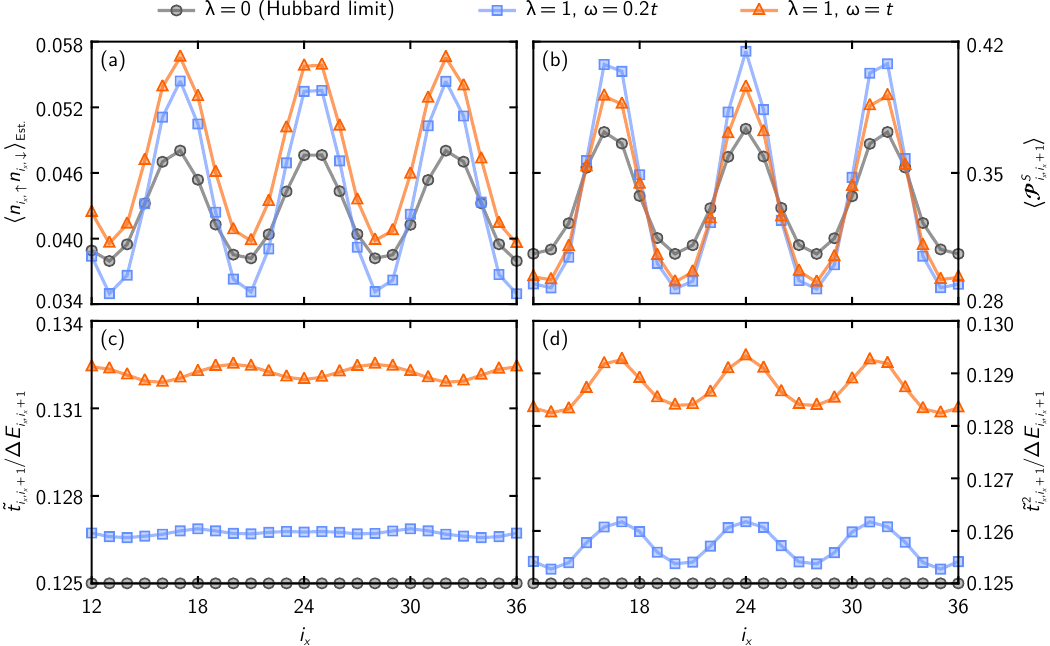}
    \caption{Microscopic decomposition of the double occupancy for the fully filled stripe phase in the $1/8$-doped Hubbard--Holstein model on a $48 \times 4$ cylinder at $u=8$. (a) Perturbative estimate of the local double occupancy $\langle n_{i_x,\uparrow}n_{i_x,\downarrow}\rangle_{\rm Est.}$ obtained from second-order virtual processes, agreeing with the double occupancy shown in Fig.~\ref{fig:combined_dopedAFM}(d). (b) Corresponding nearest-neighbor singlet weight $\langle \mathcal{P}^S_{i_x,i_x+1}\rangle$ (see main text for definition), which governs the availability of virtual doublon-holon excitations. (c) Ratio $\tilde{t}_{i_x,i_x+1}/\Delta E_{i_x,i_x+1}$ quantifying the amplitude of first-order corrections to the wavefunction. (d) The effective superexchange coupling $\tilde{t}^2_{i_x,i_x+1}/\Delta E_{i_x,i_x+1}$ controlling spin exchange. Symbols denote results for the Hubbard limit ($\lambda=0$, gray circles) and Hubbard--Holstein model ($\lambda=1$) at phonon frequencies $\omega=t$ (orange triangles) and $\omega=0.2t$ (blue squares). The modulation of panel (a) primarily follows that of the singlet weight in panel (b), reflecting its origin in virtual singlet processes on top of the stripe background, while the overall amplitude is set by panels (c) through the renormalized hopping $\tilde{t}$ and excitation energy $\Delta E$. Data are shown for bulk sites to minimize boundary effects.}
    \label{fig:effective_theory} 
\end{figure*}

In the regime of strong effective repulsion ($U-2\lambda t\gg t$), the ground state is dominated by configurations without double occupancy. We estimate the residual double occupancy by applying second-order perturbation theory to the effective electronic Hamiltonian derived in Eq.~\eqref{eq:H_eff}. The first-order correction to the wavefunction, $|\Psi^{(1)}\rangle$, arises from the hopping processes that generate high-energy doublon-holon pairs. Because of fermionic antisymmetry, doublon creation via nearest-neighbor hopping is nonzero only when the two electrons on the bond $(\mathbf{i},\mathbf{j})$ occupy a singlet spin state. It follows that the correction involves the projection of the ground state onto the neighboring-singlet manifold, $\mathcal{P}^S_{\mathbf{i},\mathbf{j}}|\Psi^{(0)}\rangle$, where $\mathcal{P}^S_{\mathbf{i},\mathbf{j}} = \frac{1}{4}n_\mathbf{i}n_{\mathbf{j}} - \mathbf{S}_\mathbf{i}\cdot\mathbf{S}_{\mathbf{j}}$. Specifically, the component of the first-order correction associated with a doublon-holon pair on the bond $(\mathbf{i}, \mathbf{i}+\delta)$ takes the form:
\begin{equation}
|\Psi^{(1)}_{\mathbf{i},\delta}\rangle = \Bigl( \frac{\sqrt{2}\tilde{t}_{\mathbf{i},\mathbf{i}+\delta}}{\Delta E_{\mathbf{i},\delta}}\eta^+_{\mathbf{i},\mathbf{i}+\delta} + \frac{\sqrt{2}\tilde{t}_{\mathbf{i},\mathbf{i}+\delta}}{\Delta E_{\mathbf{i}+\delta,-\delta}}\eta^+_{\mathbf{i}+\delta,\mathbf{i}} \Bigr) \mathcal{P}^S_{\mathbf{i},\mathbf{i}+\delta}|\Psi^{(0)}\rangle,
\end{equation}
where $\eta^+_{\mathbf{i},\mathbf{j}}$ is the operator that creates a doublon at $\mathbf{i}$ and a holon at $\mathbf{j}$ (transitioning from the singly occupied subspace). Here the virtual excitation energy $\Delta E_{\mathbf{i},\delta}$ captures the leading contribution to the doublon-holon creation,
\begin{equation}\label{eq:energy_difference_tJ}
\Delta E_{\mathbf{i},\delta}\approx U+2V_{\mathbf{i},\mathbf{i}} - \frac{1}{2}(V_{\mathbf{i}+\delta,\mathbf{i}+\delta}+V_{\mathbf{i},\mathbf{i}+\delta}+V_{\mathbf{i},\mathbf{i}}),
\end{equation}
where $V_{\mathbf{i},\mathbf{j}}$ denotes density-density interactions between site $\mathbf{i}$ and $\mathbf{j}$. The contribution of these processes to the local double occupancy is weighted by the probability of the electrons being in the singlet configurations. This weight is given by the expectation value of the singlet projector in the unperturbed ground state, $\langle \mathcal{P}^S_{\mathbf{i},\mathbf{i}+\delta} \rangle_0$. Summing over all neighboring sites $\delta$, the estimated double occupancy at site $\mathbf{i}$ reads:
\begin{equation}
    \langle n_{\mathbf{i}\uparrow}n_{\mathbf{i}\downarrow}\rangle \approx \sum_{\delta}2\Bigl(\frac{\tilde{t}_{\mathbf{i},\mathbf{i}+\delta}}{\Delta E_{\mathbf{i},\delta}}\Bigr)^2 \langle \frac{1}{4}n_\mathbf{i}n_{\mathbf{i}+\delta} - \mathbf{S}_\mathbf{i}\cdot\mathbf{S}_{\mathbf{i}+\delta}\rangle.
\end{equation}
Here, we have $\langle \mathcal{P}^S_{\mathbf{i},\mathbf{i}+\delta} \rangle_0 \simeq \langle \mathcal{P}^S_{\mathbf{i},\mathbf{i}+\delta} \rangle$ to leading order, since normalization corrections enter only at $O((\tilde{t}/\Delta E)^4)$. As illustrated in Fig.~\ref{fig:effective_theory}(a), this perturbative estimate is in excellent agreement with the full NGS-MPS calculations presented in Fig.~\ref{fig:combined_dopedAFM}(d). This consistency demonstrates the robustness of the perturbative estimate and supports the microscopic interpretation developed here.

The spatial profile of the double occupancy is fundamentally governed by two principal quantities: the effective $t$-$J$ parameters~\cite{chaoCanonicalPerturbationExpansion1978} and the singlet weight. The former involves terms proportional to $\tilde{t}/\Delta E$ (governing the double-occupancy amplitude) and $\tilde{t}^2/\Delta E$ (characterizing the superexchange interaction). The latter, $\langle \mathcal{P}^S_{\mathbf{i},\mathbf{i}+\delta} \rangle$, depends on both nearest-neighbor density and spin correlations. In the fully filled stripe phase, these correlations are dominated by the charge distribution itself, thereby inheriting its modulation (Fig.~\ref{fig:combined_dopedAFM}(a) and \ref{fig:effective_theory}(b)). To understand the stabilization of this specific charge order, we examine the effective electronic Hamiltonian [Eq.~\eqref{eq:H_eff}]. In the doped antiferromagnetic regime where $U$ dominates, the phonon-mediated onsite terms in Eq.~\eqref{eq:V_eff_final} takes the form (at site $j$):
\begin{equation}
- 2 v_j n_{j\uparrow}n_{j\downarrow} - 2(\lambda-v_j) (\langle n_j\rangle n_j - \frac{1}{2}\langle n_j\rangle^2) - v_j n_j,
\end{equation}
where $v_j=\sum_l(2g\tilde{\lambda}_{lj} - \omega\tilde{\lambda}^2_{lj})$ interpolates between $0$ ($\omega\to0$) and $\lambda$ ($\omega\to\infty$). Physically, the third term shifts the chemical potential while the first two terms drive competing tendencies. The first term reduces the Hubbard repulsion ($U_{\text{eff}} \approx U - 2v$), effectively lowering the energy difference $\Delta E$ [Eq.~\eqref{eq:energy_difference_tJ}] between the low-energy and high-energy sectors. Reducing $\Delta E$ systematically increases the effective $t$-$J$ parameters (Fig.~\ref{fig:effective_theory}(c) and \ref{fig:effective_theory}(d)), thereby promoting both charge fluctuations and spin correlations. In contrast, the second term suppresses onsite density fluctuations, pinning the charge inhomogeneity at the existing ordering vectors. In the relevant soft-phonon regime, the pinning mechanism dominates the charge sector, overriding the delocalizing tendency of the screening to lock the stripe order. Conversely, in the spin sector, the screening effect remains the primary driver, albeit a diminishing one as the phonon frequency lowers. Collectively, these two mechanisms underpin the physical picture proposed in the main text: soft phonons reinforce the stripe phase, maintaining an increasingly robust pinning field while diminishing the screening response as the frequency decreases.

Unlike the VMC studies~\cite{karakuzuStripeCorrelationsTwodimensional2022,ohgoeCompetitionSuperconductingAntiferromagnetic2017,karakuzuSuperconductivityChargedensityWaves2017} that rely on non-unitary penalty terms, our unitary transformation yields an effective electronic Hamiltonian [Eq.~\eqref{eq:H_eff}] that naturally incorporates the essential effects of phonons. As exemplified here, this framework explicitly resolves the governing local retardation mechanism, placing our physical picture on solid ground. Furthermore, the versatility of this effective Hamiltonian allows us to extend our analysis to the doped CDW regime. As discussed in the main text, we attribute the stabilization of the novel bipolaronic stripe phase to the non-local nature of the phonon-induced interaction [Eq.~\eqref{eq:V_eff_final}]. Finally, when combined with an advanced fermionic solver capable of scalable and reliable simulations in the intermediate regime, this approach offers a promising pathway to elucidate how stripe order evolves into exotic quantum phases.

\section{\label{app:additional_results}Supplementary Numerical Analysis and Benchmarks}
This appendix presents supplementary numerical data that validate and extend the findings in the main text. We begin by confirming the 1D PS instability via a Maxwell construction. Then, we present the benchmarks of the NGS-MPS method against NGS-ED on 2D clusters. Finally, we examine finite-width effects at half-filling by comparing results across 4-leg cylinders, 4-leg ladders, and 3-leg cylinders

\subsection{\texorpdfstring{Maxwell construction of 1D generalized HH model with $\text{NGS-MPS}^*$}{Maxwell construction of 1D generalized HH model with NGS-MPS*}}
\label{app:1D_PS_maxwell}
\begin{figure}[htbp]
    \includegraphics[width=\columnwidth]{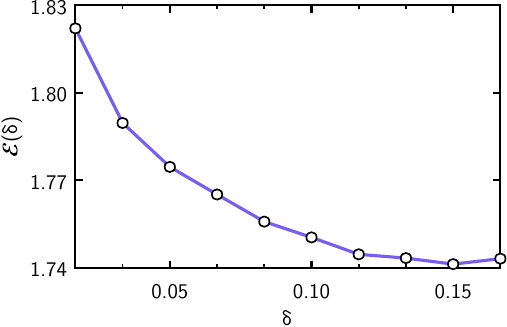}
    \caption{Energy per hole $\mathcal{E}(\delta)$ as a function of doping $\delta$ for the 1D generalized Hubbard--Holstein model ($L=120$, PBC) with parameters $u=8$, $\lambda=1$, and $\omega=0.2t$. Data are evaluated by enforcing a uniform electron density within the restricted $\text{NGS-MPS}^*$ ansatz and extrapolated to the infinite-bond-dimension limit ($D\to\infty$). The curve exhibits a global minimum at $\delta=15\%$, which thermodynamically demarcates the boundary of the phase-separation region.}
    \label{fig:1D_maxwell_construction} 
\end{figure}

As reported in Fig.~\ref{fig:1D_LBO_PS_and_corr}(a), the 1D generalized HH model exhibits PS over the doping level from $0\%$ to $15\%$. Here, we independently verify the PS via a Maxwell construction~\cite{shiVariationalApproachManyBody2020}. Under PBC, we enforce a uniform electron density and calculate the ground-state energy using the restricted $\text{NGS-MPS}^*$ ansatz. The energy per hole
\begin{equation*}
    \mathcal{E}(\delta) = \frac{E(\delta)-E(0)}{N\delta},
\end{equation*}
is plotted in Fig.~\ref{fig:1D_maxwell_construction}, where the global minimum of the energy curve is located at $\delta^* = 15\%$. The mathematical condition for this global minimum,
\begin{equation*}
    \left.\frac{d\mathcal{E}(\delta)}{d\delta}\right|_{\delta=\delta^*} = \frac{E'(\delta^*) \cdot (N\delta^*) - [E(\delta^*) - E(0)] \cdot N}{(N\delta^*)^2} = 0,
\end{equation*}
directly yields the common tangent condition of the Maxwell construction, $E'(\delta^*) = [E(\delta^*) - E(0)]/\delta^*$. This tangent line thermodynamically demarcates the phase-separation region $\delta\in(0\%,15\%)$, consistent with our NGS-MPS results of the macroscopically PS.

\subsection{\label{app:benchmark_2D} Benchmarks against NGSED in 2D torus}
\begin{figure}[htbp]
    \includegraphics[width=\columnwidth]{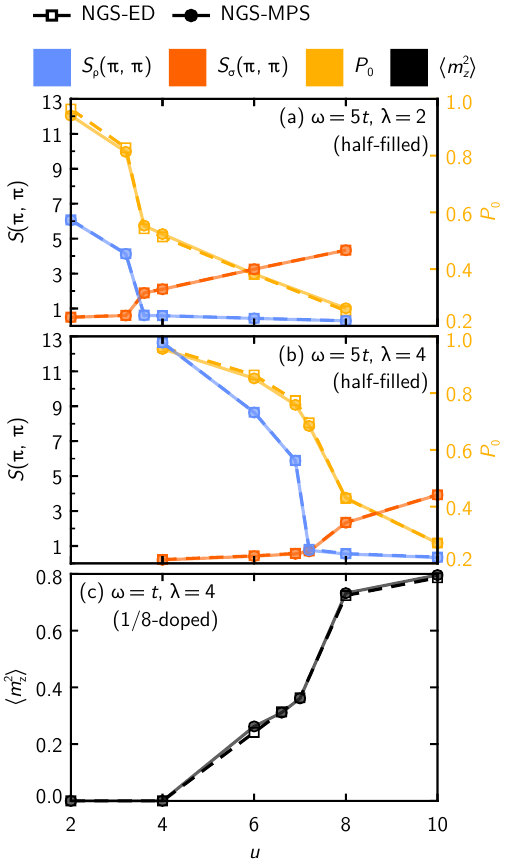}
    \caption{Benchmarks of the NGS-MPS method against NGS-ED on a $4\times 4$ torus. (a), (b) Half-filled system at phonon frequency $\omega=5t$ and electron-phonon couplings of $\lambda=2$ and $4$, respectively. Shown are the charge structure factor $S_\rho(\pi, \pi)$ (blue), the spin structure factor $S_\sigma(\pi, \pi)$ (orange), and the onsite singlet-pairing correlation $P_0$ (gold, right axis). (c) Squared local magnetic moment $\langle m_z^2 \rangle$ (black) at $1/8$ doping for $(\omega=t,\lambda=4)$. Across all panels, solid lines with filled markers denote the NGS-MPS results, and dashed lines with open markers represent the NGS-ED reference data.}
    \label{fig:benchmark} 
\end{figure}

Figure~\ref{fig:benchmark} benchmarks the NGS-MPS method against NGS-ED on a $4\times 4$ torus. We examine various physical observables, including pairing correlations, structure factors, and local moments:
\begin{subequations}
    \begin{align}
        P_\mathrm{0} &= \frac{1}{N}\sum_{\mathbf{i},\mathbf{j}}\langle c^\dagger_{\mathbf{i}\uparrow} c^\dagger_{\mathbf{i}\downarrow} c_{\mathbf{j}\downarrow} c_{\mathbf{j}\uparrow}\rangle, \\
        S_\mathrm{\sigma}(\mathbf{q}) &= \frac{1}{N}\sum_{\mathbf{i},\mathbf{j}}\langle \mathbf{S}_{\mathbf{i}} \cdot \mathbf{S}_{\mathbf{j}}\rangle e^{i\mathbf{q}\cdot(\mathbf{i}-\mathbf{j})}, \\
        S_\mathrm{\rho}(\mathbf{q}) &= \frac{1}{N}\sum_{\mathbf{i},\mathbf{j}}\langle n_\mathbf{i} n_\mathbf{j}\rangle e^{i\mathbf{q}\cdot(\mathbf{i}-\mathbf{j})}, \\
        \langle m_z^2\rangle &= \frac{1}{N}\sum_\mathbf{i}\langle(n_{\mathbf{i} \uparrow}-n_{\mathbf{i} \downarrow})^2\rangle.
    \end{align}
\end{subequations}
We find excellent agreement between the two methods at both half-filling (Figs.~\ref{fig:benchmark}(a) and \ref{fig:benchmark}(b)) and $1/8$-doping (Fig.~\ref{fig:benchmark}(c)).

\subsection{\texorpdfstring{Finite-width effects for 2D Half-filled HH model with $\omega=5t$}{Finite-width effects for 2D Half-filled HH model with omega=5t}}
\label{app:half_filled_2D}
In Sec.~\ref{sec:2D_half_filled}, we analyze pair-pair correlations in the intermediate metallic regime of the half-filled HH model at $\omega=5t$ on 4-leg cylinders. Neither the extended $s$-wave channel [Eq.~\eqref{eq:2D_s*}] nor the $d_{x^2-y^2}$-wave channel [Eq.~\eqref{eq:2D_d_x2y2}] develops quasi-long-range order.
\begin{figure}[htbp]
    \includegraphics[width=\columnwidth]{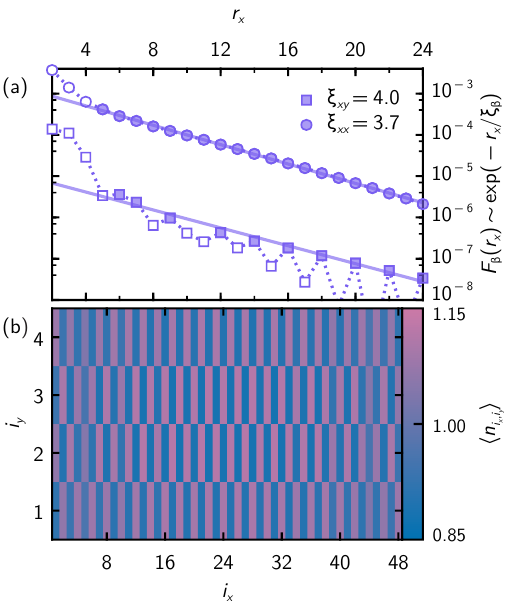}
    \caption{Supplemental results for the $4$-leg half-filled Hubbard--Holstein model at $\omega=5t$. (a) Direction-resolved pair-pair correlations in the Cooper channel, $F_{xx}$ (circles) and $F_{xy}$ (squares), extrapolated to the infinite bond dimension limit ($D\to\infty$) on a $48\times 4$ cylinder at $(\lambda=4,u=8)$. Both channels exhibit an exponential decay $F_\beta(r_x) \sim \exp(-r_x/\xi_\beta)$, indicating the absence of quasi-long-range superconducting order in the intermediate metallic regime (Fig.~\ref{fig:f1_phase_diagram}). (b) Charge density profile $\langle n_{i_x,i_y} \rangle$ in the Holstein limit ($u=0$) at weak coupling ($\lambda=0.4$) on a $48\times 4$ ladder. A robust charge-density-wave modulation is observed, contrasting with the uniform Luttinger liquid phase stabilized on the cylinder geometry under identical parameters (see Figs.~\ref{fig:corrs_and_Ks_combined}(a)--\ref{fig:corrs_and_Ks_combined}(d)).}
    \label{fig:λ4u8Corr_and_λ0.4u0CDen}
\end{figure}
\begin{figure*}[htbp]
    \includegraphics[width=\textwidth]{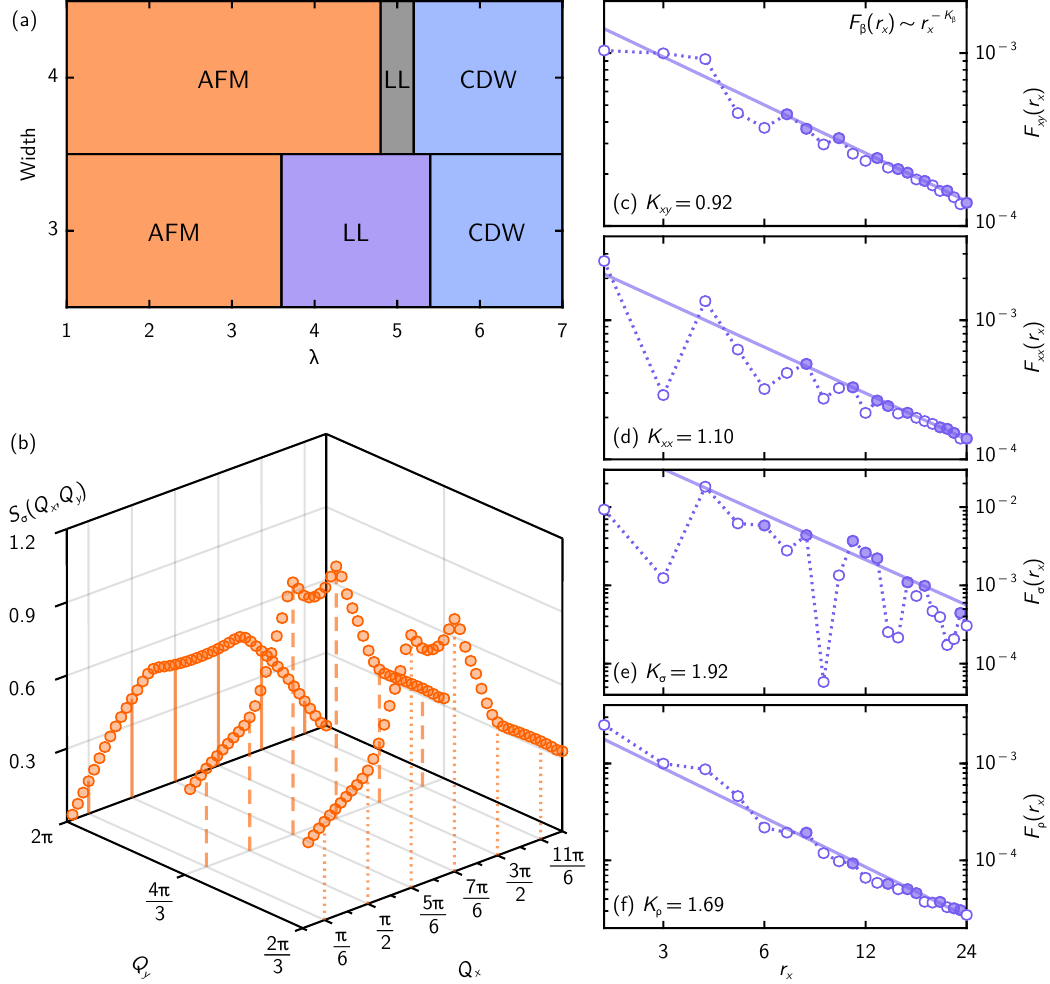}
    \caption{Impact of geometric frustration on the half-filled Hubbard--Holstein model at $u=10$ and $\omega=5t$. (a) Schematic phase diagram contrasting 3-leg and 4-leg cylinders. Geometric frustration on the 3-leg geometry suppresses conventional charge-density-wave (blue)  and antiferromagnetic (red) orders, substantially enlarging the intermediate metallic regime ($\mathrm{LL}^*$ in green, $\mathrm{LL}$ in gray) and altering its character. Phase boundaries are approximate. (b) Spin structure factor at $\lambda=4$ on a $48\times 3$ cylinder. The peaks at $\mathbf{Q}=(\pi\pm\pi/6,~\pi\pm\pi/3)$ reflect frustrated spin correlations, distinct from the commensurate $(\pi,\pi)$ order observed on 4-leg cylinders. (c)--(f) Ground-state correlations at the same parameter ($\lambda=4$) for the (c) crossed pair-pair ($F_{xy}$), (d) longitudinal pair-pair ($F_{xx}$), (e) spin ($F_\sigma$), and (f) density ($F_\rho$) channels. All correlations exhibit algebraic decay. The slowest decay and the opposite signs of the pairing correlations in (c) and (d) indicate a dominant ordinary $d$-wave pairing instability.}
    \label{fig:Ny3_combined_figure} 
\end{figure*}
To resolve the internal structure of the Cooper channel [Eqs.~\eqref{eq:plaq_d}--\eqref{eq:2D_d_x2y2}], we compute direction-resolved singlet pair-pair correlators $F_{\hat{\alpha} \hat{\beta}}(\mathbf{i},\mathbf{j})=\langle\Delta_{\hat{\alpha}}^{\dagger}(\mathbf{i}) \Delta_{\hat{\beta}}(\mathbf{j})\rangle$, where $\Delta_{\hat{\alpha}}(\mathbf{i}) = \langle c_{\mathbf{i}\uparrow}c_{\mathbf{i}+\hat{\alpha},\downarrow} - c_{\mathbf{i}\downarrow}c_{\mathbf{i}+\hat{\alpha},\uparrow}\rangle$. For a representative point $(\lambda=4,\,U=8)$ on the $48\times 4$ cylinder, as shown in Fig.~\ref{fig:f1_phase_diagram}, the crossed ($F_{xy}$) and longitudinal ($F_{xx}$) pair-pair correlations decay exponentially, $F_\beta(r_x)\sim e^{-r_x/\xi_\beta}$, as shown in Fig.~\ref{fig:λ4u8Corr_and_λ0.4u0CDen}(a).

To illustrate the finite-width effects relevant to Sec.~\ref{sec:2D_half_filled}, we also examine the Holstein limit $(U=0)$ at $\lambda=0.4$ and $\omega=5t$ on a $48\times 4$ ladder. Figure~\ref{fig:λ4u8Corr_and_λ0.4u0CDen}(b) depicts the ground state on the ladder, revealing sizable CDW modulation even at this small $\lambda$. This stands in contrast to the LL phase observed at this $\lambda$ on the $48\times 4$ cylinder (Fig.~\ref{fig:corrs_and_Ks_combined}(a)--\ref{fig:corrs_and_Ks_combined}(d)) as discussed in the main text.

While the 4-leg geometry cleanly illustrates the interplay of CDW, AFM, and LL phases, the 3-leg cylinder provides a complementary example where geometric frustration qualitatively modifies this competition. This frustration suppresses conventional ordering (CDW and AFM) and enhances the coherence of ordinary $d$-wave pairing~\cite{chungPlaquetteOrdinaryWave2020}. Similar frustration-enhanced superconductivity has been discussed in the Holstein model on a triangular lattice~\cite{liEnhancementSuperconductivityFrustrating2019}. To elucidate this mechanism, we compare the ground state of the HH model on a 3-leg cylinder with that on a 4-leg cylinder. As shown in Fig.~\ref{fig:Ny3_combined_figure}(a), for the representative parameter ($u=10,~\omega=5t$), geometric frustration substantially enlarges the intermediate metallic regime on the 3-leg cylinder and alters its character. On 4-leg cylinders, the LL regime arises from a balance between competing CDW and AFM tendencies, and it narrows at strong coupling. On 3-leg cylinders, frustration suppresses both the CDW and AFM instabilities while stabilizing a distinct $\mathrm{LL}^*$ regime where the ordinary $d$-wave pairing correlations dominate. The underlying physics differs accordingly: on 4-leg cylinders polarons tend to form singlets (on the rung) in the AFM regime and onsite bipolarons in the CDW regime. In the LL regime, these polarons additionally induce subdominant fluctuation of the plaquette $d$-wave pairing. In contrast, geometric frustration on 3-leg cylinders facilitates the coherence of mobile bipolarons along the axial direction. For the representative frustrated point $(\lambda=4,~u=10,~\omega=5t)$ on a $48\times 3$ cylinder, Fig.~\ref{fig:Ny3_combined_figure}(b) displays the spin structure factor, where the peaks at $\mathbf{Q}=(\pi\pm\pi/6,\pi\pm\pi/3)$ reflect the frustrated spin correlations. For reference, ordered phases (CDW and SDW) both exhibit $\mathbf{Q}=(\pi,\pi\pm\pi/3)$ for width 3 and the canonical N\'eel wave vector $\mathbf{Q}=(\pi,\pi)$ for width 4. Figures~\ref{fig:Ny3_combined_figure}(c)--\ref{fig:Ny3_combined_figure}(f) show that $F_{xx}(r)$ and $F_{xy}(r)$ exhibit the slowest decay among all four-point correlators. Their opposite signs identify their nature as ordinary $d$-wave pairs rather than extended $s$-wave pairs. These findings shed light on the behavior of e-ph system exhibiting true 2D geometric frustration and motivate future studies of intrinsically frustrated geometries, such as the triangular-lattice HH model.

\bibliographystyle{apsrev4-2}
\bibliography{references}

@article{wangZerotemperaturePhasesTwodimensional2020,
  title = {Zero-Temperature Phases of the Two-Dimensional {{Hubbard-Holstein}} Model: {{A}} Non-{{Gaussian}} Exact Diagonalization Study},
  shorttitle = {Zero-Temperature Phases of the Two-Dimensional {{Hubbard-Holstein}} Model},
  author = {Wang, Yao and Esterlis, Ilya and Shi, Tao and Cirac, J. Ignacio and Demler, Eugene},
  year = {2020},
  month = nov,
  journal = {Physical Review Research},
  volume = {2},
  number = {4},
  pages = {043258},
  issn = {2643-1564},
  doi = {10.1103/PhysRevResearch.2.043258},
  urldate = {2023-09-19},
  langid = {english}
}

@article{zoliNonlocalElectronphononCorrelations2005,
  title = {Nonlocal Electron-Phonon Correlations in a Dispersive {{Holstein}} Model},
  author = {Zoli, Marco},
  year = {2005},
  month = may,
  journal = {Physical Review B},
  volume = {71},
  number = {18},
  pages = {184308},
  issn = {1098-0121, 1550-235X},
  doi = {10.1103/PhysRevB.71.184308},
  urldate = {2025-09-23},
  copyright = {http://link.aps.org/licenses/aps-default-license},
  langid = {english}
}

@article{knorzerSpinHolsteinModelsTrappedIon2022,
  title = {Spin-{{Holstein Models}} in {{Trapped-Ion Systems}}},
  author = {Kn{\"o}rzer, J. and Shi, T. and Demler, E. and Cirac, J. I.},
  year = 2022,
  month = mar,
  journal = {Physical Review Letters},
  volume = {128},
  number = {12},
  pages = {120404},
  issn = {0031-9007, 1079-7114},
  doi = {10.1103/PhysRevLett.128.120404},
  urldate = {2024-01-02},
  langid = {english}
}

@article{weberTwodimensionalHolsteinHubbardModel2018,
  title = {Two-Dimensional {{Holstein-Hubbard}} Model: {{Critical}} Temperature, {{Ising}} Universality, and Bipolaron Liquid},
  shorttitle = {Two-Dimensional {{Holstein-Hubbard}} Model},
  author = {Weber, Manuel and Hohenadler, Martin},
  year = {2018},
  month = aug,
  journal = {Physical Review B},
  volume = {98},
  number = {8},
  pages = {085405},
  issn = {2469-9950, 2469-9969},
  doi = {10.1103/PhysRevB.98.085405},
  urldate = {2025-05-27},
  langid = {english}
}

@article{spencerEffectElectronphononInteraction2005,
  title = {Effect of Electron-Phonon Interaction Range on Lattice Polaron Dynamics: {{A}} Continuous-Time Quantum {{Monte Carlo}} Study},
  shorttitle = {Effect of Electron-Phonon Interaction Range on Lattice Polaron Dynamics},
  author = {Spencer, P. E. and Samson, J. H. and Kornilovitch, P. E. and Alexandrov, A. S.},
  year = {2005},
  month = may,
  journal = {Physical Review B},
  volume = {71},
  number = {18},
  pages = {184310},
  issn = {1098-0121, 1550-235X},
  doi = {10.1103/PhysRevB.71.184310},
  urldate = {2025-09-23},
  copyright = {http://link.aps.org/licenses/aps-default-license},
  langid = {english}
}

@book{alexandrovPolaronsAdvancedMaterials2007,
  title = {Polarons in {{Advanced Materials}}},
  author = {Alexandrov, A. S.},
  year = {2007},
  series = {Springer {{Series}} in {{Materials Science}}},
  number = {103},
  publisher = {Canopus Publishing Limited},
  address = {Dordrecht},
  doi = {10.1007/978-1-4020-6348-0},
  isbn = {978-1-4020-6347-3 978-1-4020-6348-0},
  langid = {english}
}

@article{nowadnickCompetitionAntiferromagneticChargeDensityWave2012,
  title = {Competition {{Between Antiferromagnetic}} and {{Charge-Density-Wave Order}} in the {{Half-Filled Hubbard-Holstein Model}}},
  author = {Nowadnick, E. A. and Johnston, S. and Moritz, B. and Scalettar, R. T. and Devereaux, T. P.},
  year = {2012},
  month = dec,
  journal = {Physical Review Letters},
  volume = {109},
  number = {24},
  pages = {246404},
  issn = {0031-9007, 1079-7114},
  doi = {10.1103/PhysRevLett.109.246404},
  urldate = {2025-09-23},
  copyright = {http://link.aps.org/licenses/aps-default-license},
  langid = {english}
}

@article{langKineticTheorySemiconductors1963,
  title = {Kinetic {{Theory}} of {{Semiconductors}} with {{Low Mobility}}},
  author = {Lang, I. G. and Firsov, {\relax Yu}. A.},
  year = {1963},
  month = jan,
  journal = {Soviet Journal of Experimental and Theoretical Physics},
  volume = {16},
  pages = {1301},
  publisher = {Springer},
  issn = {1063-7761},
  urldate = {2025-09-28}
}

@article{provilleMobileBipolaronsAdiabatic1998,
  title = {Mobile Bipolarons in the Adiabatic {{Holstein-Hubbard}} Model in One and Two Dimensions},
  author = {Proville, L. and Aubry, S.},
  year = {1998},
  month = mar,
  journal = {Physica D: Nonlinear Phenomena},
  volume = {113},
  number = {2-4},
  pages = {307--317},
  issn = {01672789},
  doi = {10.1016/S0167-2789(97)00283-2},
  urldate = {2025-09-23},
  copyright = {https://www.elsevier.com/tdm/userlicense/1.0/},
  langid = {english}
}

@article{provilleSmallBipolarons2dimensional1999,
  title = {Small Bipolarons in the 2-Dimensional {{Holstein-Hubbard}} Model. {{I}}. {{The}} Adiabatic Limit},
  author = {Proville, L. and Aubry, S.},
  year = {1999},
  month = sep,
  journal = {The European Physical Journal B},
  volume = {11},
  number = {1},
  pages = {41--58},
  issn = {1434-6028},
  doi = {10.1007/s100510050915},
  urldate = {2025-09-23},
  copyright = {http://www.springer.com/tdm},
  langid = {english}
}

@article{ohgoeCompetitionSuperconductingAntiferromagnetic2017,
  title = {Competition among Superconducting, Antiferromagnetic, and Charge Orders with Intervention by Phase Separation in the 2D Holstein-Hubbard Model},
  author = {Ohgoe, Takahiro and Imada, Masatoshi},
  journal = {Phys. Rev. Lett.},
  volume = {119},
  issue = {19},
  pages = {197001},
  numpages = {5},
  year = {2017},
  month = {Nov},
  publisher = {American Physical Society},
  doi = {10.1103/PhysRevLett.119.197001},
  url = {https://link.aps.org/doi/10.1103/PhysRevLett.119.197001}
}

@article{liEnhancementSuperconductivityFrustrating2019,
  title = {Enhancement of Superconductivity by Frustrating the Charge Order},
  author = {Li, Zi-Xiang and Cohen, Marvin L. and Lee, Dung-Hai},
  year = {2019},
  month = dec,
  journal = {Physical Review B},
  volume = {100},
  number = {24},
  pages = {245105},
  issn = {2469-9950, 2469-9969},
  doi = {10.1103/PhysRevB.100.245105},
  urldate = {2025-05-13},
  langid = {english}
}

@article{costaPhononDispersionCompetition2018,
  title = {Phonon {{Dispersion}} and the {{Competition}} between {{Pairing}} and {{Charge Order}}},
  author = {Costa, N. C. and Blommel, T. and Chiu, W.-T. and Batrouni, G. and Scalettar, R. T.},
  year = {2018},
  month = may,
  journal = {Physical Review Letters},
  volume = {120},
  number = {18},
  pages = {187003},
  issn = {0031-9007, 1079-7114},
  doi = {10.1103/PhysRevLett.120.187003},
  urldate = {2025-05-07},
  langid = {english}
}

@article{costaPhaseDiagramTwodimensional2020,
  title = {Phase Diagram of the Two-Dimensional {{Hubbard-Holstein}} Model},
  author = {Costa, Natanael C. and Seki, Kazuhiro and Yunoki, Seiji and Sorella, Sandro},
  year = {2020},
  month = may,
  journal = {Communications Physics},
  volume = {3},
  number = {1},
  pages = {80},
  issn = {2399-3650},
  doi = {10.1038/s42005-020-0342-2},
  urldate = {2024-11-03},
  langid = {english}
}

@article{karakuzuSuperconductivityChargedensityWaves2017,
  title = {Superconductivity, Charge-Density Waves, Antiferromagnetism, and Phase Separation in the {{Hubbard-Holstein}} Model},
  author = {Karakuzu, Seher and Tocchio, Luca F. and Sorella, Sandro and Becca, Federico},
  year = {2017},
  month = nov,
  journal = {Physical Review B},
  volume = {96},
  number = {20},
  pages = {205145},
  issn = {2469-9950, 2469-9969},
  doi = {10.1103/PhysRevB.96.205145},
  urldate = {2024-12-26},
  copyright = {https://link.aps.org/licenses/aps-default-license},
  langid = {english}
}

@article{hohenadlerDominantChargeDensity2019,
  title = {Dominant Charge Density Wave Correlations in the {{Holstein}} Model on the Half-Filled Square Lattice},
  author = {Hohenadler, M. and Batrouni, G. G.},
  year = {2019},
  month = oct,
  journal = {Physical Review B},
  volume = {100},
  number = {16},
  pages = {165114},
  issn = {2469-9950, 2469-9969},
  doi = {10.1103/PhysRevB.100.165114},
  urldate = {2025-05-13},
  langid = {english}
}

@article{chakravartyDimensionalCrossoverQuantum1996,
  title = {Dimensional {{Crossover}} in {{Quantum Antiferromagnets}}},
  author = {Chakravarty, Sudip},
  year = {1996},
  month = nov,
  journal = {Physical Review Letters},
  volume = {77},
  number = {21},
  pages = {4446--4449},
  issn = {0031-9007, 1079-7114},
  doi = {10.1103/PhysRevLett.77.4446},
  urldate = {2025-04-10},
  copyright = {http://link.aps.org/licenses/aps-default-license},
  langid = {english}
}

@article{wangPhononMediatedLongRangeAttractive2021,
  title = {Phonon-{{Mediated Long-Range Attractive Interaction}} in {{One-Dimensional Cuprates}}},
  author = {Wang, Yao and Chen, Zhuoyu and Shi, Tao and Moritz, Brian and Shen, Zhi-Xun and Devereaux, Thomas P.},
  year = {2021},
  month = nov,
  journal = {Physical Review Letters},
  volume = {127},
  number = {19},
  pages = {197003},
  issn = {0031-9007, 1079-7114},
  doi = {10.1103/PhysRevLett.127.197003},
  urldate = {2023-11-22},
  langid = {english}
}

@article{chenAnomalouslyStrongNearneighbor2021,
  title = {Anomalously Strong Near-Neighbor Attraction in Doped {{1D}} Cuprate Chains},
  author = {Chen, Zhuoyu and Wang, Yao and Rebec, Slavko N. and Jia, Tao and Hashimoto, Makoto and Lu, Donghui and Moritz, Brian and Moore, Robert G. and Devereaux, Thomas P. and Shen, Zhi-Xun},
  year = {2021},
  month = sep,
  journal = {Science},
  volume = {373},
  number = {6560},
  pages = {1235--1239},
  issn = {0036-8075, 1095-9203},
  doi = {10.1126/science.abf5174},
  urldate = {2024-06-21},
  langid = {english}
}

@misc{tangTracesElectronPhononCoupling2022,
  title = {Traces of {{Electron-Phonon Coupling}} in {{One-Dimensional Cuprates}}},
  author = {Tang, Ta and Moritz, Brian and Peng, Cheng and Shen, Zhi-Xun and Devereaux, Thomas},
  year = {2022},
  month = nov,
  doi = {10.21203/rs.3.rs-2196634/v1},
  urldate = {2024-07-01},
  copyright = {https://creativecommons.org/licenses/by/4.0/},
  langid = {english}
}

@article{wangFluctuatingNatureLightEnhanced2021,
  title = {Fluctuating {{Nature}} of {{Light-Enhanced}} d -{{Wave Superconductivity}}: {{A Time-Dependent Variational Non-Gaussian Exact Diagonalization Study}}},
  shorttitle = {Fluctuating {{Nature}} of {{Light-Enhanced}} d -{{Wave Superconductivity}}},
  author = {Wang, Yao and Shi, Tao and Chen, Cheng-Chien},
  year = {2021},
  month = nov,
  journal = {Physical Review X},
  volume = {11},
  number = {4},
  pages = {041028},
  issn = {2160-3308},
  doi = {10.1103/PhysRevX.11.041028},
  urldate = {2024-08-01},
  langid = {english}
}

@article{shiVariationalStudyFermionic2018,
  title = {Variational Study of Fermionic and Bosonic Systems with Non-{{Gaussian}} States: {{Theory}} and Applications},
  shorttitle = {Variational Study of Fermionic and Bosonic Systems with Non-{{Gaussian}} States},
  author = {Shi, Tao and Demler, Eugene and Ignacio Cirac, J.},
  year = {2018},
  month = mar,
  journal = {Annals of Physics},
  volume = {390},
  pages = {245--302},
  issn = {00034916},
  doi = {10.1016/j.aop.2017.11.014},
  urldate = {2023-02-28},
  langid = {english}
}

@article{shiVariationalApproachManyBody2020,
  title = {Variational {{Approach}} for {{Many-Body Systems}} at {{Finite Temperature}}},
  author = {Shi, Tao and Demler, Eugene and Cirac, J. Ignacio},
  year = {2020},
  month = oct,
  journal = {Physical Review Letters},
  volume = {125},
  number = {18},
  pages = {180602},
  issn = {0031-9007, 1079-7114},
  doi = {10.1103/PhysRevLett.125.180602},
  urldate = {2023-09-24},
  langid = {english}
}

@article{hacklGeometryVariationalMethods2020,
  title = {Geometry of Variational Methods: Dynamics of Closed Quantum Systems},
  shorttitle = {Geometry of Variational Methods},
  author = {Hackl, Lucas and Guaita, Tommaso and Shi, Tao and Haegeman, Jutho and Demler, Eugene and Cirac, Ignacio},
  year = {2020},
  month = oct,
  journal = {SciPost Physics},
  volume = {9},
  number = {4},
  pages = {048},
  issn = {2542-4653},
  doi = {10.21468/SciPostPhys.9.4.048},
  urldate = {2023-06-15},
  langid = {english}
}

@article{McLachlan01011964,
    author = {A.D. McLachlan},
    title = {A variational solution of the time-dependent Schrodinger equation},
    journal = {Molecular Physics},
    volume = {8},
    number = {1},
    pages = {39--44},
    year = {1964},
    publisher = {Taylor \& Francis},
    doi = {10.1080/00268976400100041},
    URL = { 
            https://doi.org/10.1080/00268976400100041
    },
    eprint = { 
            https://doi.org/10.1080/00268976400100041
    }
}

@article{HOLSTEIN1959325,
title = {Studies of polaron motion: Part I. The molecular-crystal model},
journal = {Annals of Physics},
volume = {8},
number = {3},
pages = {325-342},
year = {1959},
issn = {0003-4916},
doi = {https://doi.org/10.1016/0003-4916(59)90002-8},
url = {https://www.sciencedirect.com/science/article/pii/0003491659900028},
author = {T Holstein}
}

@article{fishmanCompressionCorrelationMatrices2015,
  title = {Compression of {{Correlation Matrices}} and an {{Efficient Method}} for {{Forming Matrix Product States}} of {{Fermionic Gaussian States}}},
  author = {Fishman, Matthew T. and White, Steven R.},
  year = {2015},
  month = aug,
  journal = {Physical Review B},
  volume = {92},
  number = {7},
  eprint = {1504.07701},
  primaryclass = {cond-mat},
  pages = {075132},
  issn = {1098-0121, 1550-235X},
  doi = {10.1103/PhysRevB.92.075132},
  urldate = {2024-06-12},
  archiveprefix = {arXiv},
  langid = {english}
}

@article{jinMatrixProductStates2022,
  title = {Matrix Product States for {{Hartree-Fock-Bogoliubov}} Wave Functions},
  author = {Jin, Hui-Ke and Sun, Rong-Yang and Zhou, Yi and Tu, Hong-Hao},
  year = {2022},
  month = feb,
  journal = {Physical Review B},
  volume = {105},
  number = {8},
  pages = {L081101},
  issn = {2469-9950, 2469-9969},
  doi = {10.1103/PhysRevB.105.L081101},
  urldate = {2024-12-17},
  langid = {english}
}

@article{bertschSymmetryRestorationHartreeFockBogoliubov2012,
  title = {Symmetry {{Restoration}} in {{Hartree-Fock-Bogoliubov Based Theories}}},
  author = {Bertsch, G. F. and Robledo, L. M.},
  year = {2012},
  month = jan,
  journal = {Physical Review Letters},
  volume = {108},
  number = {4},
  pages = {042505},
  issn = {0031-9007, 1079-7114},
  doi = {10.1103/PhysRevLett.108.042505},
  urldate = {2025-04-17},
  copyright = {http://link.aps.org/licenses/aps-default-license},
  langid = {english}
}

@article{haegemanUnifyingTimeEvolution2016,
  title = {Unifying Time Evolution and Optimization with Matrix Product States},
  author = {Haegeman, Jutho and Lubich, Christian and Oseledets, Ivan and Vandereycken, Bart and Verstraete, Frank},
  year = {2016},
  month = oct,
  journal = {Physical Review B},
  volume = {94},
  number = {16},
  pages = {165116},
  issn = {2469-9950, 2469-9969},
  doi = {10.1103/PhysRevB.94.165116},
  urldate = {2023-04-09},
  langid = {english}
}

@article{eisertColloquiumAreaLaws2010,
  title = {{\emph{Colloquium}} : {{Area}} Laws for the Entanglement Entropy},
  shorttitle = {{\emph{Colloquium}}},
  author = {Eisert, J. and Cramer, M. and Plenio, M. B.},
  year = {2010},
  month = feb,
  journal = {Reviews of Modern Physics},
  volume = {82},
  number = {1},
  pages = {277--306},
  issn = {0034-6861, 1539-0756},
  doi = {10.1103/RevModPhys.82.277},
  urldate = {2025-09-23},
  copyright = {http://link.aps.org/licenses/aps-default-license},
  langid = {english}
}

@article{pasqualecalabreseEntanglementEntropyQuantum2004,
  title = {Entanglement Entropy and Quantum Field Theory},
  author = {{Pasquale Calabrese} and {John Cardy}},
  year = 2004,
  month = jun,
  journal = {Journal of Statistical Mechanics: Theory and Experiment},
  volume = {2004},
  number = {06},
  pages = {P06002},
  issn = {1742-5468},
  doi = {10.1088/1742-5468/2004/06/P06002},
  urldate = {2025-12-26},
  langid = {english}
}

@article{pollmannTheoryFiniteEntanglementScaling2009,
  title = {Theory of {{Finite-Entanglement Scaling}} at {{One-Dimensional Quantum Critical Points}}},
  author = {Pollmann, Frank and Mukerjee, Subroto and Turner, Ari M. and Moore, Joel E.},
  year = {2009},
  month = jun,
  journal = {Physical Review Letters},
  volume = {102},
  number = {25},
  pages = {255701},
  issn = {0031-9007, 1079-7114},
  doi = {10.1103/PhysRevLett.102.255701},
  urldate = {2025-09-23},
  copyright = {http://link.aps.org/licenses/aps-default-license},
  langid = {english}
}

@article{verstraeteMatrixProductStates2006,
  title = {Matrix Product States Represent Ground States Faithfully},
  author = {Verstraete, F. and Cirac, J. I.},
  year = {2006},
  month = mar,
  journal = {Physical Review B},
  volume = {73},
  number = {9},
  pages = {094423},
  issn = {1098-0121, 1550-235X},
  doi = {10.1103/PhysRevB.73.094423},
  urldate = {2023-03-16},
  langid = {english}
}

@article{ciracMatrixProductStates2021,
  title = {Matrix Product States and Projected Entangled Pair States: {{Concepts}}, Symmetries, Theorems},
  shorttitle = {Matrix Product States and Projected Entangled Pair States},
  author = {Cirac, J. Ignacio and {P{\'e}rez-Garc{\'i}a}, David and Schuch, Norbert and Verstraete, Frank},
  year = {2021},
  month = dec,
  journal = {Reviews of Modern Physics},
  volume = {93},
  number = {4},
  pages = {045003},
  issn = {0034-6861, 1539-0756},
  doi = {10.1103/RevModPhys.93.045003},
  urldate = {2025-04-23},
  langid = {english}
}

@article{orusTensorNetworksComplex2019,
  title = {Tensor Networks for Complex Quantum Systems},
  author = {Or{\'u}s, Rom{\'a}n},
  year = 2019,
  month = aug,
  journal = {Nature Reviews Physics},
  volume = {1},
  number = {9},
  pages = {538--550},
  issn = {2522-5820},
  doi = {10.1038/s42254-019-0086-7},
  urldate = {2025-12-26},
  langid = {english}
}

@article{stoudenmireStudyingTwoDimensionalSystems2012,
  title = {Studying {{Two-Dimensional Systems}} with the {{Density Matrix Renormalization Group}}},
  author = {Stoudenmire, E.M. and White, Steven R.},
  year = {2012},
  month = mar,
  journal = {Annual Review of Condensed Matter Physics},
  volume = {3},
  number = {1},
  pages = {111--128},
  issn = {1947-5454, 1947-5462},
  doi = {10.1146/annurev-conmatphys-020911-125018},
  urldate = {2024-08-13},
  langid = {english}
}

@article{hastingsAreaLawOnedimensional2007,
  title = {An Area Law for One-Dimensional Quantum Systems},
  author = {Hastings, M B},
  year = 2007,
  month = aug,
  journal = {Journal of Statistical Mechanics: Theory and Experiment},
  volume = {2007},
  number = {08},
  pages = {P08024-P08024},
  issn = {1742-5468},
  doi = {10.1088/1742-5468/2007/08/P08024},
  urldate = {2025-12-26},
  langid = {english}
}

@article{caiHightemperatureSuperconductivityInduced2025,
  title = {High-Temperature Superconductivity Induced by the {{Su-Schrieffer-Heeger}} Electron-Phonon Coupling},
  author = {Cai, Xun and Li, Zi-Xiang and Yao, Hong},
  year = 2025,
  month = oct,
  journal = {Physical Review B},
  volume = {112},
  number = {14},
  pages = {144517},
  issn = {2469-9950, 2469-9969},
  doi = {10.1103/rhss-d52m},
  urldate = {2025-12-27},
  langid = {english}
}

@article{chenRoleElectronphononCoupling2023,
  title = {Role of Electron-Phonon Coupling in Excitonic Insulator Candidate {{Ta}} 2 {{NiSe}} 5},
  author = {Chen, Cheng and Chen, Xiang and Tang, Weichen and Li, Zhenglu and Wang, Siqi and Ding, Shuhan and Kang, Zhibo and Jozwiak, Chris and Bostwick, Aaron and Rotenberg, Eli and Hashimoto, Makoto and Lu, Donghui and Ruff, Jacob P. C. and Louie, Steven G. and Birgeneau, Robert J. and Chen, Yulin and Wang, Yao and He, Yu},
  year = 2023,
  month = oct,
  journal = {Physical Review Research},
  volume = {5},
  number = {4},
  pages = {043089},
  issn = {2643-1564},
  doi = {10.1103/PhysRevResearch.5.043089},
  urldate = {2025-12-27},
  langid = {english}
}

@article{thomasTheoryElectronPhononInteractions2025,
  title = {Theory of {{Electron-Phonon Interactions}} in {{Extended Correlated Systems Probed}} by {{Resonant Inelastic X-Ray Scattering}}},
  author = {Thomas, Jinu and Banerjee, Debshikha and Nocera, Alberto and Johnston, Steven},
  year = 2025,
  month = apr,
  journal = {Physical Review X},
  volume = {15},
  number = {2},
  pages = {021030},
  issn = {2160-3308},
  doi = {10.1103/PhysRevX.15.021030},
  urldate = {2025-12-27},
  langid = {english}
}

@article{zhaoChebyshevPseudositeMatrix2023,
  title = {Chebyshev Pseudosite Matrix Product State Approach for the Spectral Functions of Electron-Phonon Coupling Systems},
  author = {Zhao, Pei-Yuan and Ding, Ke and Yang, Shuo},
  year = 2023,
  month = apr,
  journal = {Physical Review Research},
  volume = {5},
  number = {2},
  pages = {023026},
  issn = {2643-1564},
  doi = {10.1103/PhysRevResearch.5.023026},
  urldate = {2025-12-27},
  langid = {english}
}

@article{zhanCooperationElectronPhononCoupling2025,
  title = {Cooperation between {{Electron-Phonon Coupling}} and {{Electronic Interaction}} in {{Bilayer Nickelates La}} 3 {{Ni}} 2 {{O}} 7},
  author = {Zhan, Jun and Gu, Yuhao and Wu, Xianxin and Hu, Jiangping},
  year = 2025,
  month = mar,
  journal = {Physical Review Letters},
  volume = {134},
  number = {13},
  pages = {136002},
  issn = {0031-9007, 1079-7114},
  doi = {10.1103/PhysRevLett.134.136002},
  urldate = {2025-12-23},
  langid = {english}
}

@article{liWhatMakesTc2016,
  title = {What Makes the {{Tc}} of Monolayer {{FeSe}} on {{SrTiO3}} so High: A Sign-Problem-Free Quantum {{Monte Carlo}} Study},
  shorttitle = {What Makes the {{Tc}} of Monolayer {{FeSe}} on {{SrTiO3}} so High},
  author = {Li, Zi-Xiang and Wang, Fa and Yao, Hong and Lee, Dung-Hai},
  year = 2016,
  month = jun,
  journal = {Science Bulletin},
  volume = {61},
  number = {12},
  pages = {925--930},
  issn = {20959273},
  doi = {10.1007/s11434-016-1087-x},
  urldate = {2025-12-27},
  langid = {english}
}

@article{wangRobustDwaveSuperconductivity2025,
  title = {Robust D-Wave Superconductivity from the {{Su-Schrieffer-Heeger-Hubbard}} Model: Possible Route to High-Temperature Superconductivity},
  shorttitle = {Robust D-Wave Superconductivity from the {{Su-Schrieffer-Heeger-Hubbard}} Model},
  author = {Wang, Hao-Xin and Jiang, Yi-Fan and Yao, Hong},
  year = 2025,
  month = jul,
  journal = {Science Bulletin},
  volume = {70},
  number = {14},
  pages = {2260--2265},
  issn = {20959273},
  doi = {10.1016/j.scib.2025.04.055},
  urldate = {2025-12-27},
  langid = {english}
}

@article{xiangHightemperatureSuperconductivityFeSe2012,
  title = {High-Temperature Superconductivity at the {{FeSe}}/{{SrTiO}} 3 Interface},
  author = {Xiang, Yuan-Yuan and Wang, Fa and Wang, Da and Wang, Qiang-Hua and Lee, Dung-Hai},
  year = 2012,
  month = oct,
  journal = {Physical Review B},
  volume = {86},
  number = {13},
  pages = {134508},
  issn = {1098-0121, 1550-235X},
  doi = {10.1103/PhysRevB.86.134508},
  urldate = {2025-12-27},
  copyright = {http://link.aps.org/licenses/aps-default-license},
  langid = {english}
}

@article{benderVariationalMonteCarlo2023,
  title = {Variational {{Monte Carlo}} Algorithm for Lattice Gauge Theories with Continuous Gauge Groups: {{A}} Study of ( 2 + 1 ) -Dimensional Compact {{QED}} with Dynamical Fermions at Finite Density},
  shorttitle = {Variational {{Monte Carlo}} Algorithm for Lattice Gauge Theories with Continuous Gauge Groups},
  author = {Bender, Julian and Emonts, Patrick and Cirac, J. Ignacio},
  year = 2023,
  month = nov,
  journal = {Physical Review Research},
  volume = {5},
  number = {4},
  pages = {043128},
  issn = {2643-1564},
  doi = {10.1103/PhysRevResearch.5.043128},
  urldate = {2026-01-07},
  langid = {english}
}

@article{HALDANE1983464,
    title = {Continuum dynamics of the 1-D Heisenberg antiferromagnet: Identification with the O(3) nonlinear sigma model},
    journal = {Physics Letters A},
    volume = {93},
    number = {9},
    pages = {464-468},
    year = {1983},
    issn = {0375-9601},
    doi = {https://doi.org/10.1016/0375-9601(83)90631-X},
    url = {https://www.sciencedirect.com/science/article/pii/037596018390631X},
    author = {F.D.M. Haldane}
}

@article{haldaneNonlinearFieldTheory1983,
  title = {Nonlinear {{Field Theory}} of {{Large-Spin Heisenberg Antiferromagnets}}: {{Semiclassically Quantized Solitons}} of the {{One-Dimensional Easy-Axis N\'eel State}}},
  shorttitle = {Nonlinear {{Field Theory}} of {{Large-Spin Heisenberg Antiferromagnets}}},
  author = {Haldane, F. D. M.},
  year = 1983,
  month = apr,
  journal = {Physical Review Letters},
  volume = {50},
  number = {15},
  pages = {1153--1156},
  issn = {0031-9007},
  doi = {10.1103/PhysRevLett.50.1153},
  urldate = {2026-01-13},
  copyright = {http://link.aps.org/licenses/aps-default-license},
  langid = {english}
}

@article{carleoSolvingQuantumManybody2017,
  title = {Solving the Quantum Many-Body Problem with Artificial Neural Networks},
  author = {Carleo, Giuseppe and Troyer, Matthias},
  year = 2017,
  month = feb,
  journal = {Science},
  volume = {355},
  number = {6325},
  pages = {602--606},
  issn = {0036-8075, 1095-9203},
  doi = {10.1126/science.aag2302},
  urldate = {2026-01-14},
  copyright = {http://www.sciencemag.org/about/science-licenses-journal-article-reuse},
  langid = {english}
}

@article{EMERY1993597,
    title = {Frustrated electronic phase separation and high-temperature superconductors},
    journal = {Physica C: Superconductivity},
    volume = {209},
    number = {4},
    pages = {597-621},
    year = {1993},
    issn = {0921-4534},
    doi = {https://doi.org/10.1016/0921-4534(93)90581-A},
    url = {https://www.sciencedirect.com/science/article/pii/092145349390581A},
    author = {V.J. Emery and S.A. Kivelson}
}

@article{fradkinColloquiumTheoryIntertwined2015,
  title = {{\emph{Colloquium}} : {{Theory}} of Intertwined Orders in High Temperature Superconductors},
  shorttitle = {{\emph{Colloquium}}},
  author = {Fradkin, Eduardo and Kivelson, Steven A. and Tranquada, John M.},
  year = 2015,
  month = may,
  journal = {Reviews of Modern Physics},
  volume = {87},
  number = {2},
  pages = {457--482},
  issn = {0034-6861, 1539-0756},
  doi = {10.1103/RevModPhys.87.457},
  urldate = {2025-06-14},
  copyright = {http://link.aps.org/licenses/aps-default-license},
  langid = {english}
}

@article{chaoCanonicalPerturbationExpansion1978,
  title = {Canonical Perturbation Expansion of the {{Hubbard}} Model},
  author = {Chao, K. A. and Spa{\l}ek, J. and Ole{\'s}, A. M.},
  year = 1978,
  month = oct,
  journal = {Physical Review B},
  volume = {18},
  number = {7},
  pages = {3453--3464},
  issn = {0163-1829},
  doi = {10.1103/PhysRevB.18.3453},
  urldate = {2026-01-21},
  copyright = {http://link.aps.org/licenses/aps-default-license},
  langid = {english}
}

@article{micnasSuperconductivityNarrowbandSystems1990,
  title = {Superconductivity in Narrow-Band Systems with Local Nonretarded Attractive Interactions},
  author = {Micnas, R. and Ranninger, J. and Robaszkiewicz, S.},
  year = 1990,
  month = jan,
  journal = {Reviews of Modern Physics},
  volume = {62},
  number = {1},
  pages = {113--171},
  issn = {0034-6861, 1539-0756},
  doi = {10.1103/RevModPhys.62.113},
  urldate = {2023-02-12},
  langid = {english}
}

@article{bakCommensuratePhasesIncommensurate1982,
  title = {Commensurate Phases, Incommensurate Phases and the Devil's Staircase},
  author = {Bak, P},
  year = 1982,
  month = jun,
  journal = {Reports on Progress in Physics},
  volume = {45},
  number = {6},
  pages = {587--629},
  issn = {0034-4885, 1361-6633},
  doi = {10.1088/0034-4885/45/6/001},
  urldate = {2026-01-28},
  langid = {english}
}

@article{schulzCriticalBehaviorCommensurateincommensurate1980,
  title = {Critical Behavior of Commensurate-Incommensurate Phase Transitions in Two Dimensions},
  author = {Schulz, H. J.},
  year = 1980,
  month = dec,
  journal = {Physical Review B},
  volume = {22},
  number = {11},
  pages = {5274--5277},
  issn = {0163-1829},
  doi = {10.1103/PhysRevB.22.5274},
  urldate = {2026-01-28},
  copyright = {http://link.aps.org/licenses/aps-default-license},
  langid = {english}
}

@article{PhysRevB.97.155156,
  title = {Exploring the anisotropic Kondo model in and out of equilibrium with alkaline-earth atoms},
  author = {Kan\'asz-Nagy, M\'arton and Ashida, Yuto and Shi, Tao and Moca, C\ifmmode \u{a}\else \u{a}\fi{}t\ifmmode \u{a}\else \u{a}\fi{}lin Pa\ifmmode \mbox{\c{s}}\else \c{s}\fi{}cu and Ikeda, Tatsuhiko N. and F\"olling, Simon and Cirac, J. Ignacio and Zar\'and, Gergely and Demler, Eugene A.},
  journal = {Phys. Rev. B},
  volume = {97},
  issue = {15},
  pages = {155156},
  numpages = {20},
  year = {2018},
  month = {Apr},
  publisher = {American Physical Society},
  doi = {10.1103/PhysRevB.97.155156},
  url = {https://link.aps.org/doi/10.1103/PhysRevB.97.155156}
}

@article{salaVariationalStudyU12018,
  title = {Variational Study of {{U}}(1) and {{SU}}(2) Lattice Gauge Theories with {{Gaussian}} States in \$1+1\$ Dimensions},
  author = {Sala, P. and Shi, T. and K{\"u}hn, S. and Ba{\~n}uls, M. C. and Demler, E. and Cirac, J. I.},
  year = 2018,
  month = aug,
  journal = {Physical Review D},
  volume = {98},
  number = {3},
  pages = {034505},
  publisher = {American Physical Society},
  doi = {10.1103/PhysRevD.98.034505},
  urldate = {2022-10-11},
  langid = {american}
}

@article{ashidaEfficientVariationalApproach2019,
  title = {Efficient Variational Approach to Dynamics of a Spatially Extended Bosonic {{Kondo}} Model},
  author = {Ashida, Yuto and Shi, Tao and Schmidt, Richard and Sadeghpour, H. R. and Cirac, J. Ignacio and Demler, Eugene},
  year = 2019,
  month = oct,
  journal = {Physical Review A},
  volume = {100},
  number = {4},
  pages = {043618},
  issn = {2469-9926, 2469-9934},
  doi = {10.1103/PhysRevA.100.043618},
  urldate = {2025-12-29},
  langid = {english}
}

@article{ashidaQuantumRydbergCentral2019,
  title = {Quantum {{Rydberg Central Spin Model}}},
  author = {Ashida, Yuto and Shi, Tao and Schmidt, Richard and Sadeghpour, H. R. and Cirac, J. Ignacio and Demler, Eugene},
  year = 2019,
  month = oct,
  journal = {Physical Review Letters},
  volume = {123},
  number = {18},
  pages = {183001},
  issn = {0031-9007, 1079-7114},
  doi = {10.1103/PhysRevLett.123.183001},
  urldate = {2025-12-29},
  langid = {english}
}

@article{shiUltrafastMolecularDynamics2020,
  title = {Ultrafast Molecular Dynamics in Terahertz-{{STM}} Experiments: {{Theoretical}} Analysis Using the {{Anderson-Holstein}} Model},
  shorttitle = {Ultrafast Molecular Dynamics in Terahertz-{{STM}} Experiments},
  author = {Shi, Tao and Cirac, J. Ignacio and Demler, Eugene},
  year = 2020,
  month = sep,
  journal = {Physical Review Research},
  volume = {2},
  number = {3},
  pages = {033379},
  issn = {2643-1564},
  doi = {10.1103/PhysRevResearch.2.033379},
  urldate = {2025-12-24},
  langid = {english}
}

@article{ashidaVariationalPrincipleQuantum2018,
  title = {Variational Principle for Quantum Impurity Systems in and out of Equilibrium: {{Application}} to {{Kondo}} Problems},
  shorttitle = {Variational Principle for Quantum Impurity Systems in and out of Equilibrium},
  author = {Ashida, Yuto and Shi, Tao and Ba{\~n}uls, Mari Carmen and Cirac, J. Ignacio and Demler, Eugene},
  year = 2018,
  month = jul,
  journal = {Physical Review B},
  volume = {98},
  number = {2},
  pages = {024103},
  issn = {2469-9950, 2469-9969},
  doi = {10.1103/PhysRevB.98.024103},
  urldate = {2023-02-08},
  langid = {english}
}

@article{ashidaSolvingQuantumImpurity2018,
  title = {Solving {{Quantum Impurity Problems}} in and out of {{Equilibrium}} with the {{Variational Approach}}},
  author = {Ashida, Yuto and Shi, Tao and Ba{\~n}uls, Mari Carmen and Cirac, J. Ignacio and Demler, Eugene},
  year = 2018,
  month = jul,
  journal = {Physical Review Letters},
  volume = {121},
  number = {2},
  pages = {026805},
  issn = {0031-9007, 1079-7114},
  doi = {10.1103/PhysRevLett.121.026805},
  urldate = {2023-02-16},
  langid = {english}
}

@article{zhangAttractiveRepulsiveAngulons2025,
  title = {Attractive and Repulsive Angulons in Superfluid Environments},
  author = {Zhang, Wei and Zeng, Zhongda and Shi, Tao},
  year = 2025,
  month = apr,
  journal = {Physical Review A},
  volume = {111},
  number = {4},
  pages = {043317},
  issn = {2469-9926, 2469-9934},
  doi = {10.1103/PhysRevA.111.043317},
  urldate = {2025-12-29},
  langid = {english}
}

@article{quVariationalApproachDynamics2025,
  title = {Variational Approach to the Dynamics of Dissipative Quantum Impurity Models},
  author = {Qu, Yi-Fan and Stefanini, Martino and Shi, Tao and Esslinger, Tilman and Gopalakrishnan, Sarang and Marino, Jamir and Demler, Eugene},
  year = 2025,
  month = apr,
  journal = {Physical Review B},
  volume = {111},
  number = {15},
  pages = {155113},
  issn = {2469-9950, 2469-9969},
  doi = {10.1103/PhysRevB.111.155113},
  urldate = {2025-12-29},
  langid = {english}
}

@article{dolgirevEmergenceSharpQuantum2021,
  title = {Emergence of a {{Sharp Quantum Collective Mode}} in a {{One-Dimensional Fermi Polaron}}},
  author = {Dolgirev, Pavel E. and Qu, Yi-Fan and Zvonarev, Mikhail B. and Shi, Tao and Demler, Eugene},
  year = 2021,
  month = oct,
  journal = {Physical Review X},
  volume = {11},
  number = {4},
  pages = {041015},
  issn = {2160-3308},
  doi = {10.1103/PhysRevX.11.041015},
  urldate = {2025-12-29},
  langid = {english}
}

@misc{quEfficientVariationalApproach2022,
  title = {Efficient Variational Approach to the {{Fermi}} Polaron Problem in Two Dimensions, Both in and out of Equilibrium},
  author = {Qu, Yi-Fan and Dolgirev, Pavel E. and Demler, Eugene and Shi, Tao},
  year = 2022,
  publisher = {arXiv},
  doi = {10.48550/ARXIV.2209.10998},
  urldate = {2025-12-29},
  copyright = {arXiv.org perpetual, non-exclusive license}
}

@article{schindlerVariationalAnsatzGround2022,
  title = {Variational {{{\emph{Ansatz}}}} for the {{Ground State}} of the {{Quantum Sherrington-Kirkpatrick Model}}},
  author = {Schindler, Paul M. and Guaita, Tommaso and Shi, Tao and Demler, Eugene and Cirac, J. Ignacio},
  year = 2022,
  month = nov,
  journal = {Physical Review Letters},
  volume = {129},
  number = {22},
  pages = {220401},
  issn = {0031-9007, 1079-7114},
  doi = {10.1103/PhysRevLett.129.220401},
  urldate = {2025-12-29},
  langid = {english}
}

@misc{weiKondoImpurityAttractive2025,
  title = {Kondo Impurity in an Attractive {{Fermi-Hubbard}} Bath: {{Equilibrium}} and Dynamics},
  shorttitle = {Kondo Impurity in an Attractive {{Fermi-Hubbard}} Bath},
  author = {Wei, Zhi-Yuan and Shi, Tao and Cirac, J. Ignacio and Demler, Eugene A.},
  year = 2025,
  publisher = {arXiv},
  doi = {10.48550/ARXIV.2501.05562},
  urldate = {2025-12-29},
  copyright = {arXiv.org perpetual, non-exclusive license}
}

@article{doi:10.1137/100788860,
    author = {Al-Mohy, Awad H. and Higham, Nicholas J.},
    title = {Computing the Action of the Matrix Exponential, with an Application to Exponential Integrators},
    journal = {SIAM Journal on Scientific Computing},
    volume = {33},
    number = {2},
    pages = {488-511},
    year = {2011},
    doi = {10.1137/100788860},
    URL = { 
        https://doi.org/10.1137/100788860
    },
    eprint = { 
        https://doi.org/10.1137/100788860
    }
}

@article{zhangDensityMatrixApproach1998,
  title = {Density {{Matrix Approach}} to {{Local Hilbert Space Reduction}}},
  author = {Zhang, Chunli and Jeckelmann, Eric and White, Steven R.},
  year = 1998,
  month = mar,
  journal = {Physical Review Letters},
  volume = {80},
  number = {12},
  pages = {2661--2664},
  issn = {0031-9007, 1079-7114},
  doi = {10.1103/PhysRevLett.80.2661},
  urldate = {2025-10-14},
  copyright = {http://link.aps.org/licenses/aps-default-license},
  langid = {english}
}

@article{whiteDensityMatrixFormulation1992,
  title = {Density Matrix Formulation for Quantum Renormalization Groups},
  author = {White, Steven R.},
  year = 1992,
  month = nov,
  journal = {Physical Review Letters},
  volume = {69},
  number = {19},
  pages = {2863--2866},
  issn = {0031-9007},
  doi = {10.1103/PhysRevLett.69.2863},
  urldate = {2025-12-26},
  copyright = {http://link.aps.org/licenses/aps-default-license},
  langid = {english}
}

@article{kochDynamicalLowRankApproximation2007,
  title = {Dynamical {{Low}}-{{Rank Approximation}}},
  author = {Koch, Othmar and Lubich, Christian},
  year = 2007,
  month = jan,
  journal = {SIAM Journal on Matrix Analysis and Applications},
  volume = {29},
  number = {2},
  pages = {434--454},
  issn = {0895-4798, 1095-7162},
  doi = {10.1137/050639703},
  urldate = {2025-12-26},
  langid = {english}
}

@article{stolppComparativeStudyStateoftheart2021,
  title = {Comparative Study of State-of-the-Art Matrix-Product-State Methods for Lattice Models with Large Local {{Hilbert}} Spaces without {{U}}(1) Symmetry},
  author = {Stolpp, Jan and K{\"o}hler, Thomas and Manmana, Salvatore R. and Jeckelmann, Eric and {Heidrich-Meisner}, Fabian and Paeckel, Sebastian},
  year = 2021,
  month = dec,
  journal = {Computer Physics Communications},
  volume = {269},
  pages = {108106},
  issn = {00104655},
  doi = {10.1016/j.cpc.2021.108106},
  urldate = {2025-12-26},
  langid = {english}
}

@article{guoCriticalStrongCouplingPhases2012,
  title = {Critical and {{Strong-Coupling Phases}} in {{One-}} and {{Two-Bath Spin-Boson Models}}},
  author = {Guo, Cheng and Weichselbaum, Andreas and {von~Delft}, Jan and Vojta, Matthias},
  year = 2012,
  month = apr,
  journal = {Physical Review Letters},
  volume = {108},
  number = {16},
  pages = {160401},
  issn = {0031-9007, 1079-7114},
  doi = {10.1103/PhysRevLett.108.160401},
  urldate = {2025-11-13},
  copyright = {http://link.aps.org/licenses/aps-default-license},
  langid = {english}
}

@article{brocktMatrixproductstateMethodDynamical2015,
  title = {Matrix-Product-State Method with a Dynamical Local Basis Optimization for Bosonic Systems out of Equilibrium},
  author = {Brockt, C. and Dorfner, F. and Vidmar, L. and {Heidrich-Meisner}, F. and Jeckelmann, E.},
  year = 2015,
  month = dec,
  journal = {Physical Review B},
  volume = {92},
  number = {24},
  pages = {241106},
  issn = {1098-0121, 1550-235X},
  doi = {10.1103/PhysRevB.92.241106},
  urldate = {2025-12-29},
  copyright = {http://link.aps.org/licenses/aps-default-license},
  langid = {english}
}

@article{schollwoeckDensitymatrixRenormalizationGroup2011,
  title = {The Density-Matrix Renormalization Group in the Age of Matrix Product States},
  author = {Schollwoeck, Ulrich},
  year = 2011,
  month = jan,
  journal = {Annals of Physics},
  volume = {326},
  number = {1},
  eprint = {1008.3477},
  primaryclass = {cond-mat},
  pages = {96--192},
  issn = {00034916},
  doi = {10.1016/j.aop.2010.09.012},
  urldate = {2022-10-08},
  archiveprefix = {arXiv},
  langid = {english}
}

@article{tangInfluenceExtendedInteractions2024,
  title = {Influence of Extended Interactions on Spin Dynamics in One-Dimensional Cuprates},
  author = {Tang, Ta and Jost, Daniel and Moritz, Brian and Devereaux, Thomas P.},
  year = 2024,
  month = oct,
  journal = {Physical Review B},
  volume = {110},
  number = {16},
  pages = {165118},
  issn = {2469-9950, 2469-9969},
  doi = {10.1103/PhysRevB.110.165118},
  urldate = {2025-02-26},
  langid = {english}
}

@article{evenblyGaugeFixingCanonical2018,
  title = {Gauge Fixing, Canonical Forms, and Optimal Truncations in Tensor Networks with Closed Loops},
  author = {Evenbly, Glen},
  year = 2018,
  month = aug,
  journal = {Physical Review B},
  volume = {98},
  number = {8},
  pages = {085155},
  issn = {2469-9950, 2469-9969},
  doi = {10.1103/PhysRevB.98.085155},
  urldate = {2026-01-09},
  langid = {english}
}

@misc{verstraeteRenormalizationAlgorithmsQuantumMany2004,
  title = {Renormalization Algorithms for {{Quantum-Many Body Systems}} in Two and Higher Dimensions},
  author = {Verstraete, F. and Cirac, J. I.},
  year = 2004,
  month = jul,
  number = {arXiv:cond-mat/0407066},
  eprint = {cond-mat/0407066},
  publisher = {arXiv},
  doi = {10.48550/arXiv.cond-mat/0407066},
  urldate = {2026-01-14},
  archiveprefix = {arXiv},
  langid = {english}
}

@article{dagottoCorrelatedElectronsHightemperature1994,
  title = {Correlated Electrons in High-Temperature Superconductors},
  author = {Dagotto, Elbio},
  year = 1994,
  month = jul,
  journal = {Reviews of Modern Physics},
  volume = {66},
  number = {3},
  pages = {763--840},
  issn = {0034-6861, 1539-0756},
  doi = {10.1103/RevModPhys.66.763},
  urldate = {2024-07-13},
  copyright = {http://link.aps.org/licenses/aps-default-license},
  langid = {english}
}

@article{keimerQuantumMatterHightemperature2015,
  title = {From Quantum Matter to High-Temperature Superconductivity in Copper Oxides},
  author = {Keimer, B. and Kivelson, S. A. and Norman, M. R. and Uchida, S. and Zaanen, J.},
  year = 2015,
  month = feb,
  journal = {Nature},
  volume = {518},
  number = {7538},
  pages = {179--186},
  issn = {0028-0836, 1476-4687},
  doi = {10.1038/nature14165},
  urldate = {2025-12-23},
  langid = {english}
}

@article{qinHubbardModelComputational2022,
  title = {The {{Hubbard Model}}: {{A Computational Perspective}}},
  shorttitle = {The {{Hubbard Model}}},
  author = {Qin, Mingpu and Sch{\"a}fer, Thomas and Andergassen, Sabine and Corboz, Philippe and Gull, Emanuel},
  year = {2022},
  month = mar,
  journal = {Annual Review of Condensed Matter Physics},
  volume = {13},
  number = {1},
  pages = {275--302},
  issn = {1947-5454, 1947-5462},
  doi = {10.1146/annurev-conmatphys-090921-033948},
  urldate = {2024-07-13},
  langid = {english}
}

@article{arovasHubbardModel2022,
  title = {The {{Hubbard Model}}},
  author = {Arovas, Daniel P. and Berg, Erez and Kivelson, Steven A. and Raghu, Srinivas},
  year = {2022},
  month = mar,
  journal = {Annual Review of Condensed Matter Physics},
  volume = {13},
  number = {1},
  pages = {239--274},
  issn = {1947-5454, 1947-5462},
  doi = {10.1146/annurev-conmatphys-031620-102024},
  urldate = {2025-04-08},
  langid = {english}
}

@article{jiangGroundStatePhase2020,
  title = {Ground State Phase Diagram of the Doped {{Hubbard}} Model on the Four-Leg Cylinder},
  author = {Jiang, Yi-Fan and Zaanen, Jan and Devereaux, Thomas P. and Jiang, Hong-Chen},
  year = {2020},
  month = jul,
  journal = {Physical Review Research},
  volume = {2},
  number = {3},
  pages = {033073},
  issn = {2643-1564},
  doi = {10.1103/PhysRevResearch.2.033073},
  urldate = {2025-03-25},
  langid = {english}
}

@article{chungPlaquetteOrdinaryWave2020,
  title = {Plaquette versus Ordinary d -Wave Pairing in the t {$\prime$} -{{Hubbard}} Model on a Width-4 Cylinder},
  author = {Chung, Chia-Min and Qin, Mingpu and Zhang, Shiwei and Schollw{\"o}ck, Ulrich and White, Steven R. and {The Simons Collaboration on the Many-Electron Problem}},
  year = {2020},
  month = jul,
  journal = {Physical Review B},
  volume = {102},
  number = {4},
  pages = {041106},
  issn = {2469-9950, 2469-9969},
  doi = {10.1103/PhysRevB.102.041106},
  urldate = {2025-05-12},
  langid = {english}
}

@article{zhengStripeOrderUnderdoped2017,
  title = {Stripe Order in the Underdoped Region of the Two-Dimensional {{Hubbard}} Model},
  author = {Zheng, Bo-Xiao and Chung, Chia-Min and Corboz, Philippe and Ehlers, Georg and Qin, Ming-Pu and Noack, Reinhard M. and Shi, Hao and White, Steven R. and Zhang, Shiwei and Chan, Garnet Kin-Lic},
  year = {2017},
  month = dec,
  journal = {Science},
  volume = {358},
  number = {6367},
  pages = {1155--1160},
  issn = {0036-8075, 1095-9203},
  doi = {10.1126/science.aam7127},
  urldate = {2025-04-11},
  langid = {english}
}

@article{tranquadaCuprateSuperconductorsViewed2020,
  title = {Cuprate Superconductors as Viewed through a Striped Lens},
  author = {Tranquada, J. M.},
  year = {2020},
  month = oct,
  journal = {Advances in Physics},
  volume = {69},
  number = {4},
  pages = {437--509},
  issn = {0001-8732, 1460-6976},
  doi = {10.1080/00018732.2021.1935698},
  urldate = {2025-06-14},
  langid = {english}
}

@article{zaanenCURRENTIDEASORIGIN1998,
  title = {{{CURRENT IDEAS ON THE ORIGIN OF STRIPES}}},
  author = {Zaanen, J},
  year = {1998},
  month = oct,
  journal = {Journal of Physics and Chemistry of Solids},
  volume = {59},
  number = {10-12},
  pages = {1769--1773},
  issn = {00223697},
  doi = {10.1016/S0022-3697(98)00106-1},
  urldate = {2025-09-26},
  copyright = {https://www.elsevier.com/tdm/userlicense/1.0/},
  langid = {english}
}

@article{zhangPseudospinSymmetryNew1990,
  title = {Pseudospin Symmetry and New Collective Modes of the {{Hubbard}} Model},
  author = {Zhang, Shoucheng},
  year = {1990},
  month = jul,
  journal = {Physical Review Letters},
  volume = {65},
  number = {1},
  pages = {120--122},
  issn = {0031-9007},
  doi = {10.1103/PhysRevLett.65.120},
  urldate = {2025-09-22},
  copyright = {http://link.aps.org/licenses/aps-default-license},
  langid = {english}
}

@article{yangPairingOffdiagonalLongrange1989,
  title = {{\emph{{$\eta$}}} Pairing and Off-Diagonal Long-Range Order in a {{Hubbard}} Model},
  author = {Yang, Chen Ning},
  year = {1989},
  month = nov,
  journal = {Physical Review Letters},
  volume = {63},
  number = {19},
  pages = {2144--2147},
  issn = {0031-9007},
  doi = {10.1103/PhysRevLett.63.2144},
  urldate = {2025-09-22},
  copyright = {http://link.aps.org/licenses/aps-default-license},
  langid = {english}
}

@article{CINI2001451,
    title = {Exact ground state of the two-dimensional Hubbard model at half-filling for U=0+},
    journal = {Solid State Communications},
    volume = {117},
    number = {8},
    pages = {451-454},
    year = {2001},
    issn = {0038-1098},
    doi = {https://doi.org/10.1016/S0038-1098(00)00504-4},
    url = {https://www.sciencedirect.com/science/article/pii/S0038109800005044},
    author = {Michele Cini and Gianluca Stefanucci}
}

@article{huangStripeOrderPerspective2018,
  title = {Stripe Order from the Perspective of the {{Hubbard}} Model},
  author = {Huang, Edwin W. and Mendl, Christian B. and Jiang, Hong-Chen and Moritz, Brian and Devereaux, Thomas P.},
  year = 2018,
  month = apr,
  journal = {npj Quantum Materials},
  volume = {3},
  number = {1},
  pages = {22},
  issn = {2397-4648},
  doi = {10.1038/s41535-018-0097-0},
  urldate = {2025-12-25},
  langid = {english}
}

@article{jiangSuperconductivityDopedHubbard2019,
  title = {Superconductivity in the Doped {{Hubbard}} Model and Its Interplay with Next-Nearest Hopping {\emph{t}} {$\prime$}},
  author = {Jiang, Hong-Chen and Devereaux, Thomas P.},
  year = 2019,
  month = sep,
  journal = {Science},
  volume = {365},
  number = {6460},
  pages = {1424--1428},
  issn = {0036-8075, 1095-9203},
  doi = {10.1126/science.aal5304},
  urldate = {2024-11-29},
  langid = {english}
}

@article{huangStrangeMetallicityDoped2019,
  title = {Strange Metallicity in the Doped {{Hubbard}} Model},
  author = {Huang, Edwin W. and Sheppard, Ryan and Moritz, Brian and Devereaux, Thomas P.},
  year = 2019,
  month = nov,
  journal = {Science},
  volume = {366},
  number = {6468},
  pages = {987--990},
  issn = {0036-8075, 1095-9203},
  doi = {10.1126/science.aau7063},
  urldate = {2025-12-25},
  langid = {english}
}

@article{kokaljBadmetallicBehaviorDoped2017,
  title = {Bad-Metallic Behavior of Doped {{Mott}} Insulators},
  author = {Kokalj, Jure},
  year = 2017,
  month = jan,
  journal = {Physical Review B},
  volume = {95},
  number = {4},
  pages = {041110},
  issn = {2469-9950, 2469-9969},
  doi = {10.1103/PhysRevB.95.041110},
  urldate = {2025-12-25},
  copyright = {http://link.aps.org/licenses/aps-default-license},
  langid = {english}
}

@article{xuCoexistenceSuperconductivityPartially2024,
  title = {Coexistence of Superconductivity with Partially Filled Stripes in the {{Hubbard}} Model},
  author = {Xu, Hao and Chung, Chia-Min and Qin, Mingpu and Schollw{\"o}ck, Ulrich and White, Steven R. and Zhang, Shiwei},
  year = 2024,
  month = may,
  journal = {Science},
  volume = {384},
  number = {6696},
  pages = {eadh7691},
  issn = {0036-8075, 1095-9203},
  doi = {10.1126/science.adh7691},
  urldate = {2024-11-29},
  langid = {english}
}

@article{liTangentSpaceApproach2023,
  title = {Tangent {{Space Approach}} for {{Thermal Tensor Network Simulations}} of the {{2D Hubbard Model}}},
  author = {Li, Qiaoyi and Gao, Yuan and He, Yuan-Yao and Qi, Yang and Chen, Bin-Bin and Li, Wei},
  year = 2023,
  month = jun,
  journal = {Physical Review Letters},
  volume = {130},
  number = {22},
  pages = {226502},
  issn = {0031-9007, 1079-7114},
  doi = {10.1103/PhysRevLett.130.226502},
  urldate = {2025-12-26},
  langid = {english}
}

@article{gunnarssonBreakdownTraditionalManyBody2017,
  title = {Breakdown of {{Traditional Many-Body Theories}} for {{Correlated Electrons}}},
  author = {Gunnarsson, O. and Rohringer, G. and Sch{\"a}fer, T. and Sangiovanni, G. and Toschi, A.},
  year = 2017,
  month = aug,
  journal = {Physical Review Letters},
  volume = {119},
  number = {5},
  pages = {056402},
  issn = {0031-9007, 1079-7114},
  doi = {10.1103/PhysRevLett.119.056402},
  urldate = {2025-12-26},
  copyright = {http://link.aps.org/licenses/aps-default-license},
  langid = {english}
}

@article{kozikNonexistenceLuttingerWardFunctional2015,
  title = {Nonexistence of the {{Luttinger-Ward Functional}} and {{Misleading Convergence}} of {{Skeleton Diagrammatic Series}} for {{Hubbard-Like Models}}},
  author = {Kozik, Evgeny and Ferrero, Michel and Georges, Antoine},
  year = 2015,
  month = apr,
  journal = {Physical Review Letters},
  volume = {114},
  number = {15},
  pages = {156402},
  issn = {0031-9007, 1079-7114},
  doi = {10.1103/PhysRevLett.114.156402},
  urldate = {2025-12-28},
  copyright = {http://link.aps.org/licenses/aps-default-license},
  langid = {english}
}

@article{reitnerAttractiveEffectStrong2020,
  title = {Attractive {{Effect}} of a {{Strong Electronic Repulsion}}: {{The Physics}} of {{Vertex Divergences}}},
  shorttitle = {Attractive {{Effect}} of a {{Strong Electronic Repulsion}}},
  author = {Reitner, M. and Chalupa, P. and Del Re, L. and Springer, D. and Ciuchi, S. and Sangiovanni, G. and Toschi, A.},
  year = 2020,
  month = nov,
  journal = {Physical Review Letters},
  volume = {125},
  number = {19},
  pages = {196403},
  issn = {0031-9007, 1079-7114},
  doi = {10.1103/PhysRevLett.125.196403},
  urldate = {2025-12-29},
  langid = {english}
}

@article{leblancSolutionsTwoDimensionalHubbard2015,
  title = {Solutions of the {{Two-Dimensional Hubbard Model}}: {{Benchmarks}} and {{Results}} from a {{Wide Range}} of {{Numerical Algorithms}}},
  shorttitle = {Solutions of the {{Two-Dimensional Hubbard Model}}},
  author = {LeBlanc, J. P. F. and Antipov, Andrey E. and Becca, Federico and Bulik, Ireneusz W. and Chan, Garnet Kin-Lic and Chung, Chia-Min and Deng, Youjin and Ferrero, Michel and Henderson, Thomas M. and {Jim{\'e}nez-Hoyos}, Carlos A. and Kozik, E. and Liu, Xuan-Wen and Millis, Andrew J. and Prokof'ev, N. V. and Qin, Mingpu and Scuseria, Gustavo E. and Shi, Hao and Svistunov, B. V. and Tocchio, Luca F. and Tupitsyn, I. S. and White, Steven R. and Zhang, Shiwei and Zheng, Bo-Xiao and Zhu, Zhenyue and Gull, Emanuel and {Simons Collaboration on the Many-Electron Problem}},
  year = 2015,
  month = dec,
  journal = {Physical Review X},
  volume = {5},
  number = {4},
  pages = {041041},
  issn = {2160-3308},
  doi = {10.1103/PhysRevX.5.041041},
  urldate = {2025-12-29},
  copyright = {http://creativecommons.org/licenses/by/3.0/},
  langid = {english}
}

@article{corbozStripesTwodimensionalModel2011,
  title = {Stripes in the Two-Dimensional t - {{J}} Model with Infinite Projected Entangled-Pair States},
  author = {Corboz, Philippe and White, Steven R. and Vidal, Guifr{\'e} and Troyer, Matthias},
  year = 2011,
  month = jul,
  journal = {Physical Review B},
  volume = {84},
  number = {4},
  pages = {041108},
  issn = {1098-0121, 1550-235X},
  doi = {10.1103/PhysRevB.84.041108},
  urldate = {2025-12-29},
  copyright = {http://link.aps.org/licenses/aps-default-license},
  langid = {english}
}

@article{changSpinChargeOrder2010,
  title = {Spin and {{Charge Order}} in the {{Doped Hubbard Model}}: {{Long-Wavelength Collective Modes}}},
  shorttitle = {Spin and {{Charge Order}} in the {{Doped Hubbard Model}}},
  author = {Chang, Chia-Chen and Zhang, Shiwei},
  year = 2010,
  month = mar,
  journal = {Physical Review Letters},
  volume = {104},
  number = {11},
  pages = {116402},
  issn = {0031-9007, 1079-7114},
  doi = {10.1103/PhysRevLett.104.116402},
  urldate = {2025-12-29},
  copyright = {http://link.aps.org/licenses/aps-default-license},
  langid = {english}
}

@article{darmawanStripeSuperconductingOrder2018,
  title = {Stripe and Superconducting Order Competing in the {{Hubbard}} Model on a Square Lattice Studied by a Combined Variational {{Monte Carlo}} and Tensor Network Method},
  author = {Darmawan, Andrew S. and Nomura, Yusuke and Yamaji, Youhei and Imada, Masatoshi},
  year = 2018,
  month = nov,
  journal = {Physical Review B},
  volume = {98},
  number = {20},
  pages = {205132},
  issn = {2469-9950, 2469-9969},
  doi = {10.1103/PhysRevB.98.205132},
  urldate = {2025-12-29},
  langid = {english}
}

@article{pavariniBandStructureTrendHoleDoped2001,
  title = {Band-{{Structure Trend}} in {{Hole-Doped Cuprates}} and {{Correlation}} with {{T}} c Max},
  author = {Pavarini, E. and Dasgupta, I. and {Saha-Dasgupta}, T. and Jepsen, O. and Andersen, O. K.},
  year = 2001,
  month = jul,
  journal = {Physical Review Letters},
  volume = {87},
  number = {4},
  pages = {047003},
  issn = {0031-9007, 1079-7114},
  doi = {10.1103/PhysRevLett.87.047003},
  urldate = {2026-01-15},
  copyright = {http://link.aps.org/licenses/aps-default-license},
  langid = {english}
}

@article{linTwodimensionalHubbardModel1987,
  title = {Two-Dimensional {{Hubbard}} Model with Nearest- and next-Nearest-Neighbor Hopping},
  author = {Lin, H. Q. and Hirsch, J. E.},
  year = 1987,
  month = mar,
  journal = {Physical Review B},
  volume = {35},
  number = {7},
  pages = {3359--3368},
  issn = {0163-1829},
  doi = {10.1103/PhysRevB.35.3359},
  urldate = {2026-01-15},
  copyright = {http://link.aps.org/licenses/aps-default-license},
  langid = {english}
}

@article{quSpintripletPairingInduced2022,
  title = {Spin-Triplet Pairing Induced by near-Neighbor Attraction in the Extended {{Hubbard}} Model for Cuprate Chain},
  author = {Qu, Dai-Wei and Chen, Bin-Bin and Jiang, Hong-Chen and Wang, Yao and Li, Wei},
  year = 2022,
  month = oct,
  journal = {Communications Physics},
  volume = {5},
  number = {1},
  pages = {257},
  issn = {2399-3650},
  doi = {10.1038/s42005-022-01030-x},
  urldate = {2024-07-01},
  langid = {english}
}

@article{karakuzuStripeCorrelationsTwodimensional2022,
  title = {Stripe Correlations in the Two-Dimensional {{Hubbard-Holstein}} Model},
  author = {Karakuzu, Seher and Tanjaroon Ly, Andy and Mai, Peizhi and Neuhaus, James and Maier, Thomas A. and Johnston, Steven},
  year = {2022},
  month = dec,
  journal = {Communications Physics},
  volume = {5},
  number = {1},
  pages = {311},
  issn = {2399-3650},
  doi = {10.1038/s42005-022-01092-x},
  urldate = {2025-06-14},
  langid = {english}
}

@article{zacharStripesFormationAntiphase2002,
  title = {Stripes Formation: {{Antiphase}} and in-Phase Domain Walls},
  shorttitle = {Stripes Formation},
  author = {Zachar, Oron},
  year = {2002},
  month = apr,
  journal = {Physical Review B},
  volume = {65},
  number = {17},
  pages = {174411},
  issn = {0163-1829, 1095-3795},
  doi = {10.1103/PhysRevB.65.174411},
  urldate = {2025-06-18},
  copyright = {http://link.aps.org/licenses/aps-default-license},
  langid = {english}
}

@article{lanzaraEvidenceUbiquitousStrong2001,
  title = {Evidence for Ubiquitous Strong Electron--Phonon Coupling in High-Temperature Superconductors},
  author = {Lanzara, A. and Bogdanov, P. V. and Zhou, X. J. and Kellar, S. A. and Feng, D. L. and Lu, E. D. and Yoshida, T. and Eisaki, H. and Fujimori, A. and Kishio, K. and Shimoyama, J.-I. and Noda, T. and Uchida, S. and Hussain, Z. and Shen, Z.-X.},
  year = {2001},
  month = aug,
  journal = {Nature},
  volume = {412},
  number = {6846},
  pages = {510--514},
  issn = {1476-4687},
  doi = {10.1038/35087518}
}

@article{reznikElectronPhononCoupling2006,
  title = {Electron--Phonon Coupling Reflecting Dynamic Charge Inhomogeneity in Copper Oxide Superconductors},
  author = {Reznik, D. and Pintschovius, L. and Ito, M. and Iikubo, S. and Sato, M. and Goka, H. and Fujita, M. and Yamada, K. and Gu, G. D. and Tranquada, J. M.},
  year = {2006},
  month = apr,
  journal = {Nature},
  volume = {440},
  number = {7088},
  pages = {1170--1173},
  issn = {0028-0836, 1476-4687},
  doi = {10.1038/nature04704},
  urldate = {2025-09-11},
  copyright = {http://www.springer.com/tdm},
  langid = {english}
}

@article{leeInterplayElectronLattice2006,
  title = {Interplay of Electron--Lattice Interactions and Superconductivity in {{Bi2Sr2CaCu2O8}}+{$\delta$}},
  author = {Lee, Jinho and Fujita, K. and McElroy, K. and Slezak, J. A. and Wang, M. and Aiura, Y. and Bando, H. and Ishikado, M. and Masui, T. and Zhu, J.-X. and Balatsky, A. V. and Eisaki, H. and Uchida, S. and Davis, J. C.},
  year = {2006},
  month = aug,
  journal = {Nature},
  volume = {442},
  number = {7102},
  pages = {546--550},
  issn = {0028-0836, 1476-4687},
  doi = {10.1038/nature04973},
  urldate = {2025-09-11},
  copyright = {http://www.springer.com/tdm},
  langid = {english}
}

@article{shenMissingQuasiparticlesChemical2004,
  title = {Missing {{Quasiparticles}} and the {{Chemical Potential Puzzle}} in the {{Doping Evolution}} of the {{Cuprate Superconductors}}},
  author = {Shen, K. M. and Ronning, F. and Lu, D. H. and Lee, W. S. and Ingle, N. J. C. and Meevasana, W. and Baumberger, F. and Damascelli, A. and Armitage, N. P. and Miller, L. L. and Kohsaka, Y. and Azuma, M. and Takano, M. and Takagi, H. and Shen, Z.-X.},
  year = {2004},
  month = dec,
  journal = {Physical Review Letters},
  volume = {93},
  number = {26},
  pages = {267002},
  issn = {0031-9007, 1079-7114},
  doi = {10.1103/PhysRevLett.93.267002},
  urldate = {2025-09-11},
  copyright = {http://link.aps.org/licenses/aps-default-license},
  langid = {english}
}

@article{tallonIsotopeEffectSuperfluid2005,
  title = {Isotope {{Effect}} in the {{Superfluid Density}} of {{High-Temperature Superconducting Cuprates}}: {{Stripes}}, {{Pseudogap}}, and {{Impurities}}},
  shorttitle = {Isotope {{Effect}} in the {{Superfluid Density}} of {{High-Temperature Superconducting Cuprates}}},
  author = {Tallon, J. L. and Islam, R. S. and Storey, J. and Williams, G. V. M. and Cooper, J. R.},
  year = 2005,
  month = jun,
  journal = {Physical Review Letters},
  volume = {94},
  number = {23},
  pages = {237002},
  issn = {0031-9007, 1079-7114},
  doi = {10.1103/PhysRevLett.94.237002},
  urldate = {2025-12-23},
  copyright = {http://link.aps.org/licenses/aps-default-license},
  langid = {english}
}

@article{cukReviewElectronPhonon2005,
  title = {A Review of Electron--Phonon Coupling Seen in the High- {{{\emph{T}}}}{\textsubscript{c}} Superconductors by Angle-resolved Photoemission Studies ({{ARPES}})},
  author = {Cuk, T. and Lu, D. H. and Zhou, X. J. and Shen, Z.-X. and Devereaux, T. P. and Nagaosa, N.},
  year = 2005,
  month = jan,
  journal = {physica status solidi (b)},
  volume = {242},
  number = {1},
  pages = {11--29},
  issn = {0370-1972, 1521-3951},
  doi = {10.1002/pssb.200404959},
  urldate = {2025-12-23},
  copyright = {http://onlinelibrary.wiley.com/termsAndConditions\#vor},
  langid = {english}
}

@article{heRapidChangeSuperconductivity2018,
  title = {Rapid Change of Superconductivity and Electron-Phonon Coupling through Critical Doping in {{Bi-2212}}},
  author = {He, Y. and Hashimoto, M. and Song, D. and Chen, S.-D. and He, J. and Vishik, I. M. and Moritz, B. and Lee, D.-H. and Nagaosa, N. and Zaanen, J. and Devereaux, T. P. and Yoshida, Y. and Eisaki, H. and Lu, D. H. and Shen, Z.-X.},
  year = 2018,
  month = oct,
  journal = {Science},
  volume = {362},
  number = {6410},
  pages = {62--65},
  issn = {0036-8075, 1095-9203},
  doi = {10.1126/science.aar3394},
  urldate = {2025-12-23},
  langid = {english}
}

@article{thomsenUntwinnedSingleCrystals1988,
  title = {Untwinned Single Crystals of {{Y Ba}} 2 {{Cu}} 3 {{O}} 7 - {$\delta$} : {{An}} Optical Investigation of the a - b Anisotropy},
  shorttitle = {Untwinned Single Crystals of {{Y Ba}} 2 {{Cu}} 3 {{O}} 7 - {$\delta$}},
  author = {Thomsen, C. and Cardona, M. and Gegenheimer, B. and Liu, R. and Simon, A.},
  year = 1988,
  month = jun,
  journal = {Physical Review B},
  volume = {37},
  number = {16},
  pages = {9860--9863},
  issn = {0163-1829},
  doi = {10.1103/PhysRevB.37.9860},
  urldate = {2025-12-23},
  copyright = {http://link.aps.org/licenses/aps-default-license},
  langid = {english}
}

@article{faustiLightInducedSuperconductivityStripeOrdered2011,
  title = {Light-{{Induced Superconductivity}} in a {{Stripe-Ordered Cuprate}}},
  author = {Fausti, D. and Tobey, R. I. and Dean, N. and Kaiser, S. and Dienst, A. and Hoffmann, M. C. and Pyon, S. and Takayama, T. and Takagi, H. and Cavalleri, A.},
  year = 2011,
  month = jan,
  journal = {Science},
  volume = {331},
  number = {6014},
  pages = {189--191},
  issn = {0036-8075, 1095-9203},
  doi = {10.1126/science.1197294},
  urldate = {2025-12-23},
  langid = {english}
}

@article{buzziPhotomolecularHighTemperatureSuperconductivity2020,
  title = {Photomolecular {{High-Temperature Superconductivity}}},
  author = {Buzzi, M. and Nicoletti, D. and Fechner, M. and {Tancogne-Dejean}, N. and Sentef, M. A. and Georges, A. and Biesner, T. and Uykur, E. and Dressel, M. and Henderson, A. and Siegrist, T. and Schlueter, J. A. and Miyagawa, K. and Kanoda, K. and Nam, M.-S. and Ardavan, A. and Coulthard, J. and Tindall, J. and Schlawin, F. and Jaksch, D. and Cavalleri, A.},
  year = 2020,
  month = aug,
  journal = {Physical Review X},
  volume = {10},
  number = {3},
  pages = {031028},
  issn = {2160-3308},
  doi = {10.1103/PhysRevX.10.031028},
  urldate = {2025-12-23},
  langid = {english}
}

@article{leeInterfacialModeCoupling2014,
  title = {Interfacial Mode Coupling as the Origin of the Enhancement of {{Tc}} in {{FeSe}} Films on {{SrTiO3}}},
  author = {Lee, J. J. and Schmitt, F. T. and Moore, R. G. and Johnston, S. and Cui, Y.-T. and Li, W. and Yi, M. and Liu, Z. K. and Hashimoto, M. and Zhang, Y. and Lu, D. H. and Devereaux, T. P. and Lee, D.-H. and Shen, Z.-X.},
  year = 2014,
  month = nov,
  journal = {Nature},
  volume = {515},
  number = {7526},
  pages = {245--248},
  issn = {0028-0836, 1476-4687},
  doi = {10.1038/nature13894},
  urldate = {2025-12-23},
  langid = {english}
}

@article{talantsevDebyeTemperatureElectronphonon,
  title = {Debye Temperature, Electron-Phonon Coupling Constant, and Three-Dome Shape of Crystalline Strain as a Function of Pressure in Highly Compressed {La3Ni2O7-$\delta$}},
  author = {Talantsev, Evgeny Fedorovich and Chistyakov, Vasiliy Vladimirovich},
  journal = {Letters on Materials},
  year = {2024},
  volume = {14},
  number = {3},
  pages = {262--268},
  doi = {10.48612/LETTERS/2024-3-262-268},
  urldate = {2025-12-23},
  langid = {english}
}

@article{huOpticallyEnhancedCoherent2014,
  title = {Optically Enhanced Coherent Transport in {{YBa2Cu3O6}}.5 by Ultrafast Redistribution of Interlayer Coupling},
  author = {Hu, W. and Kaiser, S. and Nicoletti, D. and Hunt, C. R. and Gierz, I. and Hoffmann, M. C. and Le Tacon, M. and Loew, T. and Keimer, B. and Cavalleri, A.},
  year = 2014,
  month = jul,
  journal = {Nature Materials},
  volume = {13},
  number = {7},
  pages = {705--711},
  issn = {1476-1122, 1476-4660},
  doi = {10.1038/nmat3963},
  urldate = {2025-12-24},
  langid = {english}
}

@article{zhangPhotoinducedMetastableState2018,
  title = {Photoinduced Metastable State with Modulated {{Josephson}} Coupling Strengths in {{Pr}} 0.88 {{LaCe}} 0.12 {{CuO}} 4},
  author = {Zhang, S. J. and Wang, Z. X. and Wu, D. and Liu, Q. M. and Shi, L. Y. and Lin, T. and Li, S. L. and Dai, P. C. and Dong, T. and Wang, N. L.},
  year = 2018,
  month = dec,
  journal = {Physical Review B},
  volume = {98},
  number = {22},
  pages = {224507},
  issn = {2469-9950, 2469-9969},
  doi = {10.1103/PhysRevB.98.224507},
  urldate = {2025-12-27},
  langid = {english}
}

@article{liDistinctUltrafastDynamics2025,
  title = {Distinct Ultrafast Dynamics of Bilayer and Trilayer Nickelate Superconductors Regarding the Density-Wave-like Transitions},
  author = {Li, Yidian and Cao, Yantao and Liu, Liangyang and Peng, Pai and Lin, Hao and Pei, Cuiying and Zhang, Mingxin and Wu, Heng and Du, Xian and Zhao, Wenxuan and Zhai, Kaiyi and Zhang, Xuefeng and Zhao, Jinkui and Lin, Miaoling and Tan, Pingheng and Qi, Yanpeng and Li, Gang and Guo, Hanjie and Yang, Luyi and Yang, Lexian},
  year = 2025,
  month = jan,
  journal = {Science Bulletin},
  volume = {70},
  number = {2},
  pages = {180--186},
  issn = {20959273},
  doi = {10.1016/j.scib.2024.10.011},
  urldate = {2025-12-26},
  langid = {english}
}

@article{sunTransientTrappingMetastable2020,
  title = {Transient {{Trapping}} into {{Metastable States}} in {{Systems}} with {{Competing Orders}}},
  author = {Sun, Zhiyuan and Millis, Andrew J.},
  year = 2020,
  month = may,
  journal = {Physical Review X},
  volume = {10},
  number = {2},
  pages = {021028},
  issn = {2160-3308},
  doi = {10.1103/PhysRevX.10.021028},
  urldate = {2025-12-26},
  langid = {english}
}

@article{zhangLightinducedNewCollective2018,
  title = {Light-Induced New Collective Modes in the Superconductor {{La}} 1.905 {{Ba}} 0.095 {{CuO}} 4},
  author = {Zhang, S. J. and Wang, Z. X. and Shi, L. Y. and Lin, T. and Zhang, M. Y. and Gu, G. D. and Dong, T. and Wang, N. L.},
  year = 2018,
  month = jul,
  journal = {Physical Review B},
  volume = {98},
  number = {2},
  pages = {020506},
  issn = {2469-9950, 2469-9969},
  doi = {10.1103/PhysRevB.98.020506},
  urldate = {2025-12-27},
  langid = {english}
}

@article{nicolettiMagneticFieldTuningLightInduced2018,
  title = {Magnetic-{{Field Tuning}} of {{Light-Induced Superconductivity}} in {{Striped La}} 2 - x {{Ba}} x {{CuO}} 4},
  author = {Nicoletti, D. and Fu, D. and Mehio, O. and Moore, S. and Disa, A. S. and Gu, G. D. and Cavalleri, A.},
  year = 2018,
  month = dec,
  journal = {Physical Review Letters},
  volume = {121},
  number = {26},
  pages = {267003},
  issn = {0031-9007, 1079-7114},
  doi = {10.1103/PhysRevLett.121.267003},
  urldate = {2025-12-28},
  langid = {english}
}

@article{creminPhotoenhancedMetastableCaxis2019,
  title = {Photoenhanced Metastable C-Axis Electrodynamics in Stripe-Ordered Cuprate {{La}}{\textsubscript{1.885}} {{Ba}}{\textsubscript{0.115}} {{CuO}}{\textsubscript{4}}},
  author = {Cremin, Kevin A. and Zhang, Jingdi and Homes, Christopher C. and Gu, G. D. and Sun, Zhiyuan and Fogler, Michael M. and Millis, Andrew J. and Basov, D. N. and Averitt, Richard D.},
  year = 2019,
  month = oct,
  journal = {Proceedings of the National Academy of Sciences},
  volume = {116},
  number = {40},
  pages = {19875--19879},
  issn = {0027-8424, 1091-6490},
  doi = {10.1073/pnas.1908368116},
  urldate = {2025-12-28},
  langid = {english}
}

@article{boschiniCollapseSuperconductivityCuprates2018,
  title = {Collapse of Superconductivity in Cuprates via Ultrafast Quenching of Phase Coherence},
  author = {Boschini, F. and Da Silva Neto, E. H. and Razzoli, E. and Zonno, M. and Peli, S. and Day, R. P. and Michiardi, M. and Schneider, M. and Zwartsenberg, B. and Nigge, P. and Zhong, R. D. and Schneeloch, J. and Gu, G. D. and Zhdanovich, S. and Mills, A. K. and Levy, G. and Jones, D. J. and Giannetti, C. and Damascelli, A.},
  year = 2018,
  month = may,
  journal = {Nature Materials},
  volume = {17},
  number = {5},
  pages = {416--420},
  issn = {1476-1122, 1476-4660},
  doi = {10.1038/s41563-018-0045-1},
  urldate = {2025-12-28},
  langid = {english}
}

@article{lemonikTransportSpectralSignatures2019,
  title = {Transport and Spectral Signatures of Transient Fluctuating Superfluids in the Absence of Long-Range Order},
  author = {Lemonik, Yonah and Mitra, Aditi},
  year = 2019,
  month = sep,
  journal = {Physical Review B},
  volume = {100},
  number = {9},
  pages = {094503},
  issn = {2469-9950, 2469-9969},
  doi = {10.1103/PhysRevB.100.094503},
  urldate = {2025-12-28},
  langid = {english}
}

@article{michaelParametricResonanceJosephson2020,
  title = {Parametric Resonance of {{Josephson}} Plasma Waves: {{A}} Theory for Optically Amplified Interlayer Superconductivity in {{YBa}} 2 {{Cu}} 3 {{O}} 6 + x},
  shorttitle = {Parametric Resonance of {{Josephson}} Plasma Waves},
  author = {Michael, Marios H. and Von Hoegen, Alexander and Fechner, Michael and F{\"o}rst, Michael and Cavalleri, Andrea and Demler, Eugene},
  year = 2020,
  month = nov,
  journal = {Physical Review B},
  volume = {102},
  number = {17},
  pages = {174505},
  issn = {2469-9950, 2469-9969},
  doi = {10.1103/PhysRevB.102.174505},
  urldate = {2025-12-28},
  langid = {english}
}

@article{patelLightinducedEnhancementSuperconductivity2016,
  title = {Light-Induced Enhancement of Superconductivity via Melting of Competing Bond-Density Wave Order in Underdoped Cuprates},
  author = {Patel, Aavishkar A. and Eberlein, Andreas},
  year = 2016,
  month = may,
  journal = {Physical Review B},
  volume = {93},
  number = {19},
  pages = {195139},
  issn = {2469-9950, 2469-9969},
  doi = {10.1103/PhysRevB.93.195139},
  urldate = {2025-12-28},
  copyright = {http://link.aps.org/licenses/aps-default-license},
  langid = {english}
}

@article{suzukiPhotoinducedPossibleSuperconducting2019,
  title = {Photoinduced Possible Superconducting State with Long-Lived Disproportionate Band Filling in {{FeSe}}},
  author = {Suzuki, Takeshi and Someya, Takashi and Hashimoto, Takahiro and Michimae, Shoya and Watanabe, Mari and Fujisawa, Masami and Kanai, Teruto and Ishii, Nobuhisa and Itatani, Jiro and Kasahara, Shigeru and Matsuda, Yuji and Shibauchi, Takasada and Okazaki, Kozo and Shin, Shik},
  year = 2019,
  month = sep,
  journal = {Communications Physics},
  volume = {2},
  number = {1},
  pages = {115},
  issn = {2399-3650},
  doi = {10.1038/s42005-019-0219-4},
  urldate = {2025-12-28},
  langid = {english}
}

@article{troyerComputationalComplexityFundamental2005,
  title = {Computational {{Complexity}} and {{Fundamental Limitations}} to {{Fermionic Quantum Monte Carlo Simulations}}},
  author = {Troyer, Matthias and Wiese, Uwe-Jens},
  year = 2005,
  month = may,
  journal = {Physical Review Letters},
  volume = {94},
  number = {17},
  pages = {170201},
  issn = {0031-9007, 1079-7114},
  doi = {10.1103/PhysRevLett.94.170201},
  urldate = {2026-01-14},
  copyright = {http://link.aps.org/licenses/aps-default-license},
  langid = {english}
}

@article{lohNUMERICALSTABILITYSIGN2005,
  title = {{{NUMERICAL STABILITY AND THE SIGN PROBLEM IN THE DETERMINANT QUANTUM MONTE CARLO METHOD}}},
  author = {Loh, E. Y. and Gubernatis, J. E. and Scalettar, R. T. and White, S. R. and Scalapino, D. J. and Sugar, R. L.},
  year = 2005,
  month = aug,
  journal = {International Journal of Modern Physics C},
  volume = {16},
  number = {08},
  pages = {1319--1327},
  issn = {0129-1831, 1793-6586},
  doi = {10.1142/S0129183105007911},
  urldate = {2025-12-26},
  langid = {english}
}

@article{blankenbeclerMonteCarloCalculations1981,
  title = {Monte {{Carlo}} Calculations of Coupled Boson-Fermion Systems. {{I}}},
  author = {Blankenbecler, R. and Scalapino, D. J. and Sugar, R. L.},
  year = 1981,
  month = oct,
  journal = {Physical Review D},
  volume = {24},
  number = {8},
  pages = {2278--2286},
  issn = {0556-2821},
  doi = {10.1103/PhysRevD.24.2278},
  urldate = {2025-12-26},
  copyright = {http://link.aps.org/licenses/aps-default-license},
  langid = {english}
}

@article{weberDirectedLoopQuantumMonte2017,
  title = {Directed-{{Loop Quantum Monte Carlo Method}} for {{Retarded Interactions}}},
  author = {Weber, Manuel and Assaad, Fakher F. and Hohenadler, Martin},
  year = 2017,
  month = aug,
  journal = {Physical Review Letters},
  volume = {119},
  number = {9},
  pages = {097401},
  issn = {0031-9007, 1079-7114},
  doi = {10.1103/PhysRevLett.119.097401},
  urldate = {2025-09-11},
  copyright = {https://link.aps.org/licenses/aps-default-license},
  langid = {english}
}

@article{itensor,
	title={{The ITensor Software Library for Tensor Network Calculations}},
	author={Matthew Fishman and Steven R. White and E. Miles Stoudenmire},
	journal={SciPost Phys. Codebases},
	pages={4},
	year={2022},
	publisher={SciPost},
	doi={10.21468/SciPostPhysCodeb.4},
	url={https://scipost.org/10.21468/SciPostPhysCodeb.4}
}

@article{Devos_TensorKit_jl_A_Julia_2025,
    author = {Devos, Lukas and Haegeman, Jutho},
    doi = {10.48550/arXiv.2508.10076},
    journal = {arXiv},
    title = {{TensorKit.jl: A Julia package for large-scale tensor computations, with a hint of category theory}},
    year = {2025}
}

@software{Li_FiniteMPS_jl,
    author = {Li, Qiaoyi},
    title = {{FiniteMPS.jl}},
    url = {https://github.com/Qiaoyi-Li/FiniteMPS.jl}
}

\end{document}